\documentclass[aps,pra,onecolumn,nofootinbib,a4paper]{revtex4}
\usepackage[hmargin=2cm,vmargin=2.5cm]{geometry}

\usepackage{amsmath,amssymb,amsthm}
\usepackage{bm}
\usepackage{graphicx}
\usepackage{xcolor}
\usepackage{array}
\usepackage{float} 
\usepackage{algorithm}
\usepackage{algpseudocode}
\algrenewcommand\algorithmicrequire{\textbf{Input:}}
\algrenewcommand\algorithmicensure{\textbf{Output:}}

\definecolor{refblue}{rgb}{0.05,0.25,0.55}
\usepackage[colorlinks=true,allcolors=refblue,bookmarksopen=true]{hyperref}

\newcolumntype{?}{!{\vrule width 1pt}}
\newcommand{\thickhline}{\noalign{\hrule height 1pt}}

  {\end{itemize}\endgroup\par\medskip}
\makeatletter
\def\@dotsep{0.6em}
\def\tocdim@min{3em}
\def\l@subsubsection#1#2{}
\makeatother

\newtheorem{proposition}{Proposition}

\newcommand{\E}{\mathbb{E}}
\newcommand{\R}{\mathbb{R}}
\newcommand{\diag}{\mathrm{diag}}
\newcommand{\x}{\bm{x}}
\newcommand{\1}{\mathbf{1}}
\newcommand{\cN}{\mathcal{N}}

\begin{document}

\title{Functional Tensor Trains for High-Order Greeks and Exposure Tails of Multi-Asset Structured Products}
\author{Javier Lopez-Piqueres}
\email[Corresponding author: ]{javier.lopezpiqueres@jpmchase.com}
\author{Atithi Acharya}
\author{Yue Sun}
\author{Niraj Kumar}
\email[]{niraj.x7.kumar@jpmchase.com}
\affiliation{Global Technology Applied Research, JPMorganChase, New York, NY}

\begin{abstract}
Risk management of worst-of autocallables needs price sensitivities (Greeks) up to third order and the potential future exposure, a tail quantile of the note's future value.
In Monte Carlo, each derivative multiplies the sampling noise by a factor that grows with the observation frequency, and in nested simulation the noise of the inner paths biases the quantile.
We instead solve the backward pricing recursion on an hp functional tensor train (hp-FTT) with polynomial pieces split at the barriers, so that worst-of barrier events, rank-one products of one-asset events, act exactly. Under geometric Brownian motion, Gaussian averaging lets these polynomials approximate the continuation value faster than exponentially.
One solve then gives spot Greeks at any spot without sampling noise, by integrating derivatives of the first-step density against the stored value, and its future-date surfaces replace inner simulation.
Against the strongest Monte Carlo estimators we could build or find published, on notes observed monthly on three to five assets, the order of the Greek and the accuracy decide which method is cheaper.
At $1\%$ accuracy Monte Carlo is more than ten times cheaper for deltas, second-order Greeks go either way, and hp-FTT is cheaper for both third-order Greeks tested, by $50$ to $400$ times on the mixed one.
At $0.1\%$ accuracy hp-FTT beats Monte Carlo across derivative orders under geometric Brownian motion and local volatility alike, by up to three orders of magnitude.
For the exposure alone Monte Carlo is cheaper at an error of about ten basis points of notional and hp-FTT at one, and a solve run for the Greeks yields it for little more than scenario sampling.
Sampling pays in variance for each derivative and hp-FTT in rank for each asset, so Monte Carlo keeps low-order Greeks at percent accuracy and rarely observed contracts, and each added asset narrows hp-FTT's lead.
\end{abstract}

\maketitle


\section{Introduction}
\label{sec:intro}

Equity-linked autocallables are structured products whose payoff depends on the paths of
several assets. Their annual volumes have been reported at over a hundred
billion dollars~\cite{Salon2019}, and they are growing: in the United States alone,
issuance averaged $28$ billion dollars a year over 2019--2023, nine times its level a
decade earlier~\cite{McCannYan2024}. An autocallable combines a bond with an option.
Like a bond, it pays coupons and repays the notional, but the investor also implicitly
sells the bank a put on the underlyings: its premium raises the coupons above those of
an ordinary bond, and in exchange the investor carries the risk of a
fall~\cite{DengMallettMcCann2011,CelerierVallee2017}. In the worst-of form, the weakest of
several stocks or indices decides everything: whether a coupon is paid, whether the
note is repaid early, and whether the investor takes a loss at maturity. Banks sell
these products as notes, certificates and deposits, in markets that include Europe
and Asia~\cite{BouzoubaaOsseiran2010,YoonJung2018,NguyenKoKwon2020}.

\begin{figure}[t]
\centering
\includegraphics[width=\linewidth]{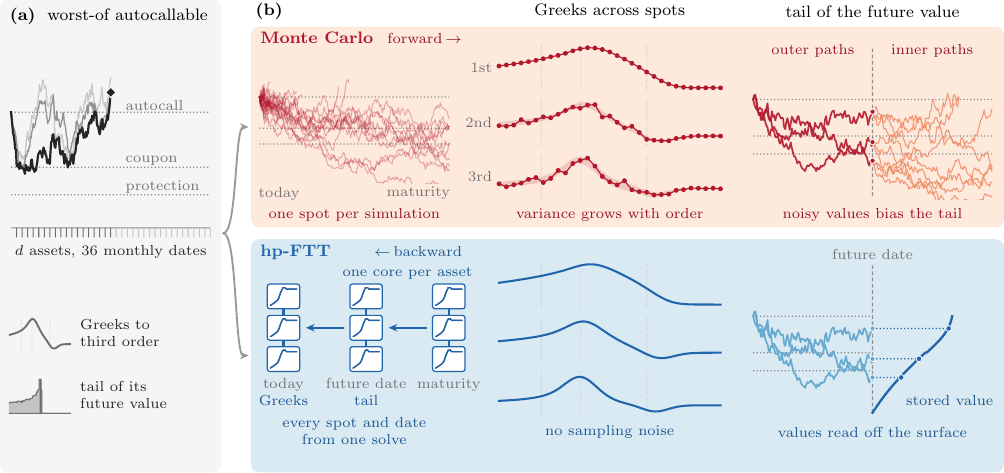}
\caption{\textbf{Two hard risk computations for a worst-of autocallable.}
\textbf{(a)} Simulated path of three assets (grey) and of their worst performer (black), on which the note misses two coupons and is called in month $18$. Dotted lines mark the autocall, coupon and
protection levels. Our focus here is on Greeks to third order and the tail of
its future value.
\textbf{(b)} Monte Carlo (top) samples paths forward, and each simulation answers for one
spot, whereas hp-FTT (bottom) solves backward once, from maturity to today, and stores the
continuation value at every date as a train of one-asset cores: the
train at today gives the Greeks at any spot, and the train at the future date the value
of every scenario. Middle: delta,
cross-gamma and cross-speed along the diagonal of spots (barriers dotted). Sampling
variance grows with the order of the Greek, while the read-out of the stored value has
none. (The Monte Carlo points are that read-out plus
Gaussian noise at our measured Monte Carlo standard error, for one solve's compute
shared among the spots shown.) Right: the note's value at a future date, scenario by
scenario, whose $97.5\%$ quantile is the potential future exposure. Both routes
simulate the scenarios up to that date, three of which are drawn. Monte Carlo values
each one with inner paths, whose noise biases the tail, and hp-FTT reads it off the
value stored for that date, drawn sideways along the diagonal of spots.}
\label{fig:overview}
\end{figure}

For the bank on the other side, these few terms make a hard risk problem. On every
observation date the note may pay, end or carry on, depending on the joint path of
several correlated assets (see Fig.~\ref{fig:overview}(a)). The bank hedges with the assets
and options on them, guided by the Greeks: delta sets the hedge, and to second order a
delta-hedged book gains or loses in proportion to its gammas and cross-gammas times the
gap between realized and implied variance and covariance~\cite{Bergomi2015}. On a
worst-of note the cross terms can weigh as much as the own ones. Near a barrier the
gammas themselves change quickly with the spots, and the third-order Greeks say how
quickly. Accuracy matters as well as order: a hedge sized on an estimated Greek leaves
the error of the estimate unhedged, so the accuracy a Greek needs is set by the risk the
bank is willing to leave open.

Beyond the hedge, market-risk capital and initial margin call for prices and Greeks at
many spots, future dates and stress scenarios~\cite{RuizZeronFRTB,RuizZeronDIM}.
Counterparty risk needs a second computation: whoever is owed the payments would lose
the note's value if the other side defaulted, so this risk is managed through a high
quantile of that value at each future date, the potential future
exposure~\cite{BCBS_CRE50,Gregory2020}. Risk systems compute these numbers for every note
in a book, every day, so whatever a single Greek or exposure costs is paid many times
over.

Monte Carlo simulation is the natural first tool~\cite{Glasserman2004}. Its error falls
as one over the square root of the number of paths whatever the number of assets,
randomized quasi-Monte Carlo can do better~\cite{WangSloan2005}, and a new model needs
only a new path generator. First derivatives remain within reach despite the jumps of
the payoff at the barriers, through likelihood-ratio weights~\cite{BroadieGlasserman1996},
or through pathwise derivatives and finite differences once the payoff is conditioned or
smoothed~\cite{FriesJoshi2011,AlmHarrachHarrachKeller2013,KosterRehmet2018,Rakhmonov2019,Hu2022}.
For prices and deltas at moderate accuracy, Monte Carlo is often the cheaper tool.

Higher-order Greeks tip the balance and expose the first of two weaknesses of sampling:
every derivative a sampler takes costs it \textit{variance}. Differentiating the payoff
runs into its jumps. Bumping and repricing amplifies the noise and leaves a bias that
more paths cannot remove, and pathwise, adjoint and vibrato derivatives need the jumps
smoothed or treated separately~\cite{Pages2016}. Even on paths conditioned to survive
each date, the pathwise delta of a worst-of note jumps where the worst performer
changes, and Hu, Xue and Yu~\cite{Hu2022} conclude that computing second-order Greeks
efficiently remains a challenging problem. None of the estimators we found treats the
third order. Differentiating the density of each path's first step instead, as the
likelihood-ratio estimator does, avoids the jumps but not the variance: its weight for a
Greek of order $k$ scales as $s^{-k}$, with $s$ the typical move of an asset over that
step, so the number of paths it needs grows as $s^{-2k}$, and frequent observation, or
the sub-steps of a model such as local volatility, make $s$ small. Conditioning on a
factor common to the assets helps only when there are few observation dates, since the
autocall couples the assets again on every date.

The second weakness is \textit{bias}. Many risk questions concern a whole function rather
than one number: the potential future exposure needs the note's value at every sampled
future state. Nested or multilevel simulation estimates that value with inner paths,
whose noise biases the quantile~\cite{GordyJuneja2010,Giles2008,GilesHajiAli2019} (see
Fig.~\ref{fig:overview}(b)), and regression avoids the inner paths but leaves an
approximation bias of its own~\cite{LongstaffSchwartz2001}. Both weaknesses have one
source: a sampler averages the cash flows of whole paths and never has the continuation
value, the value of the payments still to come as a function of the asset prices at each
date. A Greek's weight could be integrated against that function, and the value of each
sampled future state read off it.

A deterministic solver computes exactly this, carrying it backward from maturity: on a
grid~\cite{HilberReichmannSchwabWinter2013}, in Fourier space for suitable
payoffs~\cite{CarrMadan1999,FangOosterlee2008,RuijterOosterlee2012}, or in closed form
for special models~\cite{Guillaume2015a,Guillaume2015b,Tong2019}. For general models the
trouble is the grid: a hundred points per asset make a million values at three assets
and ten billion at five, beyond a laptop's memory. The central question of this paper is
whether a deterministic solver can beat sampling on high-order Greeks and exposure tails
without paying the exponential cost of a grid. Sparse
grids~\cite{ReisingerWittum2007} need far fewer points, but their error grows with the
mixed derivatives of the value~\cite{BungartzGriebel2004}. These are large near the
barriers, where the value changes steeply across a narrow \emph{layer} about as wide as
the typical move $s$ of an asset between dates. Adaptive refinement can follow such
features by adding points where the layers lie~\cite{BungartzGriebel2004}, but the number
of points still grows quickly with the number of assets.

The same exponential wall appears in various other domains, including quantum many-body physics, where tensor networks
are a way around it~\cite{Schollwock2011,Orus2014}. A tensor train, known to physicists
as a matrix product state, writes a function of $d$ variables as a product of $d$ small
matrices, one per variable~\cite{Oseledets2011,Oseledets2013}. In a functional tensor
train~\cite{GorodetskyKaramanMarzouk2019} the entries of each matrix are functions of its
variable, which we write as piecewise polynomials. At a fixed size of these matrices, the
rank, the storage grows only linearly with $d$. The rank measures
how strongly the variables on either side of a cut depend on each other, the analogue of
entanglement, and for a worst-of note, whose payoff ties all the assets together, it is
far from obvious that it stays small. Three features of the problem make it tractable,
and our solver is built around them.

First, the weakest asset is above a barrier exactly when every asset is, so each
barrier event is a product of one-asset events, a tensor train of rank one. With the
edges of each asset's polynomial pieces on the barriers, the events act exactly, one
asset at a time. Second, between observation dates the transitions smooth the jumps
that the events leave behind, allowing low-degree Legendre polynomials to resolve the
continuation value on each piece. Under geometric Brownian motion, the transitions are
Gaussian convolutions, making the continuation value entire and permitting
faster-than-exponential polynomial approximation~\cite{Trefethen2019}. Third, the
correlation matrix of stock returns has one
eigenvalue for the whole market and a few for its sectors standing out of a bulk that
random-matrix theory attributes to noise~\cite{Laloux1999,Plerou2002}, so a few common
factors carry the correlation. Given the factors, the assets move independently, and
each transition acts asset by asset.

We carry this representation backward from maturity, one date at a time~\cite{Bellman1957},
and compress it after every step to the smallest rank that a tolerance allows. We call it
an $hp$ functional tensor train (hp-FTT), borrowing the name from $hp$ finite elements,
which choose both the width $h$ of the polynomial pieces and their degree
$p$~\cite{BabuskaSuri1994,Schwab1998}. Because today's spots enter only the first
transition, we differentiate that transition's density and integrate it against the
stored future value.
Each derivative still brings a factor $1/s$ that amplifies numerical error instead of
sampling noise. Rapid polynomial approximation keeps the mode budget small; over our
tested accuracy range, extra digits come from tighter rounding at a fixed polynomial
budget, where a sampler needs a hundred times more paths per added digit.

One solve therefore gives the price and every Greek at any spot, without sampling noise
though not without numerical error. Its stored future values also give exposure tails
without inner simulation, for little more than the cost of sampling scenarios. What the
solver pays instead is rank, which rises as the tolerance tightens and assets are added.
Our comparison asks when that cost is less than the variance and bias of sampling. On
frequently observed notes with three to five assets, the usual basket size for
autocallables, the solver is cheaper for third-order Greeks and, at tight accuracy, for
every Greek and the exposure tail. Each added asset narrows its lead. Sparse observation
and deltas at percent accuracy remain favorable to Monte Carlo.

\subsection{Summary of results}

\textbf{1.} We build the hp-FTT solver for discretely monitored worst-of contracts. Given
the common factor, its transition is a matrix product operator that smooths each asset
separately, with the quadrature nodes of the factor as its bond. For Gaussian
transitions we show why accuracy costs it little: the degree that suffices for an error
$\varepsilon$ grows at most as $\log(1/\varepsilon)$, and on pieces wide compared with the typical move $s$
only as $\sqrt{\log(1/\varepsilon)}$.

\textbf{2.} We set the solver against the strongest Monte Carlo estimators for the full
set of Greeks that we could build or find published, at matched accuracy, and charge it
a full solve for a single Greek at a single spot. The order of the Greek and the
accuracy decide which method is cheaper. On notes observed monthly on baskets of the
usual size (three to five assets) under geometric Brownian motion, at $1\%$ accuracy,
Monte Carlo is more than ten times cheaper for deltas, second-order Greeks
go either way, and the solver is cheaper for both third-order Greeks tested, by $50$ to $400$
times on the mixed one. At $0.3\%$ Monte Carlo keeps only the deltas, and at $0.1\%$ the
solver beats Monte Carlo across derivative orders, by up to three orders of magnitude.
Section~\ref{sec:main_result} gives the comparisons.

\textbf{3.} The picture holds under local volatility near its continuum limit, where the
ratios stay of the same order, and under both models the tightest solves agree with deep
Monte Carlo on the price and all five Greeks to within its statistical error.

\textbf{4.} Monte Carlo keeps low-order Greeks at percent accuracy and notes observed
only a few times, where sampling pays little variance. Each added asset moves the balance
toward it, since the solver pays in rank, as Secs.~\ref{sec:lr} and~\ref{sec:freq} explain.

\textbf{5.} For the $97.5\%$ quantile of the note's value at a future date we compare
the solver with tuned nested, multilevel and regression Monte Carlo. The solver, its
build included, is about an order of magnitude costlier than the best of them at an
error of about ten basis points of notional and cheaper at one. A solve already run for
the Greeks gives the tail one to two orders of magnitude more cheaply from a few basis
points down to one, as shown in Sec.~\ref{sec:exposure}.

\subsection{Relation to previous work}

The Dynamic Chebyshev method carries the value function of a backward
recursion as a Chebyshev polynomial and reads delta and gamma off
it~\cite{GlauMahlstedtPoetz2019}, and the polynomials it stores at future dates give the
expected and potential future exposure of barrier and early-exercise options without
inner simulation~\cite{GlauPachonPoetz2021}. Both papers price options on a single
underlying, and the second leaves more risk factors as a next step, with Chebyshev
interpolation on a full tensor grid for two or three and sparse grids or low-rank tensors
beyond~\cite{GlauPachonPoetz2021}.

In many dimensions, tensor methods have entered pricing and risk in two roles. As surrogates, they
compress pricing maps built from valuations made elsewhere, for parametric
pricing~\cite{GlauKressnerStatti2020}, margin and market-risk
capital~\cite{RuizZeronDIM,RuizZeronFRTB,RuizZeronBook}, and portfolio
revaluation~\cite{GribbenSTNGPR2026}, as do neural surrogates
trained on simulated payoffs or prices together with their
derivatives~\cite{HugeSavine2020,Alonso2026}. As solvers, they propagate a value
function over the joint state of the assets~\cite{KastoryanoArenstein2026} or represent
the continuation value of least-squares Monte Carlo as a tensor train of polynomial
coefficients~\cite{BayerEigel2021}. Fourier-domain tensor methods have priced
multi-asset options that depend on a single date, Greeks
included~\cite{KastoryanoPancotti2022,ArensteinKastoryano2026COS,SakuraiGreeks2025}.
The challenge we take up is a contract whose payoff jumps on each of many dates. We
carry a monthly worst-of note through its whole backward recursion, with every event
applied exactly, read from one solve its Greeks up to third order at any spot and the
tail of its future value, and compare the cost with that of the strongest Monte Carlo
estimators at matched accuracy.

Our format is the functional tensor train of Gorodetsky, Karaman and
Marzouk~\cite{GorodetskyKaramanMarzouk2019}, which is a continuous analogue of the tensor
train~\cite{Oseledets2011,Oseledets2013,Bigoni2016} and whose cores are matrices of
one-variable functions, rounded in the $L^2$ norm of the function itself. Methods built
on grids compress an array of values instead, as a quantized tensor
train~\cite{KastoryanoArenstein2026} or at Chebyshev nodes~\cite{GlauKressnerStatti2020}.
Our surfaces have local features in known places, a layer about as wide as a monthly move
$s$ at each barrier, so we fix the space once, as piecewise polynomials with edges on the
barriers.

Functional tensor trains have also simulated multidimensional quantum
dynamics~\cite{SoleyBergoldGorodetskyBatista2022} and solved high-dimensional PDEs by
evolving the solution on a manifold of tensors of adaptive
rank~\cite{DektorVenturi2021,DektorRodgersVenturi2021,DektorVenturi2024}. Our backward
propagation integrates no differential equation: it applies each transition and event
as an operator and then rounds, as in the time evolution of matrix product
states~\cite{Schollwock2011}. 

\section{Multi-Asset structured products}
\label{sec:formulation}

\subsection{The product and the models}
\label{sec:product}

Our benchmark is a worst-of Phoenix autocallable with the
usual conventions~\cite{BouzoubaaOsseiran2010,Guillaume2015b}, written on $d$ assets
with prices $S_i$ and observed on $N$ equally spaced dates $t_k=k\Delta t$ up to
maturity $T=N\Delta t=3$ years. On every date the worst performance,
$m_k=\min_iS_i(t_k)/S_i(0)$, is compared with three levels, the autocall level
$B_{\rm a}$, the coupon barrier $B_{\rm c}$ and the protection barrier $B_{\rm p}$, with
$B_{\rm p}\le B_{\rm c}\le B_{\rm a}$. Before maturity the note pays a coupon $c$ if
$m_k\ge B_{\rm c}$, and if moreover $m_k\ge B_{\rm a}$ it is called: it repays the
notional and ends. At maturity it pays the coupon if $m_N\ge B_{\rm c}$ and repays the
notional if $m_N\ge B_{\rm p}$. Below that it repays only $m_N$, and the investor bears
the worst performer's loss. Our benchmark has $B_{\rm a}=1$, $B_{\rm c}=0.8$ and
$B_{\rm p}=0.7$. We work in units of the notional, with a rate of $r=3\%$ and no
dividends. The coupon is $c=0.36/N$, or $12\%$ a year, so that $N=6$, $12$ and $36$
describe the same note observed semi-annually, quarterly and monthly. The monthly note
is our main case. The benchmark omits coupon memory and a call lockout, which
Appendix~\ref{app:gbm} adds on an indicative three-stock term sheet.

In log-spots, $x_i=\log\bigl(S_i/S_i(0)\bigr)$, the protection, coupon and autocall
barriers sit at $b_{\rm p}=\log B_{\rm p}$, $b_{\rm c}=\log B_{\rm c}$ and
$b_{\rm a}=\log B_{\rm a}$ on every axis,
and one identity does much of the work in what follows. The worst performer is above a
barrier exactly when every asset is,
\begin{equation}
\1_{\{\min_ix_i\ge b\}}=\prod_{i=1}^{d}\1_{\{x_i\ge b\}},
\label{eq:factorize}
\end{equation}
so every event is a cut parallel to the axes, a product of one-asset events. The one
oblique feature is the loss at maturity, $e^{\min_ix_i}$, whose kink lies where two
assets tie, and Sec.~\ref{sec:events} shows how the solver avoids storing it. The same
identity decides which other terms of issued notes the solver handles, as discussed in
Appendix~\ref{app:gbm}.

The assets follow geometric Brownian motion with one common
factor. Under the pricing measure their log-spots obey
\begin{equation}
dx_i=\bigl(r-\tfrac12\sigma^2\bigr)\,dt+\sigma\bigl(\sqrt{\rho}\,dB_0+\sqrt{1-\rho}\,dB_i\bigr),
\qquad i=1,\dots,d,
\label{eq:gbm}
\end{equation}
with $B_0,B_1,\dots,B_d$ independent Brownian motions: every pair of assets has
correlation $\rho$, carried by the factor $B_0$, and given the factor the assets move
independently. We take $\sigma=0.2$ and $\rho=0.5$. The transition over one period is
then Gaussian and known exactly. Its width, $s=\sigma\sqrt{\Delta t}$, is about $0.06$
at monthly observation, less than half the spacing of successive barriers in
log-spot. It sets the scale on which the value changes near a barrier, and the size
of a sampler's weight for a Greek of order $k$, which grows as $s^{-k}$.

One factor is the simplest correlated model and captures the dominant market mode of
equity correlations~\cite{Laloux1999,Plerou2002}. We use equal loadings, making the
assets exchangeable with pair correlation $\rho$. At three assets, one factor with unequal
loadings fits most observed correlations exactly, as Appendix~\ref{app:factors_data}
shows. They change the one-asset kernels, not the one-factor structure or its propagation
cost at fixed ranks and resolution. One factor is also the case most favorable to our
solver. Appendix~\ref{app:factors} adds sector factors, which baskets of four or five
names typically need.

Our second model keeps the factor, the correlation, the
rate and the contract, and lets the volatility of each asset depend on its own
log-spot,
\begin{equation}
\sigma(x)=\min\bigl\{\max\{0.2\,e^{-\theta x},\,0.05\},\,1\bigr\},\qquad\theta=0.6.
\label{eq:lv_sigma}
\end{equation}
The volatility is $0.2$ at the initial level, as under geometric Brownian motion,
$0.23$ at the coupon barrier and $0.25$ at the protection barrier, and the clamps bind
only beyond $x=-2.7$ and $x=2.3$. The transition over a period is then not known in
closed form, so both methods use the same approximation of it, one Gaussian step per
month, built as follows. Because each asset's
volatility depends only on its own log-spot and the correlation is constant, the change
of variables $\tilde x_i=\int_0^{x_i}du/\sigma(u)$ gives every asset unit volatility, and by
It\^o's formula,
\begin{equation}
d\tilde x_i=\tilde\mu(\tilde x_i)\,dt+\sqrt{\rho}\,dB_0+\sqrt{1-\rho}\,dB_i,\qquad
\tilde\mu=\frac{r}{\sigma}-\frac{\sigma}{2}-\frac{\sigma'}{2},
\label{eq:lamperti}
\end{equation}
with $\sigma$ and $\sigma'=d\sigma/dx$ taken at $x_i$. The common factor still enters
additively. Inside the clamps $\sigma'=-\theta\sigma$, so the drift,
$\tilde\mu=r/\sigma-(1-\theta)\sigma/2$, depends on the log-spot only through the
volatility, which changes by a factor $e$ over a log-spot distance of $1/\theta\approx1.7$.
An asset's typical move in a month, $\sigma\sqrt{\Delta t}$, is $0.06$ to $0.07$ near the
barriers, a twentieth of that distance or less, and even at the left clamp, where the
volatility is largest, less than a fifth of it. Over a month the drift is thus nearly
linear, and linearizing it at each asset's level at the start of each month makes the
month one Gaussian step in $\tilde x$ given the factor.
Unlike under geometric Brownian motion, where the drift is constant, this step is not
exact, but the resulting chain of monthly steps lies within
$0.2\%$ of extrapolations of Euler chains to the continuous model, on the price and
every Greek, as Appendix~\ref{app:lv_routes} shows. The main results under local
volatility use it: Monte Carlo samples it, and the tensor solver applies it as an
operator, with the events at the observation dates only. Models that no change of
variables makes Gaussian over a period, such as local correlation, need Euler sub-steps
instead, each of which, given the common factor, moves an asset by a Gaussian increment
whose mean and width follow $\sigma(x_i)$. We also run both methods on a chain of two such
sub-steps per month, to check that the outcome of the comparison does not depend on the
chain (see Appendix~\ref{app:lv}).

The note is priced backward from maturity, one date at a time.\footnote{The price can
also be computed forward, carrying the law of the log-spots, which Monte Carlo samples,
through the transitions. The two directions are adjoint, and are analogous to the Schr\"odinger and
Heisenberg pictures in quantum mechanics, $\mathrm{Tr}[\hat\rho(t)\hat A]=\mathrm{Tr}[\hat\rho\hat A(t)]$: a forward solve fixes the initial spot and prices any cash flow paid on the surviving paths, while a backward solve fixes the contract and values it at every spot. We
go backward because that direction returns the value as a function of the spots on every
date, hence the Greeks at any spot and the future values of Sec.~\ref{sec:exposure} from
one solve.} Let
$V_k(\x)$ be the value of a live note on date $t_k$, that date's payments
included, and $C_k(\x)$ its continuation value: the value on the same date of the
payments still to come, if the note survives. Under the pricing measure, the
continuation value is the discounted expectation of the next date's value, and since
the log-spots are a Markov process, that expectation depends only on the current
log-spots,
\begin{align}
C_k(\x)&=e^{-r\Delta t}\,\E\bigl[V_{k+1}(\bm X_{k+1})\,\big|\,\bm X_k=\x\bigr],
\qquad k=0,\dots,N-1.
\label{eq:recursion}\\
\intertext{On a date before maturity the note pays the coupon if the worst performance
is at or above the coupon barrier, and the notional if it is at or above the autocall
level, in which case it ends. Otherwise it continues, worth its continuation value, so
that}
V_k&=\mathcal E_kC_k=c\,\1_{\{\min_ix_i\ge b_{\rm c}\}}+\1_{\{\min_ix_i\ge b_{\rm a}\}}
+\bigl(1-\1_{\{\min_ix_i\ge b_{\rm a}\}}\bigr)\,C_k,\qquad k=1,\dots,N-1.
\label{eq:event}\\
\intertext{At maturity it pays the coupon if the worst performance is at or above the
coupon barrier, and the notional if it is at or above the protection barrier, or else
the worst performance $e^{\min_ix_i}$ itself,}
V_N&=c\,\1_{\{\min_ix_i\ge b_{\rm c}\}}+\1_{\{\min_ix_i\ge b_{\rm p}\}}
+e^{\min_ix_i}\,\1_{\{\min_ix_i<b_{\rm p}\}},
\label{eq:payoff}
\end{align}
and the price is $C_0(\x_0)$, since nothing happens on the valuation date. From maturity
back to today the recursion alternates two operations: the transition of
Eq.~\eqref{eq:recursion} averages the value over the moves of one period, which smooths
it, and on each date the event map $\mathcal E_k$ of Eq.~\eqref{eq:event} cuts it along
the barriers and adds that date's payments. Plain Monte Carlo never forms $C_k$ and
averages the cash flows of simulated paths, and regression Monte Carlo fits $C_k$ on
those paths. A grid-based solver obtains $C_k$ from $V_{k+1}$ by time-stepping the
pricing partial differential equation on a grid over all $d$ log-spots, whose number of
points grows exponentially with $d$. Our solver instead computes the expectation of
Eq.~\eqref{eq:recursion} directly, as an integral against the transition density, and
stores $C_k$ as a tensor train, whose size grows only linearly with $d$ at a given rank
(see Sec.~\ref{sec:solver}).

\subsection{Greeks}
\label{sec:greek_defs}

The Greeks are the derivatives of $C_0$ with respect to the log-spots: delta
$\Delta=\partial_1C_0$, gamma $\Gamma_{11}=\partial^2_{11}C_0$, cross-gamma
$\Gamma_{12}=\partial^2_{12}C_0$, speed $\partial^3_{111}C_0$ and cross-speed
$\partial^3_{123}C_0$, the mixed third derivative, which needs three assets. We do not
compute the remaining kind, $\partial^3_{112}C_0$. The solver returns the Greeks at any
spot, and we compare the methods at spots on the diagonal, $\x_0=(x_0,\dots,x_0)$,
where every asset has the same performance. There the assets are exchangeable, so each
of these Greeks stands for all derivatives of its kind. Costs are scored at
$x_0=-0.10$, every asset about $10\%$ below its initial level, between the coupon
barrier and the autocall level, for relative errors of $1\%$, $0.3\%$ and $0.1\%$. The
spot $x_0=0$, where every asset is at its initial level, lies on the autocall barrier,
and we use it to check the solver against Monte Carlo.

The Greeks beyond delta measure what a delta hedge
leaves exposed. A note hedged with $\partial C/\partial S_i$ units of each asset and
financed at the rate $r$ gains, over a short interval $dt$ in which the spots move by
$dS_i$,
\begin{equation}
\tfrac12\sum_{i,j}\frac{\partial^2C}{\partial S_i\,\partial S_j}\,
\bigl(dS_i\,dS_j-\sigma^2\rho_{ij}S_iS_j\,dt\bigr),
\qquad\rho_{ij}=\rho\ (i\neq j),\quad\rho_{ii}=1,
\label{eq:hedged_pnl}
\end{equation}
to second order. This is the multi-asset form of the classic decomposition of a
delta-hedged option's profit and loss into gamma times the gap between realized and
implied variance~\cite{Bergomi2015}. On a
worst-of note the own and the cross terms can pull in opposite directions: on three
assets, between the protection and the coupon barrier, the hedged note loses when one
asset makes a large move alone but gains when all three move together. Third
derivatives measure how these coefficients shift as the spots move, and they are where a
sampler struggles most, since its weight gains a factor with every derivative.

\subsection{Potential future exposure}
\label{sec:pfe_def}

The tail of the note's future value measures a credit risk: whoever is owed the note's
payments would lose its value on a future date $t_k$ if the other side defaulted then.
That value is the random variable $V_k(\bm X_k)=\mathcal E_kC_k(\bm X_k)$,
with $\bm X_k$ the log-spots at $t_k$, and the exposure is its positive
part~\cite{BCBS_CRE50}, here the value itself, which is positive. Counterparty risk is
managed through a high quantile of the exposure at each date, the potential future
exposure, or peak exposure~\cite{BCBS_CRE50,Gregory2020}. At a level $\alpha$ it is the
generalized inverse that defines value at risk~\cite{McNeilFreyEmbrechts2015},
\begin{equation}
\mathrm{PFE}_k=\inf\bigl\{q:\ \Pr[V_k(\bm X_k)\le q]\ge\alpha\bigr\},
\label{eq:pfe}
\end{equation}
the value exceeded with probability $1-\alpha$, and Sec.~\ref{sec:exposure} takes
$\alpha=97.5\%$. A sampler has no $C_k$: it must estimate it at each sampled state, by
inner simulation or by regression, and the errors of those estimates bias the quantile.
The backward solve has already computed $C_k$ as a function, at every date it passes.

\section{The functional tensor-train solver}
\label{sec:solver}

\begin{figure}[t]
\centering
\includegraphics[width=\linewidth]{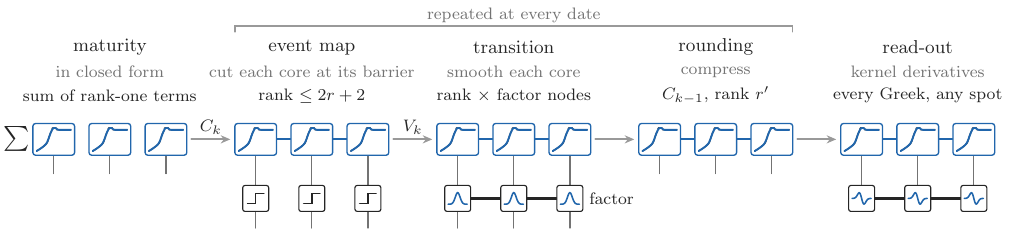}
\caption{\textbf{The backward solve as a tensor network.} Three assets are shown:
boxes are cores holding one asset's functions and encoded as matrix functions, horizontal
lines are bonds whose size is the rank, and vertical lines are log-spots. From a
closed-form sum of rank-one terms at maturity, each earlier date applies three
operations: the event map multiplies every core by its own step, the transition
smooths every core with its own kernel (the kernels share one bond for the common
factor), and rounding compresses the train. Contracting kernel derivatives with the
cores reads out the Greeks. Appendix~\ref{app:propagator} shows one core up close.}
\label{fig:solver}
\end{figure}

The solver carries the recursion defined in Sec.~\ref{sec:formulation} backward from
maturity on a tensor train whose size per asset is fixed in advance and whose rank
adapts. Each date takes three operations (see Fig.~\ref{fig:solver}): the event map
cuts the value along the barriers, the transition smooths it, and a rounding
compresses the train. In tensor-network language the transition is a matrix product
operator and the rounding a truncation of the bonds by their singular values, as in
the time evolution of matrix product states~\cite{Schollwock2011}.

\subsection{Representation and transition}
\label{sec:repr}

A continuation value is stored as a functional tensor
train (FTT)~\cite{Oseledets2011,GorodetskyKaramanMarzouk2019}, a matrix product state whose
physical index is a continuous log-spot:
\begin{equation}
C(\x)=G_1(x_1)\,G_2(x_2)\cdots G_d(x_d).
\label{eq:ftt}
\end{equation}
Each core $G_i(x_i)$ is a $\chi_{i-1}\times\chi_i$ matrix of functions of one log-spot,
with $\chi_0=\chi_d=1$, and the largest $\chi_i$ is the rank. The log-spots are measured from
the contract's reference level, so that the autocall barrier sits at $x_i=0$, and live
on a box $[-L,L]^d$. We take $L=3$ under geometric Brownian motion, so that an asset
would have to move more than seven standard deviations of its three-year log-return to
get from a barrier to the edge of the box, and $L=4.5$ under local volatility, which
rises into the left tail. Widening either box changes no price or Greek at the
accuracies we score (Appendix~\ref{app:propagator}).

Each FTT core is a matrix function, with entries corresponding to one dimensional functions expanded in some basis:
\begin{equation}
G_i(x)=\begin{pmatrix}
G_i^{(1,1)}(x)&\cdots&G_i^{(1,\chi_i)}(x)\\
\vdots&\ddots&\vdots\\
G_i^{(\chi_{i-1},1)}(x)&\cdots&G_i^{(\chi_{i-1},\chi_i)}(x)
\end{pmatrix},
\qquad
G_i^{(u,v)}(x)=\sum_{j=1}^{n_i^{(u,v)}}a_{i,j}^{(u,v)}\,\psi_{i,j}^{(u,v)}(x).
\label{eq:hp_core}
\end{equation}
Each entry could in principle carry its own number of terms $n_i^{(u,v)}$, or even its
own basis functions $\psi_{i,j}^{(u,v)}$, and Ref.~\cite{GorodetskyKaramanMarzouk2019}
adapts both entry by entry. For simplicity we give all entries one basis, of piecewise
polynomials. The axis is cut into panels $P$ whose edges include the three log-barriers,
where the event-mapped value can jump, and every entry is a sum of Legendre polynomials
on the panels,
$G_i^{(u,v)}(x)=\sum_P\sum_{q=0}^{p_P-1}a_{i,P,q}^{(u,v)}\,\varphi_{P,q}(x)$, with
$\varphi_{P,q}$ the Legendre polynomial of degree $q$ mapped onto $P$ and zero outside
it, which we call a mode. The widths of the panels and their numbers of modes $p_P$ are the $h$ and the $p$ of
the name. On either side of the barriers the panels are at most $0.5$ wide. There a
transition spreads the jump that an event left at a barrier over a layer about $s$
wide, and the layer rule of Sec.~\ref{sec:resolution} gives each such panel a number of
modes proportional to its width in units of $s$. Away from the barriers the value is smooth, and each stretch
that touches no barrier is a single panel of $16$
modes. At monthly observation an axis carries $6$ panels and $82$ modes under geometric
Brownian motion, and $8$ panels and $114$ modes under local volatility, whatever the
number of assets. Appendix~\ref{app:propagator}
gives the details of the mesh.

The rank is the only part of the representation that adapts. After every operator the
train is rounded~\cite{Oseledets2011}: the singular values of each bond are truncated so
that the discarded part stays below a fraction $\varepsilon$ of the $L^2$ norm of the
whole surface.

Over one observation period of length $\Delta t$ every log-spot moves by a common
drift, by a shock that all the assets share, and by a shock of its own:
\begin{equation}
\x'=\x+\mu\Delta t\,\1+\beta Z\,\1+s_\perp\bm{\xi},
\qquad \beta=\sigma\sqrt{\rho\,\Delta t},\quad s_\perp=\sigma\sqrt{(1-\rho)\Delta t},
\label{eq:onefactor_step}
\end{equation}
with $\mu=r-\tfrac12\sigma^2$, $Z\sim\cN(0,1)$ the common factor and
$\bm{\xi}\sim\cN(0,I_d)$ the assets' own shocks. The variance of a step, $s^2$, thus
splits into a shared part, $\beta^2=\rho s^2$, and an own part, $s_\perp^2=(1-\rho)s^2$.

Once the factor is known the assets move independently, so only the average over the
factor couples them. We take it by Gauss--Hermite quadrature with nodes $z_g$ and
weights $w_g$:
\begin{equation}
(\mathcal K f)(\x)=\sum_{g=1}^{n_z}w_g\int\prod_{i=1}^{d}k_g(x_i,y_i)\,f(\bm y)\,
d\bm y,\qquad
k_g(x,y)=\frac{1}{s_\perp}\,\phi\Big(\frac{y-x-\mu\Delta t-\beta z_g}{s_\perp}\Big),
\label{eq:kernel}
\end{equation}
with $\phi$ the standard normal density, so that $C_{k-1}=e^{-r\Delta t}\mathcal K
V_k$. The correlated transition density is exactly a mixture over the factor of
products of one-asset densities and
Eq.~\eqref{eq:kernel} is a quadrature of that mixture. See also Appendix~\ref{app:kernel} for more details.

On a tensor train this transition is a matrix product operator
(see Fig.~\ref{fig:solver}), and applying it is a tensor contraction. At a factor node
the kernel is a product over the assets, as the train is, so the $d$-dimensional
integral of Eq.~\eqref{eq:kernel} splits into one-dimensional integrals, one per core:
\begin{equation}
\int\prod_{i=1}^{d}k_g(x_i,y_i)\,G_1(y_1)\cdots G_d(y_d)\,d\bm y
=(\mathcal K_{1,g}G_1)(x_1)\cdots(\mathcal K_{d,g}G_d)(x_d),\qquad
(\mathcal K_{i,g}G_i)(x)=\int k_g(x,y)\,G_i(y)\,dy.
\label{eq:contraction}
\end{equation}
Each core is thus smoothed by its own asset's kernel, an integral over that asset's next
log-spot, and the factor node $g$ becomes one extra bond shared by all the cores,
$n_z=12$ nodes whatever the number of assets. Two quadrature rules are at work:
Gauss--Hermite nodes for the average over the factor, and Gauss--Legendre nodes on the
panels for each one-dimensional integral of Eq.~\eqref{eq:contraction}
(see Appendix~\ref{app:propagator}). Summing over the factor nodes
can multiply the rank by as much as $n_z$, and rounding brings it back down. Every operation acts on one asset's axis at a time, so no array over all the assets jointly is ever formed. Under geometric Brownian motion the solver crosses a whole period in one
operator application, with no time-stepping error.

With several common factors under geometric Brownian motion, the shared part of a step
is a sum of independent Gaussian moves, one per factor. The solver smooths the stored
train over one factor, rounds the result, again a train, and smooths that over the next,
as a separable blur smooths an image along one axis and then along the other. The staged
average uses the nodes and weights of a product rule over all the factors at once, and the
implementations differ through the roundings and projections between stages and their
boundary treatment. The checks in Appendix~\ref{app:factors_staged} detect no additional
error from staging. But the product rule smooths the train once for
every combination of nodes, $n_1n_2$ times for two factors with $n_1$ and $n_2$ nodes,
whereas the solver smooths it $n_1$ times over the first factor and then shifts the
result once for each of the $n_2$ nodes of the second, $n_1+n_2$ operations in all. What
the extra factors still cost is rank. A bond splits the assets into two groups, and its
rank grows with the number of independent directions in which the factors correlate one
group with the other, one with a single factor, as Appendix~\ref{app:factors_rank}
explains.

Under local volatility the solve proceeds in the same way, and only the kernel $k_g$, the
density of an asset's step given the factor, changes. On the chain of our main results that step spans a month and is Gaussian in the transformed
log-spot $\tilde x_i$, with a mean and a variance that follow from linearizing the drift
of Eq.~\eqref{eq:lamperti} at the asset's level at the start of the month. The solver applies it in
$x$, as that Gaussian times the Jacobian $1/\sigma$, once per month, and reads the Greeks
off the whole first month with $24$ factor nodes rather than $12$, as
Appendix~\ref{app:lv_routes} details. On an Euler chain with $n_{\rm sub}$ sub-steps of
length $\delta t=\Delta t/n_{\rm sub}$ per period,
$C_{k-1}=e^{-r\Delta t}\mathcal K^{n_{\rm sub}}V_k$, and given the factor each asset's
move over a sub-step is Gaussian, with a mean and a width that depend on where the asset
is:
\begin{equation}
k_g(x,y)=\frac{1}{s(x)}\,\phi\Big(\frac{y-m_g(x)}{s(x)}\Big),\quad
m_g(x)=x+\big(r-\tfrac12\sigma(x)^2\big)\delta t+\sigma(x)\sqrt{\rho\,\delta t}\,z_g,
\quad s(x)=\sigma(x)\sqrt{(1-\rho)\,\delta t}.
\label{eq:lv_kernel}
\end{equation}
Either way the solver applies the kernel exactly as under geometric Brownian motion: each one-dimensional integral is computed by Gauss–Legendre quadrature on the panels and projected back onto their modes, as described in Appendix~\ref{app:propagator}.

A step from near the edge of the box can end outside it, where no value is stored.
Cutting the integrals at the edge, as if the value were zero beyond it, would leave a
false step there, which costs rank. The solver instead continues each core beyond the
edge by its value at the edge, which is nearly exact because the value is nearly flat
out there, deep below the protection barrier and high above the autocall barrier. This
lowers the rank and makes a solve two to ten times faster
(see Appendix~\ref{app:events_detail}).

\subsection{Events, terminal step and read-out}
\label{sec:events}

On every observation date the note pays a coupon, is called
or carries on, according to whether every asset is above a barrier. By
Eq.~\eqref{eq:factorize} such an event is a product of one-asset indicators, a tensor
train of rank one (see Fig.~\ref{fig:solver}), so multiplying the stored value by it acts
on each core separately, and the event map of Eq.~\eqref{eq:event} turns a continuation
value of rank $\chi$ into four terms whose sum has rank at most $2\chi+2$. Because every
barrier is a panel edge, multiplying a core by a one-asset indicator only switches whole
panels off, so the event map is applied exactly. With a barrier inside a panel, the
quadrature of the next transition, which sees the cut core only at its nodes, would
misplace the barrier by up to the gap between two nodes, an error that every later date
would inherit. The transition then smooths each of the four terms, and only the smooth
results are added and rounded.

At maturity, below the protection barrier, the note repays the worst performance
$m=e^{\min_ix_i}$ itself. This is the one feature of the payoff that panels along the
axes cannot follow, since $m$ has a kink wherever two assets tie, along diagonals that
cross the panels. Rather than store the payoff, the solver writes it with indicators
alone. It splits the repayment, $1$ above the barrier and $m$ below it, into
$(1-B_{\rm p})\,\1_{\{m\ge B_{\rm p}\}}+\min\{m,B_{\rm p}\}$, and writes the second term
as an integral over levels $\ell$,
$\min\{m,B_{\rm p}\}=\int_0^{B_{\rm p}}\1_{\{m>\ell\}}\,d\ell$ (the layer-cake
representation), since the indicator is one exactly at the levels below $m$. The payoff
of Eq.~\eqref{eq:payoff} is therefore
\begin{equation}
V_N=c\,\1_{\{m\ge B_{\rm c}\}}+(1-B_{\rm p})\,\1_{\{m\ge B_{\rm p}\}}
+\int_0^{B_{\rm p}}\1_{\{m>\ell\}}\,d\ell.
\label{eq:layer_cake}
\end{equation}
Every indicator in it is a product of one-asset indicators,
$\1_{\{m>\ell\}}=\prod_i\1_{\{x_i>\log\ell\}}$, whose expectation over the last step
is, at each factor node, a product of one-asset probabilities. The continuation value
one step before maturity, the first function the solver stores, is therefore a sum of
trains of rank one in closed form, and no stored function carries the kink along the
ties.

The price and the Greeks at a spot $\x_0$ are $C_0(\x_0)$ and
its derivatives. The solver stops one sub-step short of inception and, instead of
applying that last step as an operator, evaluates it at $\x_0$ together with its
derivatives,
\begin{equation}
\partial^{\alpha}C_0(\x_0)=\sum_{g=1}^{n_z}w_g\int\prod_{i=1}^{d}
\partial^{\alpha_i}_{x_{0,i}}k_g(x_{0,i},y_i)\;\bar C(\bm y)\,d\bm y,
\label{eq:readout}
\end{equation}
where $\bar C$ is the surface stored one sub-step after inception. A shift of
the spots moves only this first step, so each Greek integrates a derivative of the
first step's density, the weight a likelihood-ratio estimator puts on its sampled paths,
against the stored surface, by quadrature, one asset at a time given the factor. Every
Greek at every spot comes from the same surface, and takes a few milliseconds. Even under geometric Brownian motion the first period is split into two
half-month sub-steps, so that the differentiated kernel acts on a smooth surface rather
than on the jumps of the first date's events. Algorithm~\ref{alg:solve} collects the
steps of the solve.

Differentiating the polynomial cores instead would amplify their error most at the ends
of their panels, which sit on the barrier. Equation~\eqref{eq:readout} puts
every derivative on the kernel, the one factor known in closed form: each Greek is the
exact derivative of the computed price with respect to the spots, and the error of the
stored surface enters only through its integral against the kernel's derivative, which
Sec.~\ref{sec:resolution} bounds. Projected onto a mesh with no
edge near the spots and rounded more tightly, the surface at inception can also be
differentiated directly, so that a query only evaluates a train, but at a higher rank
(see Appendix~\ref{app:readout}).

\begin{algorithm}[t]
\caption{Backward solve and read-out.}
\label{alg:solve}
\begin{algorithmic}[1]
\Require dates $t_1<\dots<t_N=T$, log-barriers $b_{\rm p},b_{\rm c},b_{\rm a}$, coupon
$c$, kernel $k_g$ and the sub-steps of each period (under geometric Brownian motion one
per period, two in the first), tolerance $\varepsilon$ and spots $\x_0$
\Ensure price and Greeks at every $\x_0$, and continuation surfaces at every date
\State mesh, with panel edges at the log-barriers and modes set by the layer rule, and
quadrature rules (Sec.~\ref{sec:repr})
\State $C\gets\mathrm{round}_\varepsilon(e^{-r\Delta t}\,\mathcal KV_N)$ for the last
sub-step, in closed form from Eq.~\eqref{eq:layer_cake}
\State $C\gets\mathrm{round}_\varepsilon(\mathcal KC)$ for the other sub-steps of the
last period \Comment{now $C=C_{N-1}$}
\For{$k=N-1,N-2,\dots,1$} \Comment{on entry $C=C_k$}
\State $C\gets e^{-r\Delta t}\,\mathcal E_kC$ \Comment{events of date $t_k$,
Eq.~\eqref{eq:event}}
\State $C\gets\mathrm{round}_\varepsilon(\mathcal KC)$ for each sub-step of the period
but the one from inception
\Comment{now $C=C_{k-1}$, or $\bar C$ if $k=1$}
\EndFor
\State evaluate Eq.~\eqref{eq:readout} at every $\x_0$: kernel derivatives of the
sub-step from inception, contracted with the cores of $\bar C$
\end{algorithmic}
\end{algorithm}

\subsection{Resolution and error control}
\label{sec:resolution}

How many modes a panel needs is set by the width $s$ of a step. Near each barrier the value
changes across a layer about $s$ wide, where the transition has smoothed the jump
that the event left. The layer rule gives each
panel of width $h$ that touches a barrier a number of Legendre modes proportional to
$h/s$, between a floor $p_{\rm min}$ and a cap $p_{\rm max}$:
\begin{equation}
  p_{\rm layer}(h,s)=\min\{p_{\rm max},\max\{p_{\rm min},\lceil2h/s\rceil\}\},
  \label{eq:layer_rule}
\end{equation}
where $s=\sigma\sqrt{\Delta t}$ under geometric Brownian motion and, under local
volatility, the lowest volatility \emph{at the barriers} times the square root of the
time step, a month on the chain of our main results and a sub-step on an Euler chain. In our experiments, we choose the parameters $p_{\rm min}=8$ and $p_{\rm max}=48$.
Layer-adapted $hp$ resolution is a classical idea, for example for the boundary layers
of reaction--diffusion problems~\cite{MelenkSchwab1998}. Proposition~\ref{prop:resolution}
below explains why the number of modes should grow in proportion to $h/s$. The
constants, $2$, $p_{\rm min}$ and $p_{\rm max}$, are our choices, which we check by
refining them together with the rest of the tuning knobs (see Appendix~\ref{app:err}).

Over one observation period the log-spots move as in
Eq.~\eqref{eq:onefactor_step}, and the analysis uses the idiosyncratic width
$s_\perp$. A continuation value is $C(x)=e^{-r\Delta t}\,\E[V(X')\mid X=x]$, where
$V$, the event-mapped value at the next date, jumps at the barriers but is bounded,
$|V|\le M$. Fix every coordinate but $x_i$, let $P=[a,a+h]$ be a panel,
$\kappa=h/s_\perp$ its width in units of $s_\perp$, and $E_p$ the smallest uniform
error on $P$ that a polynomial in $x_i$ of degree at most $p$ can reach. The solver
stores not this best polynomial but the Legendre projection of $C$, whose uniform
error is at most $O(\sqrt p)\,E_p$.

\begin{proposition}[Resolution of a continuation value]
\label{prop:resolution}
\leavevmode
\begin{enumerate}
\item[(a)] $x_i\mapsto C(x)$ extends to an entire function, and
$|C(x+\mathrm{i}\eta\,e_i)|\le M\,e^{\eta^2/(2s_\perp^2)}$ for all real $x$ and $\eta$.
\item[(b)] On every panel,
\begin{equation}
  E_p\le M\,\frac{\kappa^2}{p}\,\exp\!\Big(-\frac{13}{9}\,\frac{p^2}{\kappa^2}\Big)
  \quad\text{for }1\le p\le\tfrac13\kappa^2,
  \qquad
  E_p\le 2M\,e^{-0.48\,p}\quad\text{for }p>\tfrac13\kappa^2 .
  \label{eq:resolution}
\end{equation}
For $0<\varepsilon<M\kappa^2$ and $1\le p\le\tfrac13\kappa^2$, the first bound
ensures $E_p\le\varepsilon$ when
$p\ge\kappa\,\big(\tfrac{9}{13}\log(M\kappa^2/\varepsilon)\big)^{1/2}$:
within this regime a sufficient degree grows linearly in $h/s_\perp$ and only as
the square root of $\log(1/\varepsilon)$.
\end{enumerate}
\end{proposition}

The mechanism is that of a heat kernel. Conditioning on the common factor leaves a
Gaussian convolution of width $s_\perp$ in each coordinate, which makes the value an
entire function of each log-spot: an imaginary shift by $\mathrm{i}\eta$ multiplies
the kernel's modulus by $e^{\eta^2/(2s_\perp^2)}$, and the Bernstein-ellipse estimate
of polynomial approximation~\cite{Trefethen2019}, with its radius chosen, gives (b).
The proof, and a bound for differentiating the best polynomial approximant, are in
Appendix~\ref{app:resolution}.

The bounds fix the scale of the mode budget, not its exact size. At fixed panel width
and smoothing scale, a sufficient degree grows at most as $\log(1/\varepsilon)$, with
the stronger $\sqrt{\log(1/\varepsilon)}$ dependence in the quadratic-exponent regime.
In that regime it also grows in proportion to $h/s_\perp$, as the layer rule does.
In our runs one mode budget serves every target we use, so a tighter target is paid in
rank instead: at three assets under geometric Brownian motion, a solve accurate to $0.1\%$
on a Greek costs only $1.3$ to $1.5$ times one accurate to $1\%$
as measured in Appendix~\ref{app:gbm_ratios}. The $k$-th
derivative of a Gaussian kernel of width $s_\perp$ has $L^1$ norm
$\E|{\rm He}_k(Z)|\,s_\perp^{-k}$, with ${\rm He}_k$ the $k$-th Hermite polynomial, so
the read-out passes the error of the surface to a Greek of order $k$ multiplied by at
most a constant times $s_\perp^{-k}$, the factor that the likelihood-ratio weight
carries on each path. Monte Carlo pays for this factor squared, in a number of samples
that grows as $s_\perp^{-2k}$, and the solver for the factor itself, in about
$k\log_{10}(1/s_\perp)$ more digits of the surface, which by (b) need only a few modes
each.

The proposition describes the worst case, a layer anywhere in the panel. The layers that
the events create sit near the panel edges, where they are far cheaper to resolve. The
analysis is also per coordinate and per date, and it covers neither the quadrature and
rounding errors nor local volatility, whose kernel has a mean and a width that depend on
the state.
Appendix~\ref{app:resolution} measures the first point and states the scope of the
analysis.

To compare costs at a given accuracy, we need to know when each method has reached it. Monte Carlo reports its own error, the standard error of the estimator. The solver has two knobs of
different character: the discretization (panels, modes, quadrature, domain), which the
layer rule fixes in advance, and the rounding tolerance. The discretization is a spectral element method, and refining every one of its knobs at
once moves no Greek by more than $0.02\%$ in the three cases checked in
Appendix~\ref{app:err}. Rounding offers no comparable guarantee: it bounds the error of
the whole surface in the $L^2$ norm, not a derivative at a point, and the parts it
discards carry both signs, so that a Greek's error can grow as the tolerance tightens.
The tolerance is therefore a solver tolerance, like that of a Krylov solve inside a
finite-element code, set below the discretization error and then left alone, and the
cost comparisons charge the cheapest tolerance from which a target holds at every
tighter one.

Without a reference, two solves a decade apart in tolerance give an empirical
error estimate, the decade estimate, which sees the compression only. The rounding
error of a Greek can instead be estimated directly, by weighting each discarded part
with a dual function, the read-out carried forward by the transposed steps. This
goal-oriented estimate matched the actual rounding error to within $3\%$ in every
case where we built it, at about the cost of a solve, as detailed in
Appendix~\ref{app:cert}. No comparison in this paper uses either estimate: every
Greek is scored against deep Monte Carlo instead, and every exposure tail against a
refined surface that a nested Monte Carlo oracle checks.

\section{High-Order Greeks}
\label{sec:greeks}

This section compares our tensor-network solver on the Greeks, at matched accuracy and
under both models, with the strongest Monte Carlo estimators we could build or find
published, and then traces the outcome to the order of the Greek, the observation dates,
the sub-steps of local volatility and the number of assets.

\subsection{How we compare the methods}
\label{sec:baselines}

We set the tensor against two Monte Carlo estimators that address the full tested Greek
set, the strongest we could build and the strongest we could find published for this
problem. The first is a likelihood-ratio estimator, which moves every derivative from the
payoff onto the density of the first step, sampled with scrambled Sobol
points~\cite{Matousek1998}, four reflected paths per point and a control variate. Under
local volatility it takes the exact weight of the whole first month, which the change of
variables of Sec.~\ref{sec:product} makes available. The
second is bump-and-reprice with common random numbers on paths conditioned to survive
each date~\cite{Rakhmonov2019}, which makes the value of a path continuous in the spots.
We run it as published, on pseudo-random paths with central bumps of $1\%$ in log-spot,
and charge it the exact bias of its bump. Neither dominates, so we show both and, in the
text, compare the tensor with the cheaper of the two.

We also make a supplementary comparison under geometric Brownian motion with the
pathwise estimator of Hu, Xue and Yu~\cite{Hu2022}. It improves on the likelihood ratio
for delta at four and five assets, but its second-order finite differences cannot reach
the tighter targets and it supplies no third-order estimator. It therefore does not
serve the full tested Greek set, though it remains a valid competitor for a single
delta. Appendix~\ref{app:published} sets the three side by side, and
Appendix~\ref{app:mc_other} lists the estimators we discarded or did not build.

We compare the methods through the ratio of their costs: the compute that Monte Carlo
needs for one Greek at one spot, to a relative target $\varepsilon$ of $1\%$, $0.3\%$ or
$0.1\%$, over the compute that the tensor needs for the same, both in core-seconds on
one laptop. The ratio is set up to be hard for the tensor, which is charged a full solve
for that single Greek at that single spot, although the Greeks can be read off one solve
at any spot for a few milliseconds each: for instance, at three assets, the price and five Greeks at
$151$ spots took $0.6$ core-seconds, after a solve of $13$.

For Monte Carlo, the target is a standard error $\varepsilon|v|$ on a Greek of value
$v$. We infer the required work $W(\varepsilon)$ from the deepest runs. Let their total
work be $W_{\rm ref}$ and the standard error of their inverse-variance-weighted average
be $e_{\rm ref}$. The inverse-square-root law gives
\begin{equation}
W(\varepsilon)=W_{\rm ref}\Big(\frac{e_{\rm ref}}{\varepsilon|v|}\Big)^{2}.
\label{eq:mc_cost}
\end{equation}
The conditioned bumps are charged the compute at which their noise and the exact bias of
their bump together reach the target. The pooled runs of the likelihood-ratio estimator
are also the reference for the tensor's error, and a cell (one Greek at one target,
number of assets and model) is scored only where two of their standard errors lie below
the target. These runs therefore went past every scored target, so
Eq.~\eqref{eq:mc_cost} only scales them down, to at most a quarter of their compute. There
the law is exact for pseudo-random sampling and understates the cost of our randomized
quasi-Monte Carlo, whose error falls slightly faster than the inverse square root as a run
lengthens, and so rises faster as it is shortened. Appendix~\ref{app:scoring}
checks the resulting cost estimates. The tensor is charged its full
solve at the cheapest tolerance from which the target holds there and at every tighter
one, and each ratio is quoted as the range between that solve and the next looser one,
half a decade apart, which misses the target.

Both charges assume an answer that meets its target. Choosing the tensor tolerance
without a reference, by comparing two solves a decade apart, multiplies its charge by
a median of about four. Requiring two Monte Carlo standard errors rather than one,
for $95\%$ rather than $68\%$ confidence, likewise multiplies its cost by four.
These similar cost multipliers do not imply equal reliability: the decade estimate is
an empirical compression check, not a confidence guarantee. Appendix~\ref{app:scoring_checks}
compares reference-free stopping rules and measures their costs and failures.

\subsection{Which method is cheaper}
\label{sec:main_result}

\begin{figure}[t]
\centering
\includegraphics[width=\linewidth]{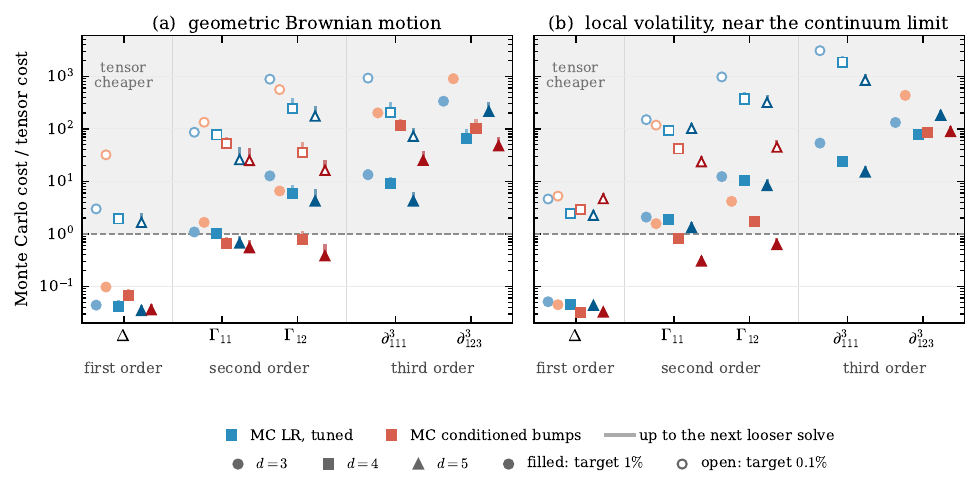}
\caption{\textbf{Monte Carlo versus hp-FTT at matched Greek accuracy.} Monte Carlo
cost divided by full tensor-solve cost at $x_0=-0.10$, monthly observation, three to
five assets. Filled: $1\%$ target; open: $0.1\%$. \textbf{(a)} Geometric Brownian
motion. \textbf{(b)} Local volatility near its continuum limit, with both methods on
the same transformed monthly chain. The tensor is cheaper above the dashed line
(shaded). Each marker uses the sustained tolerance; its bar extends to the next looser
solve, which misses the target. Blue: tuned likelihood ratio. Red: conditioned
bump-and-reprice~\cite{Rakhmonov2019}, run as published with $1\%$ log-spot bumps and
charged their exact bias. Red markers are omitted where that bias exceeds the target.
All markers require two reference standard errors below the target. Every cell is
resolved at $1\%$; cross-speed is unresolved at $0.1\%$.}
\label{fig:greeks}
\end{figure}

Figure~\ref{fig:greeks} gives the main result, Greek by Greek, for monthly observation
on three to five assets under both models.
The order of the Greek and the target decide which method is
cheaper. At $1\%$, Monte Carlo is cheaper on delta, by more than an order of magnitude.
Second-order Greeks go either way: on gamma the two are about level under geometric
Brownian motion, and on cross-gamma the tensor leads at three assets, while at five
assets the conditioned bumps are up to about three times cheaper on cross-gamma and,
under local volatility, on gamma.

The tensor leads on both third-order Greeks, by about an order of magnitude on speed and
$50$ to $400$ times on cross-speed. A tighter target raises every ratio, because it
multiplies Monte Carlo's cost by the square of the tightening factor and the tensor's
charge by far less. At $0.1\%$ it beats Monte Carlo across derivative orders under both
models: against the estimators in Fig.~\ref{fig:greeks}, by a few times on delta and by
one to three orders of magnitude on the others. The supplementary pathwise comparison
has one near-parity case, delta at five assets under geometric Brownian motion, where
the estimator of Ref.~\cite{Hu2022} is about as cheap ($0.82$--$1.2$). At $0.05\%$ the
tensor is $3$ to $18$ times cheaper on delta at every tested size, as detailed in
Appendix~\ref{app:published}. Cross-speed is resolved beyond $1\%$ only down to
$0.3\%$, where the tensor is two to three orders of magnitude cheaper.
Appendices~\ref{app:gbm_ratios} and~\ref{app:lv_routes} list the values.

\begin{figure}[t]
\centering
\includegraphics[width=\linewidth]{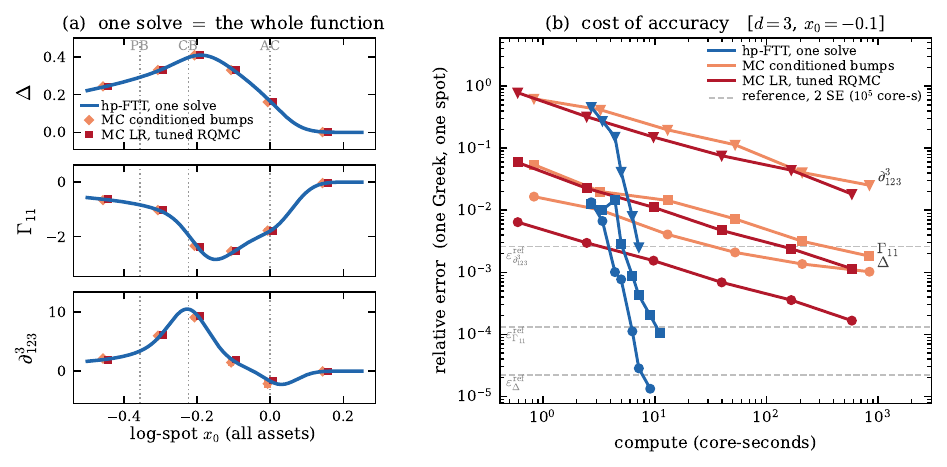}
\caption{\textbf{One tensor solve against Monte Carlo at equal compute.} Three-asset
Phoenix note under geometric Brownian motion, monthly observation ($N=36$).
\textbf{(a)} Delta, gamma and cross-speed along $x_1=x_2=x_3=x_0$, read from one solve
(curves). Each Monte Carlo marker receives about that solve's compute at one spot:
published conditioned bumps~\cite{Rakhmonov2019} (diamonds) and tuned likelihood ratio
(squares). Bars are one standard error; dotted verticals mark the barriers.
\textbf{(b)} Relative error against compute at $x_0=-0.10$. Dashed lines mark two
standard errors of the reference pooled from the deepest likelihood-ratio runs.
Likelihood ratio is scored by standard error; conditioned bumps by root-mean-square
error, which levels off at the exact bias of the $1\%$ log-spot bump. Tensor tolerances
run from $10^{-4}$ to $10^{-9}$ in half-decades, and each curve stops where its error
remains below the dashed line at every tighter tolerance. Run sizes are given in
Appendix~\ref{app:mc}.}
\label{fig:object}
\end{figure}

At three, four and five assets the tightest solve agrees with the deepest Monte Carlo
runs on the price and all five Greeks, under both models, to within their statistical
error, as Appendices~\ref{app:mc} and~\ref{app:lv_routes} detail.
Figure~\ref{fig:object}(a) shows that the agreement is not confined to the scored spot.

Figure~\ref{fig:object}(b) follows the error of each method as its compute grows, at the
scored spot. The tensor's error falls quickly as the tolerance tightens, though not at
every step, since rounding discards parts of both signs. Gamma's error grows by almost
half between the tolerances $3\times10^{-5}$ and
$10^{-5}$. From about ten core-seconds on, the tensor stays within two standard errors
of the reference on all three Greeks. The tuned Monte Carlo gets there only with more
than three orders of magnitude more compute, beyond the plotted range, and the
conditioned bumps never do on delta and cross-speed, whose bias lies above that band.

Local volatility changes the numbers but not the picture (see Fig.~\ref{fig:greeks}(b)).
On a cruder chain than the figure's, two Euler sub-steps per month instead of one
Gaussian step, the ratios at $1\%$ differ by at most a factor of about two, and only
gamma at four assets changes side, so the choice of chain does not decide the outcome.

Three quantities move the balance between the sampler's variance and the tensor's rank:
the order of the Greek, the width of an asset's move between observation dates (set by
the observation frequency and, under local volatility, by the sub-steps), and the number
of assets.

\subsection{What a derivative costs a sampler}
\label{sec:lr}

Differentiating the payoff costs a sampler both variance and bias. A
$k$-th finite difference with a bump $\delta$ multiplies the noise by $\delta^{-k}$, of
which common random numbers recover only a factor $\delta^{1/2}$, since the payoff
jumps, and leaves a bias of order $\delta^{2}$. Balancing the two, with
$\delta\propto\varepsilon^{1/2}$, makes the cost grow as $\varepsilon^{-(k+3/2)}$: one
more decimal place on a third derivative costs about $3\times10^{4}$ times more, against
$100$ times for an unbiased estimator. Conditioning each date on the note's
survival~\cite{Rakhmonov2019} cancels far more of the noise but leaves the bias: with
the published bump of $1\%$, speed is off by about $1\%$.

The likelihood-ratio estimator instead moves every derivative
from the payoff onto the density of the first step. A shift of the initial log-spots
$\x_0$ changes only the density $p_1$ of the first transition, so for a derivative of
any order, with $F$ the discounted payoff of a path~\cite{Glasserman2004},
\begin{equation}
\partial^{\alpha}C_0(\x_0)=\E[F\,w_{\alpha}],\qquad
w_{\alpha}=\frac{\partial^{\alpha}p_1}{p_1}.
\label{eq:lr_weight}
\end{equation}
Every weight has mean zero, so each Greek is the covariance of the payoff with its
weight. The estimator never differentiates the payoff, so the
barriers cost it no bias, and one set of paths returns every Greek. Under geometric
Brownian motion the weights are Hermite polynomials of the first-step normals. No other
weight does better on these payoffs: a weight that is unbiased for every payoff of the
observation dates has $w_{\alpha}$ as its conditional expectation given those dates, so
by the Rao--Blackwell argument its estimate has at least the variance of
$F\,w_{\alpha}$. Malliavin weights~\cite{FournieEtAl1999} are of this kind.

The likelihood-ratio estimator pays in variance alone. Each derivative brings down a factor
of order $1/s$, with $s=\sigma\sqrt{\Delta t}$ the width of one period's step, so a given
error on a Greek of order $k$ needs $M\propto s^{-2k}$ paths. With $\bm Y$ the state at
the end of the step that carries the weight, the law of total variance gives
\begin{equation}
\mathrm{Var}(F\,w_{\alpha})=\E\big[w_{\alpha}^2\,\mathrm{Var}(F\mid\bm Y)\big]
+\mathrm{Var}\big(w_{\alpha}\,\E[F\mid\bm Y]\big).
\label{eq:lr_variance}
\end{equation}
Both terms carry $w_{\alpha}^2\propto s^{-2k}$. Control variates built on the first step
are fixed once $\bm Y$ is, so they can trim the second term but not the first, the noise of
the rest of the path. Added to our tuned estimator, the regression control variate of
Appendix~\ref{app:mc_other} left every standard error unchanged to within $3\%$, so little
of the second is left.

The read-out of Eq.~\eqref{eq:readout} computes the same expectation without sampling.
Since the weight depends on the path only through $\bm Y$, the tower property gives
$\E[F\,w_{\alpha}]=\E\big[w_{\alpha}\,\E[F\mid\bm Y]\big]$. The inner average, which a
sampler pays for in the first term of Eq.~\eqref{eq:lr_variance}, is the stored surface up
to the discount of the step. The outer one, which the sampler pays for in the second, is
taken by quadrature, and given the common factor it splits into $d$ one-dimensional
integrals, one per core. What the factor $s^{-k}$ amplifies instead of noise is
the error of the surface, as shown in Sec.~\ref{sec:resolution}. Both methods thus face
the same factor. What differs is the price of beating it.

The two prices can be read on one scale, the number of digits by which the amplified
noise or error must be beaten, about $D=\log_{10}\big(s^{-k}/(\varepsilon|v|)\big)$ for a
Greek of order $k$ and value $v$ at a relative target $\varepsilon$. Each derivative
multiplies $s^{-k}$ by $1/s$, about $17$ at monthly observation, but the Greek itself by
less. Each decade of the target adds one digit.
A sampler's cost grows as $10^{2D}$ and the solver's charge by a small factor per digit,
larger as assets are added, so the ratio climbs geometrically with the order of the Greek
and with the digits of the target.

At three assets and $1\%$, Monte Carlo's cost grows by four
orders of magnitude from delta to cross-speed, as each derivative adds to the variance
of the weight, while the tensor's charge less than doubles, for the tighter tolerances
that higher orders need. Appendix~\ref{app:gbm_ratios} gives the costs. The ratio climbs from
more than an order of magnitude in Monte Carlo's favor on delta to more than two orders
in the tensor's on cross-speed (see Fig.~\ref{fig:greeks}(a)).

Cross-speed is the extreme because, at the scored spot, just below where it changes sign
(see Fig.~\ref{fig:object}(a)), its value is a tenth of speed's while its estimator has a
quarter of speed's variance, so by Eq.~\eqref{eq:mc_cost} a given relative accuracy costs
Monte Carlo $10^2/4=25$ times more. Tightening the target tenfold multiplies every Monte Carlo
cost by a hundred and the tensor's charge by at most one and a half, so every ratio
grows about seventyfold.

\subsection{What dates, sub-steps and assets cost each method}
\label{sec:freq}

More observation dates shorten each period, which enlarges every likelihood-ratio
weight, and lengthen each path. From six dates to $36$, at three assets, Monte Carlo's
cost to $1\%$ on the second- and third-order Greeks grows by more than two orders of
magnitude, while the tensor's charge grows less than tenfold, so the ratios of these
Greeks grow by more than an order of magnitude.
At six dates and $1\%$ Monte Carlo is the cheaper method on gamma, cross-gamma and speed,
and the tensor only on cross-speed. At twelve the tensor also leads on speed and
cross-gamma, and at $36$ on gamma as well. See Appendix~\ref{app:gbm_ratios} for details.

With few dates Monte Carlo also has a stronger estimator, which conditions on the path
of the common factor, a form of conditional Monte Carlo~\cite{Glasserman2004}. Given that
path the assets move independently, so the worst-of events become products of one-asset
probabilities and only the factor is sampled, as Appendix~\ref{app:mc_other} describes.
At six dates and $1\%$ it comes within a factor of three of the tensor at three assets
and is several times cheaper at five. Its exact conditional price, however, sums over the
$2^{N-1}$ autocall patterns, so its cost doubles with every date, and by twelve dates it
is less performant than the likelihood ratio on every Greek. Few observation dates thus favor
Monte Carlo at moderate accuracy, and the tensor gains as the dates multiply or the
target tightens.

Under local volatility the change of variables of Sec.~\ref{sec:product} lets both
methods keep what they had under geometric Brownian motion: the tensor the factorized
operator of Eq.~\eqref{eq:kernel}, with a kernel Gaussian in $\tilde x$, and the sampler
the exact weight of the whole first month. Models that no change of variables makes
Gaussian over a period, such as local correlation, need Euler sub-steps, and both
methods then pay for every sub-step.
Two sub-steps per month miss the continuous diffusion by several percent on speed, which
extrapolation in the number of sub-steps corrects, and the sampler must choose where its
weight sits, since a weight on the short first sub-step is noisy.
Appendices~\ref{app:lv_weights}, \ref{app:lv} and~\ref{app:lv_routes} give the weights, the
costs and the ratios on such chains. Only the tensor could compose the sub-steps of a period into one
operator, built once and applied at every date, as Appendix~\ref{app:lv_routes}
discusses, but we have not implemented it.

Adding assets moves the ratio the other way (see Fig.~\ref{fig:greeks}). Under geometric
Brownian motion at $1\%$ the tensor's lead on speed and cross-gamma shrinks from three
assets to five, gamma passes to Monte Carlo from four assets and cross-gamma at five,
and cross-speed keeps a lead of $50$ to $400$ times. At $0.1\%$ the tensor
gains a cost advantage across derivative orders. Against the likelihood ratio, the
shift comes from the tensor's side. From three assets to five the sampler's cost grows
at most about threefold, while the tensor's charge grows four- to tenfold: its
discretization stays the same, but its rank grows, because the worst-of payoff couples
every asset, and a solve becomes two to five times costlier with each added asset, more
at tighter tolerances. A richer correlation structure acts through the rank as well: a
second factor raises it little, three sectors nearly double it, and it stays lowest
when the assets are ordered by sector, as Appendix~\ref{app:factors_rank} shows.

A full grid on the tensor's own mesh, which rounds nothing, competes with the tensor at
three assets, but at four it costs several times to two orders of magnitude more, and at
five its arrays would need hundreds of gigabytes, as Appendix~\ref{app:fullgrid} shows.
Beyond five assets the rank grows more slowly, and each added asset makes a solve only
$1.2$ to $1.8$ times costlier. At $1\%$ the tensor's lead over the likelihood ratio
narrows to a few times on speed at ten assets but holds on cross-gamma and, at seven
assets, on cross-speed, as Appendix~\ref{app:gbm_ratios} details. With six observation
dates instead of $36$, Monte Carlo conditioned on the common factor reaches $1\%$ on
every Greek in at most a core-second at seven and ten assets, while a tensor solve
that reaches $1\%$ on every Greek takes $5$ and $16$ core-seconds.

\section{Exposure tails}
\label{sec:exposure}

The Greeks differentiate the last transition of the backward solve, the one that
reaches today, while exposure uses the surfaces that the solve passes on its way, on future
dates. On such a date the note's value in each state is an expectation over the paths
that follow, which a nested sampler estimates state by state with inner paths. A stored
surface gives these values without inner paths, which removes the inner simulation and
with it the bias that the inner noise puts into a quantile.
The surface has an error of its own, but a deterministic one: it moves the potential
future exposure by no more than the surface's largest
error~\cite{DemeterfiGlauWunderlich2025}. Value functions stored by a backward
recursion~\cite{GlauPachonPoetz2021} or interpolated from a
pricer~\cite{DemeterfiGlauWunderlich2025} have served this purpose for exposure on a single
underlying. Here the tensor stores them on three to five assets, and we compare it with nested,
multilevel and regression Monte Carlo on a high quantile of the exposure.

\subsection{How we score the tail}
\label{sec:tail_scoring}

We test the stored surface on one quantity: the potential future exposure of the monthly
note of Sec.~\ref{sec:formulation} at month $18$, half-way to maturity, at the level
$\alpha=97.5\%$ of Eq.~\eqref{eq:pfe}. The note's value at month $18$ depends on the
asset prices on that date, which are random, and the PFE is the level that this value
exceeds with probability $2.5\%$. The setting is that of the Greeks, geometric Brownian
motion on three to five assets. The month-$18$ log-spots are drawn from their law under
that model, started from today's prices, $\x=\bm 0$: a Gaussian with mean $\mu t\,\1$ and
covariance $\sigma^2t\,[(1-\rho)I+\rho\,\1\1^\top]$ at $t=1.5$ years. The draw is not
conditioned on the note's survival to month $18$, so the estimand is the exposure of the
same note with its first call date at month $18$. Appendix~\ref{app:tails} states the
estimand in full and adds six and seven assets.

We measure accuracy on the part of exposure that a sampler finds hard. The expected
exposure (EE), the mean of the month-$18$ value, needs no inner simulation and is already
close to $90\%$ of the PFE. Our targets are therefore $1\%$, $0.3\%$ and $0.1\%$ of
$\mathrm{PFE}-\mathrm{EE}$. In units of the notional these are errors of $11$ to $15$, $3$ to $4$ and $1.1$ to $1.5$
basis points at three to five assets. The tightest lies a little below the sampling error
of a PFE estimated from $10^4$ month-$18$ states, each valued exactly, as many as the
exposure benchmarks of Ref.~\cite{DemeterfiGlauWunderlich2025} use. Every method, including the tensor, samples the outer
states. We repeat each run with fresh random numbers and score the root-mean-square
error of one PFE estimate, counting bias as well as noise.

The reference uses a much tighter tensor surface and many more outer states than the
scored runs. To check a reference built by the same method, we also use nested Monte
Carlo with no tensor. The two agree at every dimension within two nested-estimator
standard errors, comparable to the tightest target. Appendix~\ref{app:exposure} gives
the reference checks and the gap at each dimension. The mechanisms behind inner-noise
bias and the shortcut for EE are explained below.

The methods differ in how they value the note in a sampled month-$18$ state. Nested
Monte Carlo simulates inner paths from that state to maturity, with the budget split
between states and inner paths for the least error and a jackknife that cancels the
leading bias of the inner noise, both due to Gordy and Juneja~\cite{GordyJuneja2010}.
Multilevel Monte Carlo~\cite{Giles2008,GilesHajiAli2019} spreads the inner simulation over levels of
growing size, and regression Monte Carlo~\cite{LongstaffSchwartz2001} replaces it by a
fitted polynomial. The tensor reads the value off the surface it stored for month $18$.

The tensor is charged in two ways. In the main comparison it pays for the whole build of
that surface, from maturity to month $18$, at the loosest rounding tolerance that meets
the target, and for its outer sampling. A solve run for the Greeks, however, passes month
$18$ on its way to today, so we also report the case in which the surface already exists
and only the sampling is charged.

\begin{figure}[t]
\centering
\includegraphics[width=\linewidth]{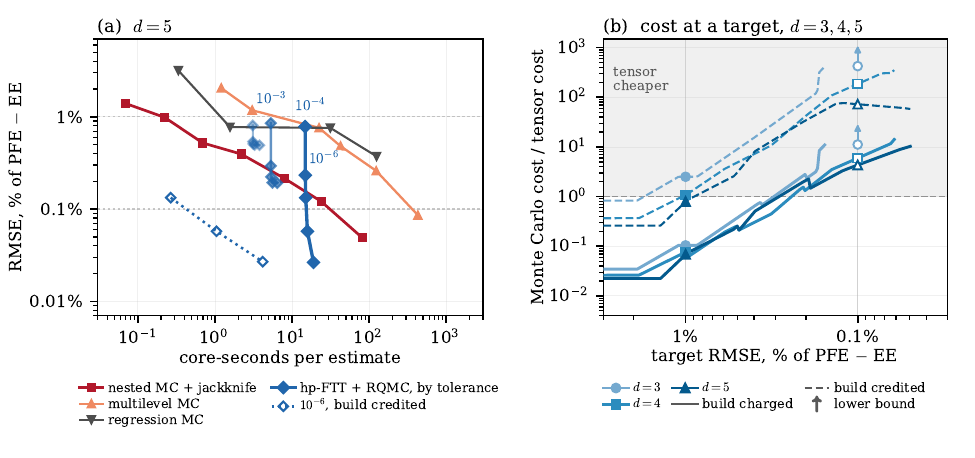}
\caption{\textbf{Exposure tails: Monte Carlo versus hp-FTT.}
Potential future exposure at $97.5\%$ at month $18$ of the monthly note under
geometric Brownian motion and the unconditional month-$18$ state law. Errors are
root-mean-square errors of the PFE that one run returns, against the tensor reference
checked by nested Monte Carlo, relative to $\mathrm{PFE}-\mathrm{EE}$.
\textbf{(a)} Five assets. Nested Monte Carlo uses the tuned split and the jackknife, and
multilevel and regression Monte Carlo are each drawn at the best of their settings at
every cost (pilot size and number of levels, polynomial degree and number of regression
paths). The tensor's month-$18$ surface is built at the rounding tolerances $10^{-3}$,
$10^{-4}$ and $10^{-6}$, each read at $512$ to $131{,}072$ randomized Sobol outer states
and charged its build. Dotted: the $10^{-6}$ surface charged its outer sampling only, as
when a solve run for the Greeks at that tolerance has already built it. Dashed lines: the
targets $1\%$ and $0.1\%$. \textbf{(b)} Cost of the cheapest Monte Carlo estimator over that of
the tensor, at its cheapest tolerance (solid) or read off a solve already run for the
Greeks (dashed), at three to five assets, shaded where the tensor is cheaper. Curves end
where Monte Carlo's cost is no longer measured, and a marker with an upward arrow is a lower bound. Filled and open markers: the targets $1\%$
and $0.1\%$.}
\label{fig:exposure}
\end{figure}

\subsection{Which method is cheaper}
\label{sec:tail_result}

As for the Greeks, Monte Carlo wins at loose targets and the tensor at tight ones, with
the crossover near $0.3\%$ of $\mathrm{PFE}-\mathrm{EE}$ (Fig.~\ref{fig:exposure}(b)).
At five assets, a nested sampler reaches $1\%$ in a fraction of a core-second, before
the tensor can build even its coarsest surface. That build must cross every date from
maturity to month $18$, whatever the tolerance. Once built, however, the surface values
each outer state without inner noise. Error then falls with outer sampling, faster at
randomized Sobol points, and the sampling costs little beside the build. This makes the
tensor curves in Fig.~\ref{fig:exposure}(a) drop almost vertically, while nested error
falls slowly. Below a few tenths of a percent the order reverses.

If the surface already exists, from a solve run for the Greeks, the tensor pays only for
its outer sampling (dashed in Fig.~\ref{fig:exposure}(b)). It is then at least ten times
cheaper than Monte Carlo at $0.3\%$ and seventy times or more at $0.1\%$, and at $1\%$
the two are about level. Appendix~\ref{app:exposure_cmp} gives the costs behind these
statements, a comparison at equal compute, and two checks that the tensor's lead does not
come from its randomized Sobol points: with pseudo-random outer states the tensor still
leads at equal compute, and nested Monte Carlo gains little from the same Sobol points.

The comparison is for one date, the unconditional state law, geometric Brownian motion
and monthly observation. A profile over dates, which we have not measured, would favor
the surface further, since the solve that reaches month $18$ stores the surface of every
later date while nested Monte Carlo repeats its inner simulation at each. The advantage
is specific to nonlinear functionals of the note's value, such as quantiles. The expected
exposure is the mean of that value, and the value is itself a conditional expectation of
the payments that follow month $18$. By the tower property, the mean of a conditional
expectation is a plain expectation, so a sampler estimates the expected exposure from
whole paths run from today to maturity, one per outer state, with no inner simulation.
This works because the note's value is positive, so that the exposure is the value
itself. A quantile depends on the value state by state and allows no such shortcut.

\subsection{What a quantile costs a sampler}
\label{sec:tail_bias}

Where a Greek costs a sampler variance, a tail of the note's future value costs it bias. At an
outer state $\bm x$ the value is an expectation over the paths that follow the exposure
date, and a nested sampler estimates it from $N_{\rm in}$ inner paths, with a noise of
mean zero and variance $\nu^2(\bm x)/N_{\rm in}$, where $\nu^2(\bm x)$ is the variance of
one inner payoff. For the expected exposure the noise averages out, but for a quantile it
does not, since spreading a distribution moves its quantiles. Because the noise has mean
zero, the shift enters through its variance, and for the quantile $q$ of the value, whose
density is $f$, it is~\cite{GordyJuneja2010}
\begin{equation}
\delta q=-\frac{1}{2N_{\rm in}\,f(q)}\,\frac{d}{dq}\Big[f(q)\,\bar\nu^2(q)\Big],
\label{eq:quantile_bias}
\end{equation}
with $\bar\nu^2(q)$ the average of $\nu^2$ over the states whose value is $q$. Here the
shift is upward: in our runs the plain nested estimator almost always overstates the
exposure (see Appendix~\ref{app:exposure}). Only more inner paths remove it, and only as
$1/N_{\rm in}$.

The bias sets the price of accuracy. A budget of $N_{\rm tot}$ inner samples covers
$N_{\rm tot}/N_{\rm in}$ outer states, so the mean-square error is the squared bias, of order
$1/N_{\rm in}^2$, plus the variance of a quantile of those states, of order
$N_{\rm in}/N_{\rm tot}$. The split that balances the two takes $N_{\rm in}\propto N_{\rm tot}^{1/3}$,
and the error falls as $N_{\rm tot}^{-1/3}$: a target ten times tighter costs a thousand times
more, where an unbiased estimator would pay a hundred. The jackknife
cancels the $1/N_{\rm in}$ term and leaves one of order
$1/N_{\rm in}^2$. The balance then gives an error falling as $N_{\rm tot}^{-2/5}$, about $300$
times the cost per decade, and over the measured range the tuned estimator's error
falls as cost$^{-0.40}$ to cost$^{-0.47}$.

The other two samplers carry biases of their own: the multilevel estimator's Newton
step from a nested pilot leaves one that more levels do not remove, and the regression
estimator's, set by its basis, is as large as its spread from five assets up. The stored
surface gives the value at every outer state without inner paths. Its own error is deterministic and
moves the quantile at first order, but no kernel derivative amplifies it, as for a
Greek, and a build a decade tighter costs on average $1.5$ to $1.7$ times more, as
Appendix~\ref{app:exposure} shows. The only sampling left is the outer draw, whose error
falls as the inverse square root of the number of outer states, and faster at randomized
Sobol points.

Per decade of accuracy the cost of nested Monte Carlo grows more than a hundredfold and
the tensor's only a few times, as for the Greeks. More assets move the balance the other
way, since they cost the tensor more, through its rank, than the sampler
(see Appendix~\ref{app:exposure_cmp}).

\section{Discussion and conclusions}
\label{sec:discussion}

In this work we carried the backward recursion of a worst-of autocallable on a
functional tensor train whose polynomial pieces end at the barriers, and set it against
the strongest Monte Carlo estimators for the tested set of spot Greeks that we could build or
find published. The two methods pay for different things. Sampling pays in variance for
every order of the derivative, more so the more often the note is observed, and in bias
for a quantile of the note's future value. The solver pays in rank for every asset.

On notes observed monthly on three to five assets, which method is cheaper depends on
the order of the Greek and on the accuracy. At $1\%$ accuracy Monte Carlo is cheaper on
the deltas, second-order Greeks go either way, and the solver is about an order of
magnitude cheaper on speed and $50$ to $400$ times cheaper on cross-speed. At
$0.1\%$ accuracy the solver beats Monte Carlo across derivative orders, under geometric
Brownian motion and local volatility alike.

A desk usually needs several spot Greeks at once. A likelihood-ratio run shares its
payoff paths among them, with a path budget largely set by the most demanding Greek, and
the tensor shares its stored surface. Additional weights and read-outs still cost work,
while conditioned bumps need the union of the requested stencils. We have not benchmarked
joint requests, so we quote costs Greek by Greek: the third-order ratios would overstate
the advantage for a desk that needs only lower orders.

The tail of the note's future value follows the same pattern: Monte Carlo is about ten
times cheaper at $1\%$ of
the tail's excess over the expected exposure, the solver is cheaper at $0.1\%$, about one
basis point of notional, even when it pays for building its surface, and a
surface already built for the Greeks gives the tail for little more than the cost of
sampling the future states.

The ratios rest on three choices of scoring. First, they charge the solver a full solve
for a single Greek at a single spot, although a surface can be read again at any spot
for a few milliseconds per Greek, and a solve run for the Greeks also serves the
exposure tail. What this reuse is worth depends on the workload, which we have not
benchmarked, and a surface is valid only for the contract terms and model parameters it
was built with. Second, they compare core-seconds on one laptop. Monte Carlo parallelizes
easily, over many cores or on a GPU, so in wall-clock time it can be faster than these
ratios suggest. Third, they set the solver's tolerance with the help of deep Monte
Carlo, which a user would not have, and hold Monte Carlo to one standard error. Without
that help, the solver's median charge would rise about fourfold, as would Monte Carlo's
charge at two standard errors. These similar cost multipliers do not imply equal
reliability: the decade estimate sometimes accepts values outside their
targets and cannot detect discretization error. Section~\ref{sec:baselines} and
Appendix~\ref{app:scoring_checks} give the checks.
Beyond the scoring, we have not tested what more accurate Greeks are worth in a hedge,
for instance by simulating the profit and loss of a hedged book.

Both methods see the model only through the step from one date to the next, which the
sampler draws and the solver averages over. Local volatility shows that a step
that depends on the state does not, by itself, change the outcome: a change of variables
makes each month one Gaussian step again, for the sampler as for the solver, and the
ratios stay of the same order. Models with no such change of variables, such as local
correlation, or stochastic local volatility with correlation between spot and variance,
need sub-steps within each month, and both methods then pay for every sub-step
as Sec.~\ref{sec:freq} explains. Stochastic volatility may favor the sampler: once the path of the
variance is known, the log-spots are again Gaussian over each month, so a sampler that
conditions on that path can put its weight on a whole month, as under geometric
Brownian motion, while the solver would have to carry the variance as one more coordinate.
Appendix~\ref{app:mc_other} discusses this possibility. We have run the solver on none of
these models.

Nor did we calibrate either of our models to market prices, so the comparisons measure
the cost of pricing under each model as specified. Calibrated local volatility misprices
the forward smile, on which the value of an autocallable
depends~\cite{Bergomi2015,FarkasFerrariUlrych2022}, so stochastic local volatility,
widely used for these notes~\cite{Sherif2019,FarkasFerrariUlrych2022,DeColKuppinger2017},
is the natural next test for the solver. A solver that handles both models would give
their Greeks free of sampling noise even without a speed-up, so that the difference
between local and stochastic local volatility could be measured more easily.

All the Greeks we scored are derivatives with respect to today's spots. These enter
only the first step after inception, so the solver computes each of these Greeks from
the surface it stores after that step (Eq.~\eqref{eq:readout}). The volatility and the
correlation, by contrast, set how the assets move at every step until maturity. Changing
them changes every stored surface, so their Greeks, such as vega and vanna, cannot be read
off one surface in this way. We have not computed them, although
they matter in practice: Salon~\cite{Salon2019} traces the steady loss, or negative
carry, of hedged autocallable books to vanna, the change of delta with volatility. These
Greeks are costly for a sampler too: a likelihood-ratio weight for them has one term for
every step of the path, so that its variance grows with the number of steps. Some risk
questions also combine a derivative with a value at a future date, such as the
sensitivity of the potential future exposure, or the Greeks in the future scenarios that
dynamic initial margin requires~\cite{RuizZeronDIM}. A sampler pays for these twice, in
variance for the derivative and in bias for the future value. All of these are promising cases for the
solver, which we leave to future work.

Where a note is observed often and hedged on Greeks of high order or managed through the
tail of its future value, a deterministic solver is worth trying before a sampler. Where
it is observed rarely or hedged only on its deltas at percent accuracy, sampling remains the
better tool. Each added asset moves the balance toward sampling, though more slowly beyond
a few assets. The features that keep the rank low, events that act one asset at a time
and smoothing between observation dates, are common to many structured products.
Wherever they are present, carrying the value function backward as a tensor network turns
what costs sampling the most, derivatives of high order and tails of future values, into
by-products of a single deterministic solve.

\subsection*{Acknowledgments}

We thank Rob Otter for executive support and valuable feedback on this project. The
authors also thank Shouvanik Chakrabarti for insightful discussions. We further
acknowledge our colleagues at the Global Technology Applied Research center of
JPMorganChase.

\subsection*{Disclaimer}

This paper was prepared for informational purposes by the Global Technology Applied Research center of JPMorgan Chase \& Co. This paper is not a merchandisable/sellable product of the Research Department of JPMorgan Chase \& Co. or its affiliates. Neither JPMorgan Chase \& Co. nor any of its affiliates makes any explicit or implied representation or warranty and none of them accept any liability in connection with this paper, including, without limitation, with respect to the completeness, accuracy, or reliability of the information contained herein and the potential legal, compliance, tax, or accounting effects thereof. This document is not intended as investment research or investment advice, or as a recommendation, offer, or solicitation for the purchase or sale of any security, financial instrument, financial product or service, or to be used in any way for evaluating the merits of participating in any transaction.

\subsection*{AI Use Disclosure}

Large language model tools were used interactively at all stages of the preparation of
this manuscript: from the earliest exploratory research stages to the final exposition
and writing steps. The final manuscript was substantially revised by the authors, who
remain solely responsible for all results.

\nocite{Trefethen2019,BouzoubaaOsseiran2010,Tong2019}
\bibliographystyle{apsrev4-1}
\bibliography{references}

\appendix

\newcommand{\apptocsec}[2]{\par\smallskip\noindent\hyperref[#1]{\textbf{Appendix~\ref*{#1}.\quad #2}}\dotfill\pageref{#1}\par}
\newcommand{\apptocsub}[2]{\noindent\hspace*{2em}\hyperref[#1]{\ref*{#1}\quad #2}\dotfill\pageref{#1}\par}

\clearpage 

\section*{Contents of the appendices}

\apptocsec{app:solver_detail}{The hp-FTT solver in detail}
\apptocsub{app:kernel}{From the dynamics to the transition kernel}
\apptocsub{app:propagator}{Applying the transition to a train}
\apptocsub{app:events_detail}{Events, terminal step and the edge of the box}
\apptocsub{app:readout}{The read-out}
\apptocsub{app:resolution}{Resolution of a Gaussian-smoothed value}
\apptocsub{app:err}{Error control: discretization, tolerance and error estimates}
\apptocsub{app:fullgrid}{The full grid}
\apptocsec{app:baselines}{The Monte Carlo baselines}
\apptocsub{app:mc}{The tuned likelihood-ratio estimator}
\apptocsub{app:published}{Estimators that differentiate the payoff}
\apptocsub{app:lv_weights}{Weights on an Euler chain}
\apptocsub{app:mc_other}{Applicability of alternative estimators}
\apptocsec{app:factors}{More than one common factor}
\apptocsub{app:factors_data}{Factors in equity data}
\apptocsub{app:factors_staged}{Staged transitions}
\apptocsub{app:factors_rank}{Cost, rank and the order of the assets}
\apptocsec{app:measurements}{Scoring and measurements}
\apptocsub{app:scoring}{Computational setup}
\apptocsub{app:gbm}{Geometric Brownian motion: agreement, contract terms and a market note}
\apptocsub{app:gbm_ratios}{Geometric Brownian motion: costs and ratios}
\apptocsub{app:lv}{Local volatility: the two-sub-step chain}
\apptocsub{app:lv_routes}{Local volatility: time steps and the continuous model}
\apptocsub{app:exposure}{Exposure tails: reference and baselines}
\apptocsub{app:exposure_cmp}{Exposure tails: comparisons and controls}

\section{The hp-FTT solver in detail}
\label{app:solver_detail}

This appendix gives the details of the solver of Sec.~\ref{sec:solver}: how the
transition follows from the dynamics and how it is applied to a stored train, how the
events, the terminal step and the edge of the box are treated, and how the Greeks are
read out. It then shows why the layer rule resolves the value, how the discretization
and rounding errors are controlled, and how a full grid checks the rounding.

\subsection{From the dynamics to the transition kernel}
\label{app:kernel}

We derive the transition of Eq.~\eqref{eq:kernel} from the dynamics of
Eq.~\eqref{eq:gbm}. Constant coefficients make the integrated step exact:
$x_i'=x_i+\mu\Delta t+\beta Z+s_\perp\xi_i$, with $Z$ and the $\xi_i$ independent
standard normals, $\beta=\sigma\sqrt{\rho\Delta t}$ and
$s_\perp=\sigma\sqrt{(1-\rho)\Delta t}$. Given $Z=z$, the asset states are independent
Gaussians with means $x_i+\mu\Delta t+\beta z$ and variance $s_\perp^2$, hence density
\begin{equation}
k_z(x_i,y)=\frac{1}{s_\perp}\,\phi\Big(\frac{y-x_i-\mu\Delta t-\beta z}{s_\perp}\Big),
\label{eq:kernel_z}
\end{equation}
the kernel of Eq.~\eqref{eq:kernel} with $z$ in place of a node, $k_g=k_{z_g}$.
Averaging over the factor $Z$, whose density is $\phi$, the law of total expectation
gives the transition exactly:
\begin{equation}
\E\bigl[f(\bm X_{k+1})\,\big|\,\bm X_k=\x\bigr]
=\int\phi(z)\int\prod_{i=1}^{d}k_z(x_i,y_i)\,f(\bm y)\,d\bm y\,dz .
\label{eq:kernel_exact}
\end{equation}
The correlated Gaussian density of the period is thus a mixture over the factor of
products of one-asset densities.

Equation~\eqref{eq:kernel} replaces the outer integral
of Eq.~\eqref{eq:kernel_exact} by the Gauss--Hermite rule for the standard normal
density: $n_z$ nodes $z_g$ and weights $w_g$ with $\sum_gw_g=1$, exact for every
polynomial in $z$ of degree below $2n_z$. In terms of the rule for the weight $e^{-t^2}$,
with nodes $t_g$ and weights $\omega_g$, the nodes are $z_g=\sqrt2\,t_g$ and the weights
$w_g=\omega_g/\sqrt\pi$. The factor quadrature is the only approximation in
Eq.~\eqref{eq:kernel}. Its application to a stored train also involves spatial
quadrature, Legendre projection and rounding, described in Appendix~\ref{app:propagator}.
These are spatial and compression errors: the GBM transition crosses a whole
observation period without time-stepping error.

The derivation does not require identical assets. Per-asset volatilities, dividend
yields and factor loadings give each kernel its own drift, factor shift and
idiosyncratic width, while preserving the one-factor quadrature bond. The market
example in Appendix~\ref{app:gbm} uses this extension. Under local
volatility the same steps apply to each month of the near-continuum chain, in the
coordinates $\tilde x$ with the drift linearized at the asset's level, and to each Euler
sub-step, with the volatility frozen at its start: given the factor, every move is again
Gaussian, with a mean and a width that depend on the asset's own level. These are the
transitions of the two chains, not of the continuous model. With $R$ factors the
assets are independent given all of them, and the outer integral runs over $\R^R$
(see Appendix~\ref{app:factors}).

Since the assets are independent given the factor, one may expect the price to reduce to an integral over the factor of
products of one-asset expectations. Over one step it does: the transition of
Eq.~\eqref{eq:kernel} and the terminal step of Sec.~\ref{sec:events} are computed this
way. Over the contract it does not, for two reasons. The factor is drawn afresh in every
period, while each asset's idiosyncratic moves accumulate from date to date, so a
recursion over the dates must carry the vector of log-spots, not the factor. And the
events couple the assets again: the contract survives a date only if not every asset is
at or above the autocall level, an event whose indicator,
$1-\prod_i\1_{\{x_i\ge b_{\rm a}\}}$, is not a product. Even given the whole factor path,
the law of the log-spots on the surviving paths is therefore a signed sum of $2^k$
products of one-asset laws after $k$ dates, one for each subset of those dates, and the
conditioning estimator of Sec.~\ref{sec:freq} pays for this sum on every sampled factor
path, at a cost that doubles with every date.

The backward recursion carries the joint state instead, the continuation value as a
function of all $d$ log-spots. Expanded without rounding, it would produce the same sum
with the factor path on a quadrature, of order $(2n_z)^N$ products of one-asset
functions after $N$ dates. Rounding after every operator recombines them, and the rank
that remains grows much more slowly. Table~\ref{tab:rank_dates} shows the measured
ranks: increasing the number of dates is less costly in rank than adding assets at the
tighter tolerance.

\begin{table}[t]
\caption{\textbf{Rank growth across observation dates.} Largest tensor rank in the
GBM backward solve, with the contract and discretization of the main comparisons.
The tolerance controls rounding; the polynomial resolution is fixed for each step length.}
\label{tab:rank_dates}
\centering
\renewcommand{\arraystretch}{1.15}
\begin{tabular}{?cc|ccc?}
\thickhline
assets & tolerance & $6$ dates & $12$ dates & $36$ dates \\
\hline
$3$ & $10^{-6}$ & $22$ & $24$ & $28$ \\
$5$ & $10^{-6}$ & $26$ & $30$ & $40$ \\
$3$ & $10^{-8}$ & $28$ & $30$ & $35$ \\
$5$ & $10^{-8}$ & $54$ & $68$ & $89$ \\
\thickhline
\end{tabular}
\end{table}
\subsection{Applying the transition to a train}
\label{app:propagator}

In every comparison of this paper the box has $L=3$ under geometric Brownian motion,
which holds each price between $e^{-3}\approx0.05$ and $e^{3}\approx20$ times its
initial level. Over the note's three years an asset's log-spot moves with a standard
deviation of $\sigma\sqrt T\approx0.35$, so the lower edge lies $7.6$ such standard
deviations below the protection barrier and the upper edge $8.7$ above the autocall
barrier. Under local volatility, which rises into the left tail, the price needs
$L=4.5$. Wider boxes change no price or Greek at the accuracies we score, as
Appendices~\ref{app:err} and~\ref{app:lv} show.

Each axis of the box is split at the three log-barriers,
$b_{\rm p}=\log0.7$, $b_{\rm c}=\log0.8$ and $b_{\rm a}=0$, and all four resulting
intervals, the outer two included, are subdivided into panels no wider than $0.5$.
Panels that touch a barrier get their modes from the layer rule, in every solve of this
paper. The others lie in the tails, where the value is smooth, and the production solves
merge each stretch of them into a single panel of $16$ modes rather than several panels
of $8$. At monthly observation this cuts an axis from $130$ modes to $82$ and makes a
solve $1.5$ to $1.9$ times faster at three to five assets, except at the tolerance
$10^{-8}$ at four and five assets, where the merged tails raise the rank and a solve is
only $1.17$ times faster at four assets and $15\%$ slower at five. Measured against a
solve with every setting refined, neither mesh is consistently the more accurate. Under local
volatility these tail panels also end at the two points where the clamps of
Eq.~\eqref{eq:lv_sigma} begin to bind, since the volatility has a kink there.

\begin{figure}[H]
\centering
\includegraphics{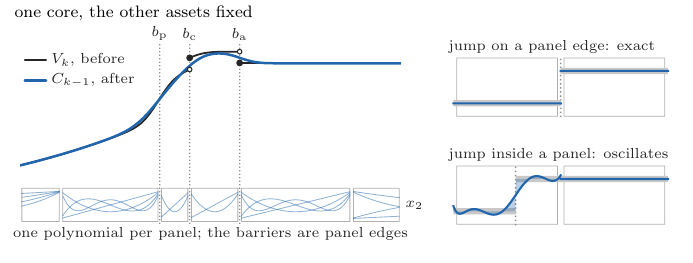}
\caption{\textbf{Barriers on panel edges keep the events exact.} One core across the
transition of Fig.~\ref{fig:solver}, one date before maturity, with the other assets far
above the barriers and the coupon enlarged so that its step shows. The value jumps at the
coupon and autocall barriers before the transition and is smooth after it. Each panel
below holds one polynomial. Right: a step is represented exactly when its jump lies on a
panel edge and oscillates when it lies inside a panel.}
\label{fig:events_mesh}
\end{figure}

When the first period is sub-stepped differently from the others, its panels get their
modes from the width of its own sub-steps. Under geometric Brownian motion its
half-month sub-steps raise the panels at the barriers from $16$, $8$, $8$ and $18$ modes
to $22$, $8$, $11$ and $25$, while the tails keep $16$. The rule thus adds modes where a
shorter step sharpens the layers, and nowhere else. The layer rule sets
neither the factor nodes nor the rank. Input quadrature instead resolves the kernel,
even where the value is nearly flat. Table~\ref{tab:mesh_resolution} separates these
budgets, and Fig.~\ref{fig:events_mesh} shows one core on the mesh.

\begin{table}[t]
\caption{\textbf{Stored resolution and spatial quadrature per asset.} Monthly
observation; GBM uses one full-period step except for its first period. The stored
panels may merge tail intervals, whereas input panels remain at most $0.5$ wide.
Node counts are within the box; boundary nodes are appended separately. The GBM
barrier panels have $8$ to $18$ modes; the first half-month steps use the
finer allocation described in the text.}
\label{tab:mesh_resolution}
\centering
\renewcommand{\arraystretch}{1.15}
\begin{tabular}{?l|ccccc?}
\thickhline
model and chain & $L$ & stored panels & modes & input nodes & nodes/input panel \\
\hline
GBM, monthly & $3$ & $6$ & $82$ & $434$ & $31$ \\
local volatility, transformed & $4.5$ & $8$ & $114$ & $1{,}176$ & $56$ \\
local volatility, two Euler steps & $4.5$ & $8$ & $129$ & $1{,}659$ & $79$ \\
\thickhline
\end{tabular}
\end{table}

The rounding of Sec.~\ref{sec:repr} bounds the part it discards in the $L^2$ norm of the
function on $[-L,L]^d$, as the continuous rounding of the functional tensor train
does~\cite{GorodetskyKaramanMarzouk2019}. The cores store Legendre coefficients, and the
basis functions do not have unit norm: on a panel $P$ of width $h$,
$\int_P\varphi_{P,q}^2\,dx=h/(2q+1)$. Singular values computed from the raw coefficients
would therefore measure the coefficients rather than the function, and give modes of
high degree and modes on narrow panels more weight than they have in the function.
Before each rounding the solver instead multiplies every coefficient by
$\sqrt{h/(2q+1)}$, the norm of its basis function, and afterwards divides by it again.
The scaled train holds the coordinates of the function in an orthonormal basis, and the
squares of these coordinates sum to the squared $L^2$ norm of the function, so singular
values computed from them measure the function itself.

At a factor node $z_g$ of Eq.~\eqref{eq:kernel} the transition smooths each core $G_i$
of a train separately, as in Eq.~\eqref{eq:contraction}, by a one-dimensional integral
over the next asset state $y$, approximated by panelwise Gauss--Legendre quadrature:
\begin{equation}
(\mathcal K_{i,g}G_i)(x)=\int k_g(x,y)G_i(y)\,dy
\;\approx\;\sum_{j=1}^{m_i}\omega_{i,j}\,k_g(x,y_{i,j})G_i(y_{i,j}).
\label{eq:spatial_quadrature}
\end{equation}
The nodes $y_{i,j}$, with weights $\omega_{i,j}$, are not factor nodes but the points of
the axis of asset $i$ at which the core is evaluated, the same at every factor node.
They lie on the \emph{input} panels,
those of the mesh before the tails are merged, all at most $0.5$ wide and each inside
one stored panel, where the core is a single polynomial. The kernel, however, is a
Gaussian as wide as an asset's own move over one step, and the input nodes must resolve
it everywhere in the box, even in the tails, where the value needs few modes. Every
input panel therefore carries the same number of nodes, set by the width of the kernel.
For the monthly GBM step the allocation is about two and a half nodes per standard
deviation, sufficient to integrate the kernel to machine precision. The node counts in
Table~\ref{tab:mesh_resolution} exceed the stored mode counts because quadrature must
resolve the narrow kernel across the whole domain, not only the value's barrier layers.

Equation~\eqref{eq:spatial_quadrature} gives the smoothed core $f=\mathcal K_{i,g}G_i$ at
any point $x$, but $f$ is not a polynomial, since smoothing a polynomial on a finite
panel produces error functions. On each \emph{output} panel $P$ of the stored mesh, the
solver therefore replaces $f$ by its projection onto the panel's modes
$\varphi_{P,q}$, the polynomial of degree below $p_P$ closest to $f$ in the $L^2$ norm.
Its coefficients are the integrals $a_{P,q}=(2q+1)\,h^{-1}\!\int_P f\,\varphi_{P,q}\,dx$,
which the solver takes by Gauss--Legendre quadrature at $p_P+2$ points of the panel.
Only these coefficients are stored. The input nodes serve to compute them, so they add
to the work of a transition but not to the size of the train. Done for every core at
every factor node, the projections give one train per factor node. Added all at once,
they could reach $n_z$ times the rank of the train they came from, so the solver adds
them one at a time, the heaviest weight first, and rounds after each addition.

\subsection{Events, terminal step and the edge of the box}
\label{app:events_detail}

The event map is applied term by term. Write $\1_{\rm c}=\1_{\{\min_ix_i\ge b_{\rm c}\}}$ and
$\1_{\rm a}=\1_{\{\min_ix_i\ge b_{\rm a}\}}$ for the coupon and call indicators of
Eq.~\eqref{eq:event}, so that $V_k=c\,\1_{\rm c}+\1_{\rm a}+C_k-\1_{\rm a}C_k$. The solver never stores
$V_k$, which jumps at the barriers. It applies the transition to each of the four
terms and adds the results,
\begin{equation}
C_{k-1}=e^{-r\Delta t}\big[c\,\mathcal K\1_{\rm c}+\mathcal K\1_{\rm a}+\mathcal KC_k
-\mathcal K(\1_{\rm a}C_k)\big],
\label{eq:four_terms}
\end{equation}
which is $e^{-r\Delta t}\mathcal KV_k$ because $\mathcal K$ is linear. $\1_{\rm c}$ and $\1_{\rm a}$
are trains of rank one whose $i$-th core is the step $\1_{\{x_i\ge b\}}$, with $b$
the coupon or the call barrier. Their transitions do not change from date to date and
are computed once for each step length.

$\1_{\rm a}C_k$ is formed core by core: each core of $C_k$ is set to zero
at the input nodes below $b_{\rm a}$. Since $b_{\rm a}$ is an edge of the input
panels, this drops whole panels, keeps the rank $\chi$, and leaves the quadrature of
Eq.~\eqref{eq:spatial_quadrature} with a smooth integrand on every remaining panel.
Before the transition the last two terms cancel on the called region, where $\1_{\rm a}=\1_{\rm c}=1$,
so that $V_k=1+c$ there: the notional and the coupon, and nothing after them. Each
transitioned term is projected onto the same output panels with the same orders, so
the sum in Eq.~\eqref{eq:four_terms} is an exact train whose cores stack those of the
four terms, of rank at most $n_z(2\chi+2)$. The transitions of $C_k$ and $\1_{\rm a}C_k$ are
rounded as they are formed, and the sum is rounded again.

It is cheaper to apply $\mathcal K$ to each of the four terms and add the results than to
form $V_k$ first and apply $\mathcal K$ to it, because applying $\mathcal K$ to a train
costs in proportion to the square of its rank: every core holds $\chi_{i-1}\chi_i$ one-asset
functions, and the kernel smooths each of them. Formed first, $V_k$ has rank up to
$2\chi+2$, and smoothing it costs about $4\chi^2$. Term by term the transition costs about
$2\chi^2$, $\chi^2$ each for $C_k$ and $\1_{\rm a}C_k$, and nothing at each date for $\1_{\rm c}$ and $\1_{\rm a}$,
whose transitions are computed once. The discount $e^{-r\Delta t}$ covers the whole
period. On an Euler chain Eq.~\eqref{eq:four_terms} takes the first sub-step back from
the date, and the other sub-steps of the period act on its smooth result.

The terminal step is in closed form. Applied to Eq.~\eqref{eq:layer_cake}, the transition
of Eq.~\eqref{eq:kernel} acts on each indicator one asset at a time, and each one-asset
integral is a normal tail probability:
\begin{equation}
\begin{aligned}
C_{N-1}(\x)&=e^{-r\Delta t}\sum_{g=1}^{n_z}w_g\Big[c\,P_g(B_{\rm c},\x)
+(1-B_{\rm p})\,P_g(B_{\rm p},\x)+\int_0^{B_{\rm p}}P_g(\ell,\x)\,d\ell\Big],\\
P_g(\ell,\x)&=\prod_{i=1}^{d}\int k_g(x_i,y)\,\1_{\{y>\log\ell\}}\,dy
=\prod_{i=1}^{d}\bar\Phi\Big(\frac{\log\ell-m_g(x_i)}{s(x_i)}\Big),
\end{aligned}
\label{eq:terminal_product}
\end{equation}
where $P_g(\ell,\x)$ is the probability that every asset ends above the level $\ell$
when the factor sits at the node $z_g$, $\bar\Phi$ is the standard normal survival
function,
and $m_g(x_i)$ and $s(x_i)$ are the mean and width of asset $i$'s move,
$x_i+\mu\Delta t+\beta z_g$ and $s_\perp$ under geometric Brownian motion
(see Eq.~\eqref{eq:kernel}). Each $P_g$ is a train of rank one.

The solver evaluates the integral by Gauss--Legendre quadrature at $Q_t=96$ levels, so
each factor node contributes $Q_t+2$ such trains, one per level plus the coupon and the
redemption, $n_z(Q_t+2)=1{,}176$ in all, which it adds up and rounds as they
accumulate. On an Euler chain the same sum, with the mean and width of a
sub-step, Eq.~\eqref{eq:lv_kernel}, takes the place of the last sub-step, and the
other sub-steps of the last period follow as operators. No stored function ever
carries the kink of the payoff along the ties.

Near $\pm L$, part of a kernel's mass falls outside the box,
where no value is stored. The solver continues the train by its value at the nearest
edge, coordinate by coordinate, which keeps it a tensor train of the same rank, and
integrates each kernel over its whole reach. Gauss--Legendre panels are appended to
every axis, out to nine idiosyncratic standard deviations beyond the largest factor
shift of a step ($0.60$ on each side for a monthly step under geometric Brownian
motion, and under local volatility $2.12$ below the box, where the volatility is largest,
and $0.12$ above it). The kernels and the event indicators, which are known
everywhere, are evaluated at the appended nodes, the train at the edge, and the
result is projected onto the panels of the box as before. The terminal step and the
read-out do not change.

Beyond the box the value is nearly flat, deep below the protection barrier and high
above the autocall barrier, so continuing it by its edge value is nearly exact. The
boxes are chosen wide enough to keep it so (see Appendix~\ref{app:lv}). Cutting the
train at the edge instead would leave a false step there, which costs rank. At
tolerance $10^{-8}$ under geometric Brownian motion, the continuation lowers the largest
rank from $54$ to $35$ at three assets, from $199$ to $80$ at four and from $233$ to $89$
at five, and makes a solve about two, five and ten times faster. Under local volatility,
at the same tolerance, it lowers the largest rank from $47$ to $40$ at three assets, from
$133$ to $56$ at four and from $155$ to $61$ at five, and makes a solve two to six times
faster.

\subsection{The read-out}
\label{app:readout}

The read-out of Sec.~\ref{sec:events} differentiates the kernel of the sub-step from
inception and integrates it against the surface stored one sub-step later,
Eq.~\eqref{eq:readout}. Below we explain the length of that sub-step, why the kernel
rather than the surface is differentiated, and how its derivatives are taken.

Under geometric Brownian motion the first period is split into two half-month
sub-steps. Without the split, the sub-step from inception would reach the first
observation date, and the differentiated kernel would act on the value at that date,
after its events, which jumps at the barriers. With the split, it acts on the
continuation value half a month in, which is smooth. The split thus changes what the
kernel acts on, not what is differentiated. The jumps do no harm to the integral over
the spots, whose panels end at the barriers, but the sum over the factor nodes
converges far more slowly against them. Read out across the whole first month with the
solve's $12$ factor nodes, a three-asset solve moves cross-gamma by $1.0\%$ and
cross-speed by $0.9\%$ at the autocall barrier, to $2.6$ and $1.9$ standard errors from
the deep Monte Carlo reference, where the half-month read-out stays within $0.5$. With
$24$ factor nodes the whole-month read-out agrees with the half-month one to within
$0.015\%$ on every Greek at both spots, so doubling the factor nodes works as well. The
near-continuum chain of local volatility, whose step spans a month, does this
(see Sec.~\ref{sec:repr}), while under geometric Brownian motion the split keeps the read-out
on the solve's own $12$ nodes.

A query at a new spot costs the kernel read-out a few milliseconds at three assets, and
more with more assets, since its work grows with the square of the rank. A possible
alternative is to evaluate a train directly, with the Greeks stored as functions: either the
surface at inception, differentiated at the new spot, or one train per Greek, built by
applying the differentiated kernel as an operator. Differentiating the stored surface
works poorly on the solver's own mesh. Its panels end at the barriers, as the event
maps require, and the $k$-th derivative of a polynomial amplifies its error by about
$(p^2/h)^k$ at the ends of its panel, against $(p/h)^k$ inside, so Greeks read from the
cores are poor at the barriers. Since $C_0$ does not jump, the last step can instead
project it onto a mesh with no edge near the spots, for example one whose central panel
holds all three barriers. With enough modes on that panel, and rounded at $10^{-12}$
instead of $10^{-8}$, the projected surface reproduces the kernel read-out to $0.02\%$
at every spot, at three and at five assets. It costs one more operator application, $7$
to $17\%$ of a solve, and a higher rank, $53$ instead of $31$ at three assets and $175$
instead of $71$ at five. Since evaluating a train also costs in proportion to the square
of its rank, whether such queries beat the kernel read-out remains to be measured. One
train per Greek would need a projection and a rounding for every Greek, and we have not
tried it.

The read-out takes the kernel's derivatives by forward-mode automatic differentiation,
which spares coding them by hand: under geometric Brownian motion they are
Hermite polynomials times the kernel, and under local volatility they also carry the
derivatives of $\sigma(x)$. This costs the read-out little, since it needs them once per spot: at three assets the
price and five Greeks at a spot take $1.9$ milliseconds with automatic differentiation
and $1.55$ coded by hand, after a solve of about ten seconds. Monte Carlo needs the same
derivatives, divided by the density, as likelihood-ratio weights, but on every path,
where automatic differentiation would make a three-asset sample under geometric
Brownian motion $20\%$ costlier, so our estimators code them by hand.

\subsection{Resolution of a Gaussian-smoothed value}
\label{app:resolution}

Here we prove Proposition~\ref{prop:resolution}, give a separate error bound for
differentiating a polynomial approximation, and measure how the position of a layer in
its panel affects its resolution. The transition is the exact Gaussian step of
Eq.~\eqref{eq:onefactor_step}, and none of these bounds includes numerical quadrature or
tensor-train rounding.

\begin{proof}[Proof of Proposition~\ref{prop:resolution}]
(a) Fix the real coordinates other than $x_i$. Conditionally on the factor $Z=z$,
integrate $V$ against the other $d-1$ conditional Gaussian densities and call the
result $G(y_i,z)$. Then $|G|\le M$, and
$C(x)=e^{-r\Delta t}\!\int\!\phi(z)\!\int\!\phi_{s_\perp}
\big(y_i-x_i-c(z)\big)\,G(y_i,z)\,dy_i\,dz$,
where $c(z)=\mu\Delta t+\beta z$ and $\phi_s$ is the centered normal density of
standard deviation $s$. Only the inner Gaussian depends on $x_i$, and replacing $x_i$ by
$x_i+\mathrm{i}\eta$ multiplies its modulus by exactly $e^{\eta^2/(2s_\perp^2)}$,
whatever the real center $c(z)$. The factor $z$ stays real, and its density $\phi(z)$ is
untouched and has total mass one. Taking moduli inside both integrals, with $|G|\le M$
and $e^{-r\Delta t}\le1$ since the rate in this model is nonnegative,
\[
|C(x+\mathrm{i}\eta\,e_i)|\le e^{\eta^2/(2s_\perp^2)}\!\int\!\phi(z)\!\int\!\phi_{s_\perp}
\big(y_i-x_i-c(z)\big)\,|G(y_i,z)|\,dy_i\,dz\le M\,e^{\eta^2/(2s_\perp^2)} .
\]
The same domination holds uniformly for $x_i$ in compact subsets of the complex plane, so
the integral converges locally uniformly, $C$ is entire in $x_i$, and (a) follows.

(b) Map $P$ to $[-1,1]$ by $x_i=a+\tfrac h2(1+t)$. On the Bernstein ellipse
$\mathcal E_\varrho$, $\varrho=e^\tau$, $|\operatorname{Im}t|\le\sinh\tau$, so (a)
bounds the mapped function by $M\exp(\tfrac18\kappa^2\sinh^2\tau)$.
The Chebyshev truncation bound~\cite{Trefethen2019} therefore gives
\begin{equation}
  E_p\le\frac{2M}{e^\tau-1}
  \exp\!\bigl(\tfrac18\kappa^2\sinh^2\tau-p\tau\bigr)
  \quad(\tau>0).
  \label{eq:resolution_ellipse}
\end{equation}
Set $c_1=\sinh^2(1)$. For $\tau\le1$, $\sinh^2\tau\le c_1\tau^2$, and the quadratic
exponent in this upper bound is minimized at $\tau=4p/(c_1\kappa^2)$, which is
less than $1$ if $p\le\kappa^2/3$. It is then at most
$-2p^2/(c_1\kappa^2)\le-\tfrac{13}{9}p^2/\kappa^2$, and
$2/(e^\tau-1)\le2/\tau=c_1\kappa^2/(2p)\le\kappa^2/p$.
For $p>\kappa^2/3$, take $\tau=1$ in Eq.~\eqref{eq:resolution_ellipse}. Since
$\kappa^2<3p$ and $1-3c_1/8>0.48$, its right-hand side is at most
$2M e^{-0.48p}$. This proves (b).
\end{proof}

Entire analyticity has the standard consequence of super-exponential polynomial
convergence~\cite{Trefethen2019}. Indeed, at fixed $h$ and $s_\perp$,
Eq.~\eqref{eq:resolution_ellipse} holds for every $\tau>0$, so
$\limsup_{p\to\infty}(E_p/M)^{1/p}=0$. The exact Legendre projection inherits this
rate, since its extra $O(\sqrt p)$ factor does not change the limit.

A separate bound covers the derivatives of a polynomial approximation. Let $q_n$ be a
best uniform
approximation to $C$ on $P$ by polynomials of degree at most $n$. For integer
$p\ge1$ and $k\ge0$,
\begin{equation}
  \big\|\partial_{x_i}^k(C-q_p)\big\|_{L^\infty(P)}
  \le 2\Big(\frac{8p^2}{h}\Big)^{k}\sum_{m\ge0}4^{mk}E_{2^m p}.
  \label{eq:resolution_deriv}
\end{equation}
Indeed, $C-q_p=\sum_{m\ge0}(q_{2^{m+1}p}-q_{2^mp})$, and the bound in
Proposition~\ref{prop:resolution}(b) ensures that this sum converges with all
derivatives. Each difference has degree at most $2^{m+1}p$ and uniform norm at
most $E_{2^{m+1}p}+E_{2^mp}\le2E_{2^mp}$. Markov's inequality on an interval of
length $h$ bounds its $k$-th derivative by $(2n^2/h)^k$ times its norm, with
$n=2^{m+1}p$, and summing proves Eq.~\eqref{eq:resolution_deriv}. This concerns a
derivative of the polynomial representation, not the kernel read-out used in the
cost comparisons.

Proposition~\ref{prop:resolution} uses only
$|V|\le M$, so it describes the worst case, a layer that may sit anywhere in the
panel. The layers that the events create sit near its ends: the jumps of the event map
lie on panel edges, and given the factor $z$ the transition centers each resulting
layer on an edge up to the shift $\mu\Delta t+\beta z$, which keeps it within two
idiosyncratic widths of the edge for about $95\%$ of the factor's mass at
$\rho=\tfrac12$.

How much the position matters is seen on one asset. One Gaussian transition,
$Y=x+sZ$ with $Z\sim\cN(0,1)$, turns a step that switches on at $y=c$ into the smooth
probability
\begin{equation}
  f_c(x)=\Pr(Y\ge c)=\Phi\big((x-c)/s\big),
  \label{eq:model_layer}
\end{equation}
with $\Phi$ the standard normal distribution function. Here $s$ is a generic width,
and $s=s_\perp$ gives the conditional layer of Proposition~\ref{prop:resolution}, not the
total width of the implemented rule.

On a panel $[a,a+h]$, with $\kappa=h/s$, we put
the jump at the edge ($c=a$), one or two widths inside ($c=a+s$ and $a+2s$, where a
factor shift can move the smoothed layer) and at the center ($c=a+h/2$).
Table~\ref{tab:layer_degree} counts the Legendre modes that resolve each of them as
$\kappa$ grows, cutting at the first mode from which every coefficient evaluated is
below $10^{-8}$. These are model layers, not the solver's barriers, so the table only measures
how the position of a layer affects its resolution.

\begin{table}[H]
\caption{\textbf{A smoothed barrier is cheaper to resolve near a panel edge.}
For $f_c(x)=\Pr(x+sZ\ge c)=\Phi((x-c)/s)$ on $[a,a+h]$, where $Z\sim\cN(0,1)$ and
$\Phi$ is its cumulative distribution function. Each entry is the number $p$ of
Legendre modes needed: every coefficient beyond them, up to degree $699$, is below
$10^{-8}$ in magnitude. The last column fits $p\propto\kappa^{\alpha}$ over
$\kappa=h/s\ge10$. This tests the layer-position effect, not the solver's accuracy.}
\label{tab:layer_degree}
\centering
\renewcommand{\arraystretch}{1.2}
\begin{tabular}{?l|ccccc|c?}
\thickhline
 & $\kappa=10$ & $20$ & $40$ & $80$ & $160$ & exponent \\
\hline
at the panel end ($c=a$) & $27$ & $41$ & $58$ & $82$ & $116$ & $0.52$ \\
\hline
one width inside ($c=a+s$) & $29$ & $44$ & $65$ & $93$ & $132$ & $0.55$ \\
\hline
two widths inside ($c=a+2s$) & $31$ & $46$ & $70$ & $99$ & $140$ & $0.55$ \\
\hline
at the panel center ($c=a+h/2$) & $32$ & $58$ & $114$ & $222$ & $440$ & $0.95$ \\
\thickhline
\end{tabular}
\end{table}

A layer at the edge is far cheaper to resolve. At $\kappa=80$ it needs $82$ modes,
against $222$ at the center, and moving the jump one or two widths inside leaves it
much closer to the edge case. The fitted powers, $0.52$ at the edge and $0.95$ at the
center, separate growth close to the square root of the panel width from nearly
linear growth. The larger values of
$\kappa$ lie beyond our production panels and only expose the trend.

Near a panel end,
then, the order needed grows roughly as $\sqrt{h/s_\perp}$, while the implemented
rule grows linearly in $h/s$, with a constant and a cap
that are empirical. With panels no wider than $0.5$ the barrier panels have
$\kappa\le12.2$ for a monthly step and $\kappa\le17.3$ for a half-month one, where the
edge and center counts of Table~\ref{tab:layer_degree} differ by less than a factor of
$1.4$, although at a fixed panel width the
gap widens as observation gets finer. Barrier alignment thus does double duty, making
the event maps exact and placing the layers where they are cheapest.

The same edges hurt a derivative of the stored
polynomials read at a barrier, because a polynomial's derivative amplifies its error
most at the ends of its panel, by the factor $(8p^2/h)^k$ of
Eq.~\eqref{eq:resolution_deriv} against about $(p/h)^k$ inside, which is one reason
the read-out differentiates the kernel instead (Sec.~\ref{sec:events}).

The analysis is per coordinate and per date. For exact tensor-product projections the
error is bounded by the sum of the one-dimensional errors, each multiplied by the norms
of the other projections, and since the exact transition is a positive contraction and
the event map is non-expansive on differences, errors in the value add at most
linearly across dates before any further projection. This does not establish the
stability of the repeated numerical projections, nor bound the quadrature and rounding
errors, which are checked separately. Nor does the proposition cover local volatility,
whose kernel has a mean and a width that depend on the state, Eq.~\eqref{eq:lv_kernel}. We
apply the
same rule there and check the result against Monte Carlo.

Finally, the analytic scale is $s_\perp$ while the rule uses $s$
($2h/s=\sqrt2\,h/s_\perp$ at $\rho=\tfrac12$). At high correlation $s_\perp$ shrinks
while $s$ does not, so the rule would need revalidation, and as $\rho\to1$ the bound in
(a) degenerates, as it must, since perfectly correlated assets make $C$ a function of
$\min_ix_i$ alone, with kinks on the diagonals.

\subsection{Error control: discretization, tolerance and error estimates}
\label{app:err}

The solver makes two kinds of error: the discretization, set by the layer rule, and the
rounding, set by the tolerance (Sec.~\ref{sec:resolution}). We first give the
measurements behind what Sec.~\ref{sec:resolution} says about each, then estimate
the rounding error without a reference, and finally turn the estimates
into rules for choosing the tolerance.

\subsubsection{The discretization and the tolerance}

Joint refinement of the discretization, specified in Table~\ref{tab:discretization},
moves no Greek by more than $0.017\%$ under local volatility (three
assets, tolerance $10^{-8}$, both spots, on the box $[-3,3]$ rather than the production
box $[-4.5,4.5]$, which is tested separately), and no Greek of the $10^{-9}$
solves under geometric Brownian motion by more than $0.002\%$ at
three assets and $0.018\%$ at four. We widened the box separately. Under geometric
Brownian motion, going from $L=3$ to $L=4.5$ moves the three-asset solve at $10^{-9}$ by
about a relative $10^{-7}$ on the price and by less than $0.0015\%$ on every Greek, at
both spots. Appendix~\ref{app:lv} widens the box under local volatility.

\begin{table}[t]
\caption{\textbf{Joint discretization-refinement test.} All five settings are changed
together; the measured differences cannot be attributed to any individual setting.
The factor-node counts are those of this test, not the $24$-node read-out of the
transformed monthly chain. Domain widening is tested separately.}
\label{tab:discretization}
\centering
\renewcommand{\arraystretch}{1.15}
\begin{tabular}{?l|cc?}
\thickhline
setting & baseline & refined \\
\hline
panel mode allocation & production mesh & $30\%$ more \\
maximum unmerged-panel width & $0.5$ & $0.4$ \\
modes per merged tail panel & $16$ & $24$ \\
common-factor nodes & $12$ & $16$ \\
terminal levels & $96$ & $128$ \\
\thickhline
\end{tabular}
\end{table}

The error falls quickly as the modes grow. We measured it at monthly observation and
tolerance $10^{-8}$ on the mesh before the tails are merged, against solves refined as
above. With six modes on every panel the three-asset error at $x_0=-0.1$ is $1.45\%$ on
speed and $0.18\%$ on gamma. With eight it is below $0.08\%$, and with ten below
$0.03\%$, about where sixteen leave it. Four and five assets behave alike. The layer
rule gives this mesh $130$ modes per axis, $8$ on most panels and $16$ and $18$ on the
two wide ones that touch a barrier, against $224$ for sixteen on every panel, and
matches sixteen to $0.0021\%$ on every Greek at three to five assets for $43$ to $45\%$
less compute.

The error is not monotone in the tolerance. At three assets under geometric
Brownian motion, gamma's error rises from $1.02\%$ to $1.47\%$ as the tolerance tightens
from $3\times10^{-5}$ to $10^{-5}$ (Fig.~\ref{fig:object}(b)). Under local volatility at
four assets the cross-speed error swings from $-2.2\%$ at $10^{-5}$ to $+3.1\%$ half a
decade tighter, and falls below $1\%$ only at $3\times10^{-8}$.

The decade estimate of Sec.~\ref{sec:resolution} usually overstates the error of the
finer of its two solves, the safe direction, but it has a blind spot. It detects
compression only, and it grows more confident as the discretization degrades, because a
lower-order solution compresses more easily: with six modes per panel, two solves a decade
apart agree far more closely than either agrees with the refined references.

\subsubsection{A goal-oriented estimate of the rounding error}
\label{app:cert}

Every tolerance in the main text was chosen against deep Monte Carlo. Here we estimate the
rounding error of a Greek from the solve itself and measure what that costs. We built the
estimate in full for three assets under geometric Brownian motion with monthly
observation, read out at $x_0=-0.1$ and at the autocall barrier, $x_0=0$, from solves at
eleven tolerances half a decade apart, from $10^{-3}$ to $10^{-8}$, and checked it under
local volatility at three to five assets and under geometric Brownian motion at four and
five.

In the theory of $hp$ finite elements the error is controlled a priori, by bounds in terms
of the mesh and the polynomial order, and a posteriori, by guaranteed bounds, by
goal-oriented estimates that weight the residual with the solution of a dual problem, or
by comparison with a refined solution. Proposition~\ref{prop:resolution} bounds the best
approximation, one ingredient of an a priori analysis of the discretization, and the
decade estimate is of the last kind, applied to the rounding. The rounding tolerance is
closer to the stopping tolerance of an iterative solver than to a mesh parameter: the solver
rounds after every operator, roundings at different tolerances are not nested, and the
tolerance controls the $L^2$ norm of the whole surface rather than a derivative at one
spot. All three kinds of estimate carry over nonetheless, because rounding is the only
nonlinear operation in the solve.

The rounding error of a Greek has an exact representation. In the order of
Algorithm~\ref{alg:solve}, let
the solve perform the roundings $Y_j=\mathrm{round}_\varepsilon(X_j)$,
$j=1,\dots,J$, where each $X_j$ is an affine function of the earlier $Y_i$:
transitions, discounting and multiplication by event indicators are linear, and the
coupon and redemption terms of Eq.~\eqref{eq:event} and the terms of the terminal
step do not depend on the surface. The price and each Greek, $G$, are linear
functionals of the last surface, Eq.~\eqref{eq:readout}. Let $G_\infty$ be the same
quantity from the recursion without rounding, on the same discretization, and
$e_j=X_j-Y_j$ the part discarded by the $j$-th rounding. Then
\begin{equation}
G-G_\infty=-\sum_{j=1}^{J}\langle \zeta_j,e_j\rangle ,
\label{eq:dwr}
\end{equation}
with $\langle\cdot,\cdot\rangle$ the inner product of $L^2([-L,L]^d)$ and $\zeta_j$ the
dual: the function whose inner product with a change in $Y_j$ gives the change it
causes in $G$, obtained from the read-out functional by the transposes of the linear
steps that follow the $j$-th rounding. Because those steps are affine,
Eq.~\eqref{eq:dwr} is an identity, not a linearization.

The dual runs forward in
time, from the spot toward maturity: it starts as the differentiated transition
density of Eq.~\eqref{eq:readout}, and each transposed step carries it one step
further, discounts it and keeps only the states on which the contract survives. One
forward sweep per Greek and spot computes it for every $j$, rounded to a tolerance
$\varepsilon_\zeta$ of its own, and it compresses well: to rank $12$ to $17$ in our case,
against $28$ for the solve at $10^{-6}$.

Equation~\eqref{eq:dwr} is the
dual-weighted error representation of goal-oriented finite-element methods~\cite{BeckerRannacher2001}, with the discarded parts in the place of the residual. Here both of
its factors are at hand: the rounding computes the discarded parts, and the dual is
the exact adjoint of the discrete steps, so the only approximation is the rounding of
the dual. Subtracting the estimate from $G$ is the corresponding adjoint correction~\cite{GilesSuli2002}. On two assets and six dates, with $\varepsilon_\zeta=10^{-13}$, the
corrected values of solves at $10^{-3}$ and $10^{-5}$ agree with a solve rounded only
at $10^{-14}$ to $3\times10^{-13}$ of their size, where the raw errors reach $49\%$.

The identity gives two error estimates. Bounding each term of Eq.~\eqref{eq:dwr} by the
Cauchy--Schwarz inequality, with $\|e_j\|\le\varepsilon\|X_j\|$ from the rounding,
gives a guaranteed bound with the exact dual and the actual run's norms,
\begin{equation}
|G-G_\infty|\le\varepsilon\sum_{j=1}^{J}\|\zeta_j\|\,\|X_j\| ,
\label{eq:dwr_bound}
\end{equation}
and evaluating Eq.~\eqref{eq:dwr} with the computed dual gives a goal-oriented
estimate. Table~\ref{tab:cert} sets both beside the decade estimate, against the truth
for every rounding error: the solve at $10^{-8}$ corrected with $\varepsilon_\zeta=10^{-4}$, a
correction of at most $2.2\times10^{-4}$ of the value.

The estimate is sharp. With
$\varepsilon_\zeta=10^{-3}$ it lies within $1.6\%$ of the actual error in all $113$ cases
where that error exceeds $10^{-6}$ of the value, and a dual sweep costs $0.9$ to $3.6$
core-seconds per Greek and spot, against $1.1$ to $10$ for a solve. The bound is rigorous
but far from sharp, because the discarded parts spread over the whole domain while a Greek
sees one spot. The decade estimate lies between the two, but it falls below the error,
the unsafe direction, in $6$ of its $50$ cases.

The estimate is not tied to this case. We repeated the dual sweeps on the production
solver under local volatility at three to five assets, with its sub-steps and a kernel
that depends on the state, and under geometric Brownian motion at four and five, with the
solve at $10^{-9}$ as the truth. Since that truth carries rounding errors of its own, we
score the $380$ cases where the actual error exceeds $0.1\%$ of the value, and in all of
them the estimate with $\varepsilon_\zeta=10^{-3}$ lies within $2.7\%$ of it. The duals
compress as before, to rank $12$ to $24$, against $34$ to $40$ for the solves at
$10^{-6}$, and a sweep costs between half and $1.2$ times as much as a solve at
$10^{-6}$.

\begin{table}[t]
\caption{\textbf{Three estimates of the rounding error.} Three assets under geometric
Brownian motion, monthly observation, spots $x_0=-0.1$ and $0$. A dual serves one
Greek at one spot. Sharpness is the estimated rounding error over the actual one, for
the bound at tolerances $10^{-3}$, $10^{-5}$ and $10^{-7}$ ($30$ cases, five
Greeks), for the estimate with $\varepsilon_\zeta=10^{-3}$ at ten tolerances from
$10^{-3}$ to $3\times10^{-8}$ ($113$ cases, the price and five Greeks), and for the
decade estimate against the error of the finer of two solves a decade apart ($50$
cases, five Greeks). Cost is that of choosing the tolerance without the reference,
as a multiple of the charge of the sustained rule applied with the truth known, over
$30$ cells (five Greeks, two spots, targets $\tau$ of $1\%$, $0.3\%$ and $0.1\%$).
The bound's sharpness uses the actual-run norms; its cost uses the pilot prescription
below. The rules are those of the text. Medians, with ranges in parentheses.}
\label{tab:cert}
\centering
\renewcommand{\arraystretch}{1.2}
\begin{tabular}{?l|l|cc?}
\thickhline
estimate & needs & sharpness & cost \\
\hline
guaranteed bound, Eq.~\eqref{eq:dwr_bound} & run norms and a dual &
$537$ ($89$--$149{,}672$) & $2.9$ ($2.0$--$6.0$) \\
\hline
goal-oriented estimate, Eq.~\eqref{eq:dwr} & a dual &
$1.000$ ($0.985$--$1.012$) & $2.6$ ($1.3$--$4.1$) \\
\hline
decade estimate & a second solve per decade &
$7.5$ ($0.07$--$360$) & $4.0$ ($2.1$--$4.7$) \\
\hline
none, $\varepsilon=10^{-5}\tau$ & nothing & no estimate & $1.5$ ($1.0$--$3.8$) \\
\thickhline
\end{tabular}
\end{table}

\subsubsection{Choosing the tolerance without a reference}
\label{app:scoring_checks}

Each estimate also gives a rule for choosing the tolerance. The last column of
Table~\ref{tab:cert} compares the cost of each rule, for a target $\tau$, with that of the
sustained rule defined in Sec.~\ref{sec:baselines}, applied with the truth known. The bound can
suggest a tolerance in advance: a pilot solve at $10^{-3}$ and the dual approximate the
sum in Eq.~\eqref{eq:dwr_bound}, using the pilot's norms, and prescribe
$\varepsilon=\tau|G|\big/\sum_j\|\zeta_j\|\,\|X_j\|$. This is a pilot-based prescription,
not a certificate for the new run. Certification would require upper bounds on that
run's intermediate norms and on the error of the rounded dual, which we do not obtain
here. The prescribed tolerance is a median $2.5$ decades tighter than the sustained
rule's, and below $10^{-8}$ in
$24$ of the $30$ cells, where we extrapolate the cost at the measured factor of $1.39$ per
decade.

The estimate supports a search instead: solve, estimate, and accept the solve if the
estimate is at most half the target, otherwise tighten by half a decade. Started at
$\varepsilon=10^{-4}\tau$, this search cost $2.6$ times the sustained charge and never
missed a target in the tested cells. Its reliability is empirical, not a guarantee
independent of the starting tolerance. The decade
estimate, accepting the finer of two solves a decade apart once they agree within the
target, cost more and once accepted a value that missed its target. The rule of thumb
$\varepsilon=10^{-5}\tau$, one solve with no check, was cheaper still and met every target
here, but we chose its factor knowing these results, and nothing would tell a user where
it fails.

A separate check over $133$ benchmark Greek cells at three to five assets gives a
median decade-rule charge about four times the reference-assisted charge. The rule accepts
$14$ values outside their targets: $10$ of $91$ on the GBM and Euler chains and
$4$ of $42$ on the transformed chain.

The goal-oriented estimate has two limits. It sees the rounding only: the discretization is fixed
by the layer rule and checked against Monte Carlo, and an estimate for it, which a method
free of references would also need, is not part of this work. And a dual belongs to one
Greek at one spot, while a solve returns every Greek at every spot and its rounding error
moves across spots, peaking near the autocall barrier, so an estimate at one spot does
not cover a curve of spots.

\subsection{The full grid}
\label{app:fullgrid}

The full grid of Sec.~\ref{sec:freq} stores the array
of Legendre coefficients over the product of the axes, $n^d$ numbers for $n$ per axis,
in place of the train, and applies the same steps to it with nothing rounded. At each
factor node the one-asset operator of Eq.~\eqref{eq:spatial_quadrature} acts as a matrix
along every axis, the event maps act on whole blocks of coefficients, since every barrier
is a panel edge, and the terminal step of Eq.~\eqref{eq:terminal_product} and the
read-out of Eq.~\eqref{eq:readout} are those of the train. At two assets
it agrees with a tensor solve rounded at
$10^{-14}$ to $1.3\times10^{-11}$, under both models.

It is therefore the limit of the
tensor's solves as the tolerance goes to zero, and measures their rounding error exactly:
at $10^{-9}$, at most $10^{-4}$ of any Greek at three and four assets. It also reproduces
the largest deviations of the tightest solves from the Monte Carlo references (speed at
three assets, delta at four), which are therefore not rounding errors.

The full array is also a competitor, and at three assets a strong one. On one thread of
the loaded machine it solves in $3.5$ core-seconds under geometric Brownian motion
($82^3=5.5\times10^{5}$ coefficients) and $35$ on the two-sub-step chain of local
volatility ($129^3=2.1\times10^{6}$, with twice as many steps). Charged in place of the
tensor, it gives a ratio on cross-speed at $1\%$ of $614$ and $95$ under the two models,
against $336$--$424$ and $155$--$214$ for the tensor. At three assets the tensor has little to
compress, and compressing costs work: its bonds need $28$ to $35$ of a possible $82$, and
rounding them back after every transition takes $45$ to $74\%$ of a solve, while the full
array spends its time on large matrix products.

At four assets the full array is no longer competitive. The tensor's
middle bond could reach $82^2=6{,}724$ but needs only $34$ to $80$. The full array takes
$309$ and $5{,}768$ core-seconds under the two models, $6$ to $45$ and $55$ to $300$ times
the tensor's charged solves, and its cross-speed ratios fall to
$7.0$ and $1.3$, against $66$--$101$ and $73$--$105$ for the tensor. At five assets its
arrays would need $269$ gigabytes.

\section{The Monte Carlo baselines}
\label{app:baselines}

This appendix describes the Monte Carlo estimators: the one we tuned, the published ones
that differentiate the payoff, the weights we built for Euler chains under local
volatility, and the other estimators we tried or set aside.

\subsection{The tuned likelihood-ratio estimator}
\label{app:mc}

Every likelihood-ratio cost in the comparisons uses a tuned version of the
estimator of Sec.~\ref{sec:baselines}, which makes three changes to the
textbook form, each kept because it lowered the cost of reaching a given accuracy.
(a)~Scrambled Sobol points (linear matrix scrambling~\cite{Matousek1998} and a digital
shift) replace pseudo-random numbers. They span all $d+(N-1)(d+1)$ dimensions of the
path, the leading $d$ coordinates driving the first step, and $32$ independent
randomizations supply the error bar.
(b)~The remaining normals are reflected as well as the first-step ones, giving four paths
per point.
(c)~A constant taken from an independent pilot run is subtracted from the payoff in the
even-order weights, a control variate that changes no expectation because every weight
has mean zero.
At three assets with monthly observation the three changes make the estimator $2$ to $3$
times cheaper on the odd orders, and $66$ to $117$ times cheaper on gamma and
cross-gamma.

Over the measured range, randomized quasi-Monte Carlo primarily improves the error
constant: fitted over
$2^{18}$ to $2^{34}$ samples, its error falls as $M^{-0.52}$ to $M^{-0.55}$ in the number
of samples $M$, against $M^{-0.48}$ to $M^{-0.51}$ for pseudo-random sampling.
The implementation uses a streamed Gray-code Sobol generator, with a per-sample cost
$3$ to $11\%$ above pseudo-random generation. This choice avoids the overhead of the
Owen-scrambled sampler in QuasiMonteCarlo.jl (version 0.3.3), which outweighed its
variance reduction in our implementation comparison.

Under geometric Brownian motion every ingredient of the tuned estimator is unbiased: the
transition is sampled exactly, the weights are exact (they agree with automatic
differentiation of the first-step density to a relative $4\times10^{-13}$), the
control-variate constant comes from an independent run, and each randomized point set is
uniformly distributed. The error of a run is therefore its standard error, and two checks
bear the error bars out. At three to five assets and two spots the deepest runs,
$4.3\times10^{9}$ to $5.3\times10^{10}$ samples, agree with the tensor's tightest solve on
the price and all five Greeks: over these $36$ comparisons the root-mean-square $z$-score
is $1.05$ and the largest is $2.2$, as expected of unbiased estimates with correct
standard errors. Being
unbiased and independent of the tensor, the deepest runs also serve as the reference for
its error.

In Fig.~\ref{fig:object}(b) that reference pools $5.3\times10^{10}$ samples,
$1.2\times10^{5}$ core-seconds of sampling. The tuned arm is drawn up to
$2.7\times10^{8}$ samples ($580$ core-seconds), and the conditioned bumps up to
$1.7\times10^{7}$ points ($835$ core-seconds). The bias of their $1\%$ bump, at which
their root-mean-square error levels off, is $0.089\%$ on delta and $0.78\%$ on
cross-speed.

\subsection{Estimators that differentiate the payoff}
\label{app:published}

Estimators that differentiate the payoff pay for its jumps. Bump-and-reprice pays in both
variance and bias, as Sec.~\ref{sec:lr} explains. At six dates and five assets, with
common random numbers, its standard error grew as $\delta^{-1.50}$ for gamma and
$\delta^{-2.49}$ for the third derivative, as predicted there, and we did not carry it to
$36$ dates. Pathwise and adjoint derivatives miss the jumps of the
payoff at every barrier on every date, and for the digital legs they vanish identically.
Smoothing restores them at the price of a bias, the bump's trade-off in another form.
Schemes designed for autocallables stabilize their sensitivities by conditioning on
survival at each date or by smoothing the
payoff~\cite{FriesJoshi2011,AlmHarrachHarrachKeller2013,KosterRehmet2018}, and the second
Monte Carlo estimator of our comparisons is of this kind.

The estimator of Rakhmonov and Rakhmonov~\cite{Rakhmonov2019}, the strongest we found
published for the second- and third-order Greeks of worst-of autocallables, conditions
each date on the note's survival along the one
direction of the Gaussian step that loads equally on every asset. Given the other
components of the step, every asset lying above a level is a half-line in that direction,
so each date's coupon and autocall indicators are replaced by their conditional
probabilities, and the path continues with that component drawn from the surviving range
by inversion. The value of a path is then continuous in the spots, and central differences
with common random numbers are stable.

Under local volatility, where each asset's step has its own width, we condition on the
component of the step weighted by $\Sigma^{-1}\bm 1$, with $\Sigma$ the covariance of the
step. The step then splits into an equal move of every asset and a remainder
independent of it, as under geometric Brownian motion. On the near-continuum chain of Sec.~\ref{sec:product} we condition
the month, one Gaussian step in the transformed coordinates, with the barriers mapped to
those coordinates, and on the Euler chain the last sub-step of each month.

As published, the paths are pseudo-random and the bumps are $1\%$ in log-spot. We add
only a closed-form integral of the payoff at maturity along the conditioned direction,
which can only lower the variance. On the near-continuum chain the loss below the
protection barrier has no closed form along that direction and is sampled by inversion
instead. Each Greek is charged its own stencil, from $2$ path values per point for delta
to $8$ for cross-speed, with the costs of a point and of a path value fitted in the same
run.

The bias of each central difference is exact: its expectation is an average of the
corresponding derivative over the bumped spots, which we read off the tensor's tightest
solve, and a Greek is charged the compute at which the root-mean-square error, noise and
bias combined, meets the target. Against these expectations the runs agree, with every
$z$-score below $2.0$ in magnitude over $55$ comparisons at three to five assets under
both models and on both local-volatility chains.

At $x_0=-0.10$ the bias of the $1\%$ bump is at most $0.15\%$ on delta, gamma and
cross-gamma but larger on the third-order Greeks, so the published bump cannot reach
$0.3\%$ on a third-order Greek, nor $1\%$ on speed under local volatility.

Table~\ref{tab:published} sets the estimators side by side under both models, with local
volatility on the near-continuum chain. The conditioned bumps are the cheaper on cross-gamma at
every size and under both models, on gamma at four and five assets, where conditioning
removes most of the noise of a second difference, and under local volatility also on
gamma at three assets and on delta. The likelihood ratio is the cheaper on speed
everywhere, on cross-speed at three and four assets, and on delta under geometric
Brownian motion. Neither dominates, and each is run as one method for every Greek.

Hu, Xue and Yu~\cite{Hu2022} differentiate the same conditioned paths, which gives
unbiased first-order Greeks. A second pathwise derivative is not valid on several assets,
because the pathwise delta jumps where the worst performer changes, so they obtain
second-order Greeks by finite differences, of conditioned prices, the estimator above, or
of pathwise deltas, with forward steps of $0.01$ or $0.001$, and they treat no third-order
Greek.

We ran their pathwise delta and their forward differences under geometric Brownian
motion. The pathwise delta is $1.4$ and $2.0$ times cheaper than the likelihood ratio's
at four and five assets, though $1.5$ times costlier at three, and at five assets and
$0.1\%$ it brings Monte Carlo level with the tensor on delta ($0.82$--$1.2$). Half that
target restores the tensor's lead at every size: at $0.05\%$, which the references resolve
to $0.005\%$ or better at two standard errors, the tensor is $3.3$ to $18$ times cheaper
than the pathwise delta. The forward differences of pathwise deltas cost far more than
the conditioned bumps on gamma and cross-gamma at $1\%$ (Table~\ref{tab:published}), and
the bias of their finer step rules out tighter targets.

\begin{table}[t]
\caption{\textbf{The Monte Carlo estimators against the tensor.} Monte Carlo cost over
tensor cost at $x_0=-0.10$, monthly observation, targets $1\%$ and $0.1\%$: the range set
by the tensor's bracket, from its charged solve to the next looser one, as in
Table~\ref{tab:gbm_ratios}. LR: the tuned likelihood-ratio estimator. CB:
bump-and-reprice on conditioned
paths~\cite{Rakhmonov2019}, as published. PW: the pathwise
variant of Ref.~\cite{Hu2022}, pathwise on delta and forward differences of pathwise
deltas on gamma and cross-gamma, the cheaper of its steps $0.01$ and $0.001$, run under
geometric Brownian motion only. b: the bias of the published step exceeds the target.
--: not defined. n.s.: not scored, because two standard errors of the reference exceed
the target. Figure~\ref{fig:greeks} draws the LR and CB entries.}
\label{tab:published}
\centering
\renewcommand{\arraystretch}{1.15}
\begin{tabular}{?l|l|l|ccccc?}
\thickhline
 & target & & delta $\Delta$ & gamma $\Gamma_{11}$ & speed $\partial^3_{111}$ &
cross-gamma $\Gamma_{12}$ & cross-speed $\partial^3_{123}$ \\
\hline
\multicolumn{8}{?l?}{geometric Brownian motion} \\
\hline
$d=3$ & $1\%$ & LR & $0.044$--$0.056$ & $1.1$--$1.2$ & $13$--$17$ & $13$--$14$ & $336$--$424$ \\
 & & CB & $0.098$--$0.12$ & $1.7$--$1.9$ & $202$--$254$ & $6.6$--$7.5$ & $901$--$1{,}135$ \\
 & & PW & $0.066$--$0.083$ & $8.2$--$9.3$ & -- & $69$--$79$ & -- \\
\cline{2-8}
 & $0.1\%$ & LR & $3.0$--$3.4$ & $86$--$109$ & $928$--$1{,}172$ & $884$--$1{,}011$ & n.s. \\
 & & CB & $32$--$36$ & $134$--$169$ & b & $560$--$640$ & n.s. \\
 & & PW & $4.5$--$5.1$ & b & -- & b & n.s. \\
\hline
$d=4$ & $1\%$ & LR & $0.042$--$0.055$ & $1.0$--$1.3$ & $9.2$--$12$ & $5.9$--$8.4$ & $66$--$101$ \\
 & & CB & $0.068$--$0.089$ & $0.65$--$0.86$ & $115$--$152$ & $0.79$--$1.1$ & $100$--$154$ \\
 & & PW & $0.030$--$0.039$ & $5.5$--$7.3$ & -- & $12$--$18$ & -- \\
\cline{2-8}
 & $0.1\%$ & LR & $1.9$--$2.5$ & $77$--$100$ & $207$--$318$ & $248$--$382$ & n.s. \\
 & & CB & b & $53$--$70$ & b & $36$--$55$ & n.s. \\
 & & PW & $1.4$--$1.8$ & b & -- & b & n.s. \\
\hline
$d=5$ & $1\%$ & LR & $0.035$--$0.045$ & $0.67$--$0.92$ & $4.2$--$6.4$ & $4.2$--$7.2$ & $215$--$323$ \\
 & & CB & $0.036$--$0.046$ & $0.54$--$0.76$ & $25$--$38$ & $0.38$--$0.65$ & $47$--$71$ \\
 & & PW & $0.017$--$0.022$ & $4.2$--$5.8$ & -- & $14$--$25$ & -- \\
\cline{2-8}
 & $0.1\%$ & LR & $1.7$--$2.5$ & $26$--$44$ & $72$--$102$ & $175$--$273$ & n.s. \\
 & & CB & b & $25$--$43$ & b & $16$--$25$ & n.s. \\
 & & PW & $0.82$--$1.2$ & b & -- & b & n.s. \\
\hline
\multicolumn{8}{?l?}{local volatility near the continuum limit (Sec.~\ref{sec:product})} \\
\hline
$d=3$ & $1\%$ & LR & $0.051$--$0.054$ & $2.1$--$2.2$ & $54$--$59$ & $12$--$14$ & $132$--$151$ \\
 & & CB & $0.045$--$0.048$ & $1.6$--$1.7$ & b & $4.2$--$4.7$ & $435$--$497$ \\
\cline{2-8}
 & $0.1\%$ & LR & $4.6$--$5.1$ & $150$--$166$ & $3{,}085$--$3{,}551$ & $976$--$1{,}114$ & n.s. \\
 & & CB & $5.3$--$5.8$ & $118$--$131$ & b & b & n.s. \\
\hline
$d=4$ & $1\%$ & LR & $0.045$--$0.052$ & $1.9$--$2.3$ & $24$--$30$ & $10$--$13$ & $78$--$98$ \\
 & & CB & $0.032$--$0.037$ & $0.83$--$1.0$ & b & $1.7$--$2.1$ & $85$--$107$ \\
\cline{2-8}
 & $0.1\%$ & LR & $2.5$--$3.1$ & $94$--$117$ & $1{,}869$--$2{,}418$ & $371$--$512$ & n.s. \\
 & & CB & $2.9$--$3.6$ & $42$--$52$ & b & b & n.s. \\
\hline
$d=5$ & $1\%$ & LR & $0.044$--$0.053$ & $1.3$--$1.7$ & $15$--$19$ & $8.3$--$11$ & $181$--$233$ \\
 & & CB & $0.033$--$0.040$ & $0.30$--$0.38$ & b & $0.63$--$0.84$ & $89$--$114$ \\
\cline{2-8}
 & $0.1\%$ & LR & $2.3$--$2.9$ & $102$--$131$ & $843$--$1{,}129$ & $319$--$442$ & n.s. \\
 & & CB & $4.7$--$6.0$ & $24$--$30$ & b & $45$--$62$ & n.s. \\
\thickhline
\end{tabular}
\end{table}

\subsection{Weights on an Euler chain}
\label{app:lv_weights}

On the near-continuum chain of the main text the first month is one Gaussian step, and
the likelihood-ratio weight sits on the whole of it, as under geometric Brownian motion.
On an Euler chain under local volatility a month has no closed-form density, and the
sampler must choose where its weight sits. We built three weights for the chain of
Sec.~\ref{sec:product} with two Euler sub-steps per month, all run with scrambled Sobol
points, four-path reflection and a control variate as in Appendix~\ref{app:mc}. The ratios
on that chain, in Appendix~\ref{app:lv}, charge each cell the cheapest of them.

The first-sub-step weight differentiates the density of the first Euler sub-step alone.
It is exact, and so unbiased for the two-sub-step chain, including the dependence of the
sub-step's covariance on the spots, and coded in closed form it agrees with automatic
differentiation to a relative $4\times10^{-13}$. But it sits on a step half as long as a
month, so its variance is larger.

The quadrature weight approximates the score of the first month's density, the
whole-month likelihood-ratio weight, by a Gauss--Hermite rule adapted to the Brownian
bridge, $6$ nodes per asset and $216$ per path at three assets. The exact score has
least variance within the unbiased weight class discussed in Sec.~\ref{sec:lr}; the
quadrature introduces a numerical approximation. With the skew switched off
it reproduces the whole-month weight of geometric Brownian motion to within $10^{-14}$,
and its Greeks agree with the tensor's tightest solve within $1.6$ standard errors. It
lowers the variance of the second- and third-order Greeks several-fold, but makes each
sample costlier by a similar factor, and the rule grows as $6^{d}$ per path. At measured
cost it is never much better than the first-sub-step weight, and we ran it at three
assets only.

The split weight moves half of each spot shift through the first sub-step and half
through the second. To build it, let $p_0$ and $p_1$ be the joint conditional densities of the two
sub-steps, $\bm x_1$ the state at the first date, and $\bm y(\bm t)$ the end point
of the first sub-step started from $\x_0+\bm t$ with the same normals, where
$\bm t=(t_1,\ldots,t_d)$ shifts the assets separately. Then
\begin{equation}
\Omega(\bm t)=\frac{p_0\big(\bm y(\bm t/2)\mid \x_0+\bm t\big)}
{p_0\big(\bm y(\bm t/2)\mid \x_0+\bm t/2\big)}\;
\frac{p_1\big(\bm x_1\mid\bm y(\bm t/2)\big)}
{p_1\big(\bm x_1\mid\bm y(\bm 0)\big)}
\label{eq:split_weight}
\end{equation}
gives $\E_{\x_0}[F\,\Omega(\bm t)]=C_0(\x_0+\bm t)$, by the change of variables from
the first sub-step's normals to $\bm y(\bm t/2)$. Thus
$\partial_{\bm t}^{\alpha}\Omega(\bm 0)$ is an unbiased weight for
$\partial^{\alpha}C_0(\x_0)$, including the mixed Greeks. It depends only on the
first sub-step's normals and on $\bm x_1$,
and costs one evaluation per path.

Under geometric Brownian motion the split weight of
Eq.~\eqref{eq:split_weight} equals the whole-month weight to a relative
$2\times10^{-13}$, and under local volatility its derivatives, which we compute by
carrying third-order Taylor expansions in each asset's shift through the formula, agree
with automatic differentiation to $6\times10^{-13}$. The four-path reflection may flip the
first sub-step alone or the whole first month. Neither pairing is better on every Greek,
so we run both and charge the cheaper per Greek.

On the same points as the first-sub-step weight at
three assets, the split weight lowers the variance of gamma, speed, cross-gamma and
cross-speed by factors of $4.2$ to $5.3$ at $x_0=-0.10$, within $4\%$ of the quadrature
weight on every one of them, and does slightly worse on delta. It costs $1.09$ times as
much per sample as the first-sub-step weight at three assets, and $1.03$ to $1.16$ times
at four and five. Its
estimates agree with the tensor's tightest solve over $30$ comparisons at three to five
assets (root-mean-square $z$-score $0.91$, largest $2.3$). It is the cheapest of the
three weights in all but five scored cells, all of them delta.

\subsection{Applicability of alternative estimators}
\label{app:mc_other}

\subsubsection*{Tested variance reductions}

Two unbiased variance reductions provided no additional cost advantage over the tuned
likelihood ratio. Conditioning
one date ahead on the whole step from Ref.~\cite{GlassermanStaum2001},
rather than along one direction as above, replaces each date's autocall and coupon
indicators by their expectations given the previous state and continues the path
conditionally on survival. Combined with the first-step likelihood ratio at $36$ dates,
the conditional expectations made each point far costlier than the variance they removed
could repay, and on most Greeks they removed none.

Regressing the payoff on Hermite polynomials of the first-step normals, and using the fit
as a control variate for every weight, left every standard error unchanged to within $3\%$,
at three assets under both models and at both spots, while costing more per sample.

\subsubsection*{Conditional Monte Carlo with few observation dates}

Conditioning on the common factor is effective with few observation dates. Given the path of the
factor the assets are independent, each asset's recursion becomes a one-dimensional
computation, and only the factor path is sampled. At six dates this is the strongest
estimator we have: at five assets it reaches $1\%$ on every Greek at $x_0=-0.10$ in at
most $1.1$ core-seconds, $2.3$ to $7.3$ times less than the tensor's charge. But the
autocall couples the assets again: the exact conditional price is an inclusion-exclusion
sum over the $2^{N-1}$ autocall patterns, so its cost doubles with every observation
date, as we measured from six to twelve dates. Avoiding the sum means carrying the joint
state of all the assets from date to date, which is the recursion the tensor solves. Conditioning therefore enters the comparison across
observation frequencies of Sec.~\ref{sec:freq}, not the monthly figures.

\subsubsection*{Methods outside the tested setting}

Under stochastic volatility the log-spots are Gaussian over each period once the path of
the variance is known, so a likelihood-ratio weight conditioned on that path sits on a
whole period again, as under geometric Brownian motion, while the solver would have to
carry the variance as one more coordinate. We therefore expect no larger advantage for
the tensor under stochastic volatility than under geometric Brownian motion.

Other alternatives were not implemented. The vibrato Monte Carlo of
Giles~\cite{Giles2009Vibrato}, extended to high orders in Ref.~\cite{Pages2016}, combines
a pathwise derivative with a likelihood ratio on the final step, but our payoff jumps at
every date, not only the last. Multilevel Monte Carlo for Greeks~\cite{BurgosGiles2012} exploits a hierarchy of time steps, which neither geometric
Brownian motion, sampled exactly, nor the near-continuum chain, one Gaussian step per
month, has.

A Malliavin or Bismut--Elworthy--Li
weight~\cite{FournieEtAl1999,ElworthyLi1994} cannot have less variance per sample than
the weight of least variance, which the split weight of Appendix~\ref{app:lv_weights}
matches under local volatility to within a few percent, at about the cost of a single
path. Unbiased
first-order estimators for discontinuous payoffs~\cite{Lakshtanov2026} appeared while this
work was in progress and are untested, and surrogates fitted to sampled prices or pathwise
derivatives, such as differential machine learning~\cite{HugeSavine2020}, approximate the
value function with no bound on their error.

\section{More than one common factor}
\label{app:factors}

Every result in the main text uses one common factor. This appendix first measures how
many factors equity correlations carry, and then shows how the solver takes more, and
what they cost, under geometric Brownian motion at monthly observation.

\subsection{Factors in equity data}
\label{app:factors_data}

\begin{figure}[t]
\centering
\includegraphics[width=\linewidth]{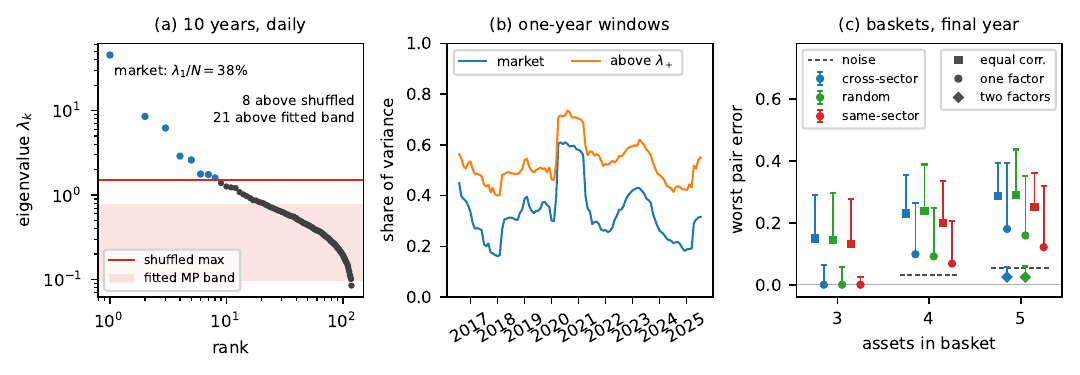}
\caption{\textbf{Factors in equity correlations.} \textbf{(a)} Eigenvalues of the correlation matrix of $119$ US large caps, ten years of daily returns to 1~August~2025, by rank. The line marks the largest eigenvalue of time-shuffled data, and the band is the Marchenko--Pastur band with its variance and sample length fitted to the bulk.
\textbf{(b)} One-year windows stepped monthly: share of the variance in the market mode, $\lambda_1/N$, and in all modes above the edge of the Marchenko--Pastur band for $252$ daily returns, $\lambda_+=2.85$. \textbf{(c)} Baskets of three to five names, with correlations over the year to 1~August~2025: of each size, $1000$ drawn at random from the $119$ (green), $1000$ with every name from a different sector (blue) and $1000$ from a single sector (red). Each marker is the median, over the $1000$, of the largest gap between a fitted and a measured pair correlation when the basket is fitted with equal correlations (squares), one factor (circles) or, at five names, two factors (diamonds), and its bar
reaches the $90$th percentile. Dashed lines: the same median for the one-factor fit on simulated baskets whose correlations are exactly one-factor, estimated from one year of Gaussian returns.}
\label{fig:rmt}
\end{figure}

Equity correlations carry a market factor, sector factors and
noise~\cite{Laloux1999,Plerou2002}. Figure~\ref{fig:rmt} shows the spectrum, its
evolution and the basket fits. To assess how many factors matter for the small
baskets on which notes are written, we use $N=119$ US large caps from all
eleven GICS sectors, with $7$ to $17$ names per sector and $T=2514$ daily returns over
the ten years to 1~August~2025, from adjusted closes on Yahoo Finance. Here $N$ and $T$
count stocks and returns, not the note's observation dates and maturity. For uncorrelated
returns the Marchenko--Pastur band ends at
$\lambda_+=(1+\sqrt{N/T})^2=1.48$~\cite{MarchenkoPastur1967}. To check whether fat
tails alone could push noise past it, we shuffle each stock's returns in time. Across
$50$ shuffles the largest eigenvalue is at most $1.50$, barely above the edge.

The market eigenvalue, $45.5$, stands far above this noise level. Every stock loads on it
with the same sign. Since the eigenvalues sum to $N$, its share of total variance is
$\lambda_1/N=38\%$. Seven more modes clear the line, setting sectors against one
another; together the eight carry $60\%$ of the variance but nearly all of the
correlation. Across the $7{,}021$ pairs, correlations average $0.37$ with standard
deviation $0.13$. A common correlation therefore misses them by $0.13$ in root-mean-square,
the market mode alone by $0.09$, and the eight modes by $0.024$. Most remaining variance
is idiosyncratic. Fitting a band to the remaining eigenvalues would count $21$ as signal,
so the number of sector modes depends on the convention, while the market mode does not.

This picture persists in one-year windows stepped monthly. With $252$ returns per window the
noise edge rises to $2.85$, and three to six modes clear it in every window. The market's
share ranges from $16$ to $61\%$, highest in the year containing March 2020. The
strength of the common movement changes, but a few factors continue to carry it.

The basket fits in Fig.~\ref{fig:rmt}(c) test whether the same low-dimensional
structure describes three to five names over the year to 1~August~2025. The caption
specifies the sampling groups and the pair-correlation error metric. A
$k$-factor model writes $\rho_{ij}=\sum_r b_{ir}b_{jr}$ and leaves each name the
idiosyncratic variance $1-\sum_r b_{ir}^2$. There are $d(d-1)/2$ correlations to fit
with $dk-k(k-1)/2$ free loadings. At three names one factor has three unknowns for
three correlations: $\rho_{ij}=b_ib_j$ gives
$b_i^2=\rho_{ij}\rho_{ik}/\rho_{jk}$, valid when these variance shares lie in $[0,1]$.
This holds for $69\%$ of the triples, including the term-sheet basket in
Appendix~\ref{app:gbm}. Equal correlations fit worst at every size. One factor is exact
in the median three-name basket, but misses the worst pair by $0.07$ to $0.10$ at four
names and $0.12$ to $0.18$ at five.

Those misses exceed sampling noise. The same fit on simulated one-factor baskets,
using one year of returns, misses by only about $0.03$ and $0.05$ at four and five names.
The real gaps are at least twice as large, so most of each reflects structure beyond
one factor. A Gaussian bootstrap rejects one factor for $62$ to $79\%$ of the random
and cross-sector baskets tested, although fat tails would lower that share. Two factors
reduce the median five-name gap to $0.025$, below the noise of a single correlation,
about $0.06$. This close fit is expected: two factors have nine free loadings for ten
correlations. At four names they have seven for six, enough to fit nearly every basket,
so we report the two-factor fit only at five names.

The data also put our benchmark correlation in context. The median basket drawn from
one sector has an average pair correlation of $0.45$, against $0.28$ for one drawn at
random, so $\rho=0.5$ is typical of related names. These data support one factor for
most three-name baskets, with unequal loadings that change each asset's kernel but not
the bond. Four- and five-name baskets generally need the additional factors considered
next.

\subsection{Staged transitions}
\label{app:factors_staged}

With $R$ factors a period moves the log-spots as
\begin{equation}
x_i'=x_i+\mu\Delta t+s\,\big(\bm b_i^\top\bm Z+\sqrt{1-|\bm b_i|^2}\,\xi_i\big),\qquad
\bm Z\sim\cN(0,I_R),\quad\bm\xi\sim\cN(0,I_d),
\label{eq:mf_step}
\end{equation}
with $s=\sigma\sqrt{\Delta t}$ as in Eq.~\eqref{eq:onefactor_step}, and the
correlation matrix is $BB^\top+\diag(1-|\bm b_i|^2)$, the rows of $B$ being the
loadings $\bm b_i$. Any correlation matrix $C$ of $d$ assets admits this form with at
most $d-1$ factors:
$C-\lambda_{\min}I$ is positive semi-definite of rank at most $d-1$, so
$C=BB^\top+\lambda_{\min}I$ with $B$ from its eigenvectors, every asset keeping the
idiosyncratic variance $\lambda_{\min}$. For positive-definite $C$ this variance is
positive, as required by the smoothing used here. The decomposition also exists for
singular $C$, but the resulting degenerate transition is outside the present analysis
and implementation tests.

Table~\ref{tab:factor_config} specifies the sector structures and factor quadratures.
For two sectors, the first contains assets $1,\dots,\lceil d/2\rceil$ and the second
the remaining assets; the three-sector case uses $\{1,2\}$, $\{3,4\}$ and $\{5\}$.
Every asset retains the systematic variance share $0.5$, so the idiosyncratic widths
match the benchmark. Factors are rotated to the principal axes of $B$, strongest
first, with fewer nodes assigned to weaker directions. The contract, volatility, rate
and scored spot $x_0=-0.10$ are unchanged.

\begin{table}[t]
\caption{\textbf{Factor configurations and quadrature.} GBM with monthly observation.
Correlations are $0.5$ within each sector; the factor-node tuple lists the
Gauss--Hermite orders in descending factor strength. Independent sectors require
different orders at three and four assets.}
\label{tab:factor_config}
\centering
\renewcommand{\arraystretch}{1.15}
\begin{tabular}{?l|ccc?}
\thickhline
structure & assets & between-sector correlation & factor nodes \\
\hline
one factor & $3$--$5$ & $0.5$ for every pair & $(12)$ \\
two sectors & $3$--$5$ & $0.25$ & $(12,7)$ \\
two independent sectors & $3$ & $0$ & $(12,9)$ \\
two independent sectors & $4$ & $0$ & $(12,12)$ \\
three sectors & $5$ & $0.25$ & $(12,7,5)$ \\
\thickhline
\end{tabular}
\end{table}

Given $\bm Z$ the assets move independently, so
the transition can be built as in Eq.~\eqref{eq:kernel} with a product rule over the
$R$ factors, a bond of $n_1n_2\cdots n_R$ nodes. Equation~\eqref{eq:mf_step} allows
more. Its moves add up, are independent of each other and do not depend on the state, so
the transition factorizes exactly into commuting smoothings, one per factor,
\begin{equation}
\mathcal K=\mathcal K_R\cdots\mathcal K_2\,\mathcal K_1,\qquad
(\mathcal K_r f)(\x)=\sum_{g=1}^{n_r}w_g\,f\big(\x+s\,z_g\,\bm b^{(r)}\big)\quad(r\ge2),
\label{eq:mf_stages}
\end{equation}
where $\bm b^{(r)}$, the $r$-th column of $B$, holds the loadings of factor $r$, and
$\mathcal K_1$ is the one-factor kernel of Eq.~\eqref{eq:kernel} with the first
factor, the whole idiosyncratic variance and the drift. Nothing here needs Gaussian
moves: independent non-Gaussian factors would stage in the same way, each with a
quadrature for its own law. Gaussianity is what lets any correlated move be written as
such a sum of independent pieces, as in Eq.~\eqref{eq:mf_step}, with factors that may be
rotated freely.

A stage $\mathcal K_r$ shifts
every asset at once along the loadings of factor $r$: each core is evaluated at its
shifted points and projected back onto its panels, so the stage is an operator of
bond $n_r$, applied, like $\mathcal K_1$, with a rounding as each node is added. A
step then costs $n_1+\dots+n_R$ node applications instead of $n_1\cdots n_R$: $19$
instead of $84$ for two sectors, $24$ instead of $420$ for three, the saving of the
separable blur described in Sec.~\ref{sec:repr}.

The stages do not make the factors free. Multiplied out, Eq.~\eqref{eq:mf_stages}
returns all $n_1\cdots n_R$ terms of the product rule, since two shifts compose into
one. Only the order of the sums changes. The product rule forms every term on its own,
while the stages store each partial sum as a train, computed once as a function of every
spot and then shifted by the nodes of the next factor rather than recomputed for each of
them. The saving thus rests on the rank of the partial sums, which, like that of the
result, grows with the number of directions in which the factors couple the assets
across a bond (Appendix~\ref{app:factors_rank}).

The event map acts
before $\mathcal K_1$, so every shift acts on a smoothed function, and the output of
$\mathcal K_1$ is represented as far beyond the box as the shifts reach, so that no
polynomial is extrapolated. The terminal step applies the shifts after the closed
form of Sec.~\ref{sec:events} with the first factor, and the read-out applies the
shifts of the first sub-step to the stored surface before it differentiates the kernel
of $\mathcal K_1$, Eq.~\eqref{eq:readout}. With one factor there is no shift stage,
and the solve is that of the main text.

Every staged solve uses the continued boundary of Sec.~\ref{sec:repr}, which extends the
train beyond the edge of the box by its values at the edge. This matters more here: with
the train cut at $\pm L$, the shifts would carry the step at the edge into the box and
bias the price. The product rule, for which we built no continued version, cuts the train.

The staged rule and the product rule use the same nodes and weights, so they differ
in only three ways: the roundings between the partial sums, the projections onto the
panels after each stage, which the product rule makes only once, and the treatment of
the edge of the box. At $10^{-8}$ the staged solves agree with the product rule to a
relative $6\times10^{-7}$ on the price and $5\times10^{-4}$ on every Greek at three to
five assets. With one factor, where nothing is staged and only the edge differs, the
two solves already differ by up to $5\times10^{-7}$ and $10^{-3}$, about as much, so
the stages add no error that we can see.

\begin{figure}[t]
\centering
\includegraphics[width=\linewidth]{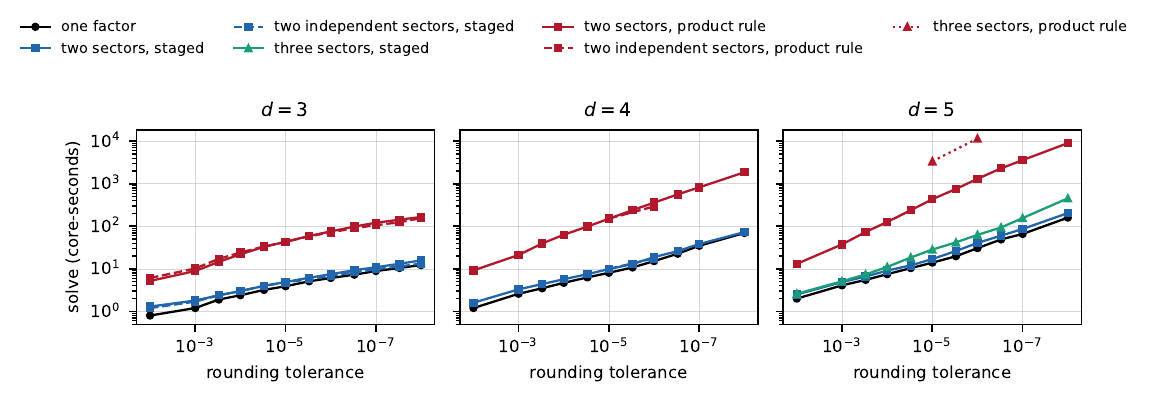}
\caption{\textbf{More than one factor.} Solve time against rounding tolerance under
geometric Brownian motion at monthly observation, three, four and five assets, with one
factor (the model of the main text), with two sectors, two independent sectors and, at
five assets, three sectors applied one factor at a time, and with the product
rule over the factors, which cuts the train at the edge of the box. All solves are
single-threaded under the full machine load of the main comparisons; the product
rule was timed separately.}
\label{fig:factors}
\end{figure}

\subsection{Cost, rank and the order of the assets}
\label{app:factors_rank}

Applying the factors in stages keeps their additional cost modest
(see Fig.~\ref{fig:factors}): over the tested tolerance range, a second factor adds less
than half to the solve time, and three sectors at five assets cost less than three
times the one-factor solve. The product rule is several to hundreds of times
costlier at the tolerances where it ran. This gap is not due to staging alone:
the product rule cuts the train at the edge of the box, which raises its rank,
whereas the staged solver continues it. Table~\ref{tab:factor_ranks} shows how
sector structure and asset ordering affect the stored rank.

\begin{table}[t]
\caption{\textbf{Factor structure and asset ordering affect the stored rank.}
Largest rank at rounding tolerance $10^{-8}$ in the staged monthly GBM solve.
Ordering is by sector unless marked interleaved. A dash denotes an untested case.}
\label{tab:factor_ranks}
\centering
\renewcommand{\arraystretch}{1.15}
\begin{tabular}{?l|ccc?}
\thickhline
structure and order & $d=3$ & $d=4$ & $d=5$ \\
\hline
one factor & $35$ & $80$ & $89$ \\
two sectors, in order & $38$ & $80$ & $117$ \\
two sectors, interleaved & -- & $178$ & $208$ \\
three sectors, in order & -- & -- & $173$ \\
\thickhline
\end{tabular}
\end{table}

The rank reflects how the factors couple the assets across a bond. For a Gaussian
kernel this coupling separates into canonical-correlation directions, with one
Hermite series per nonzero correlation (Mehler's formula~\cite{Mehler1866}).
Resolving several directions requires products of these series and can therefore
require more rank. One factor gives one direction at every bond. Two sectors laid
out in order also give at most one, since names in one sector correlate alike with
the names across a bond. Three sectors can give two directions at a bond that cuts
the middle sector, consistent with the larger ranks measured there.

Interleaving the two sectors tests this explanation without changing the model:
only the asset order changes, and the middle bonds now see two directions. At
tolerance $10^{-8}$ their ranks roughly double at four and five assets, while the
outer ranks remain unchanged. A solve costs up to about five times as much, with
the penalty growing as the tolerance tightens. Coupling strength alone is not a
rank predictor: even with one direction, the ordered two-sector surface can need
more rank than the one-factor surface despite its weaker canonical correlation.
Grouping the assets by sector is therefore preferable in these tests.

Additional factors can also cost spatial resolution. The general correlation
decomposition above leaves each asset the idiosyncratic variance $\lambda_{\min}$;
a small value narrows the kernel and may require more modes per panel. Our sector
tests keep this width fixed. We have tested at most three factors under geometric
Brownian motion and none under local volatility, where the state-dependent shifts
need not commute.

The tightest solves agree with independent, pooled likelihood-ratio Monte Carlo on
the price and all five Greeks across the six multi-factor cases; the largest
discrepancy is $2.4$ standard errors. Cost comparisons use the same multi-factor
sampler for every structure, including the one-factor baseline, and it costs about
a quarter more per path than the specialized sampler in the main text. At $1\%$
accuracy, sector factors change the comparable cost ratios by at most a factor of
four in these tests. The change reflects Monte Carlo variance as well as tensor
cost: for five-asset two-sector cross-speed, the tensor's lead narrows because the
sampler gets cheaper. With three sectors at five assets, Monte Carlo remains
cheaper on delta, gamma stays near parity, and the tensor remains cheaper on speed
and cross-gamma. Cross-speed is unscored there because two reference standard errors
exceed the $1\%$ target.

\section{Scoring and measurements}
\label{app:measurements}

This appendix specifies the numerical comparisons, checks their accuracy and
examines how their costs change with the contract, model and target. The Greek
experiments are followed by the exposure reference, estimator settings and controls.

\subsection{Computational setup}
\label{app:scoring}

Both methods run in Julia on an Apple M4 Max. We report core-seconds, elapsed time
multiplied by the number of threads, so that parallel Monte Carlo runs and
single-threaded tensor solves are compared on the same basis. Timings include
compilation but not code loading, and repeated reference runs are used to normalize
timings measured separately. For GBM, sampling costs use the constant-volatility
implementation, which returns identical results about $1.5$ to $2$ times faster
than the general implementation. We omit tensor setup and the Monte Carlo
control-variate pilot, which cost less than $0.02$ seconds and at most $0.2$
core-seconds, respectively. Evaluating the Greeks is included for the transformed
local-volatility chain and omitted elsewhere, where it adds at most $2.9\%$ to
the solve time per spot.

The reference for a Greek is itself a Monte Carlo estimate, so its uncertainty can
change which tensor solve appears to meet the target. Moving that estimate up or
down by one standard error changes the selected tolerance by at most one step on
the tested grid. We also ran shorter Monte Carlo simulations, of the lengths the
targets need, to check the costs inferred from the deep runs. For the $52$ GBM cells
whose Monte Carlo cost lies below $1{,}000$ core-seconds, the measured costs are a
median $1.34$ times the inferred ones (range $0.87$ to $2.48$), so the inference mostly
favors Monte Carlo. Using the measured costs changes which method is cheaper in only
one tested case: gamma for five assets under GBM at $1\%$, where the tensor becomes
cheaper. Two further choices err against the tensor: charging the cheapest solve that
meets a target, rather than the sustained one, would lower its charge in $13$ of the
$91$ cells on the GBM and Euler chains, by up to a factor of six, and leaving out
compilation would raise the ratios by at most $20\%$.

\subsection{Geometric Brownian motion: agreement, contract terms and a market note}
\label{app:gbm}

\subsubsection*{Reference agreement}

At three assets and $x_0=-0.10$ the deepest Monte Carlo runs
($5.3\times10^{10}$ samples) resolve the Greeks to between $0.002\%$ (delta) and $0.26\%$
(cross-speed) at two standard errors. The solve's largest deviation from them is
$0.056\%$, on speed, or $2.2$ standard errors, and every other Greek lies within $1.6$.
Joint discretization refinement and the unrounded full-grid calculation leave
the small deviations beyond two standard errors essentially unchanged, as the
checks in Appendices~\ref{app:err} and~\ref{app:fullgrid} show. Agreement also holds
away from the scored spot: all $60$ estimates of the five Greeks from the two
estimators at six spots between $x_0=-0.45$ and $0.15$ lie within $2.5$ standard
errors of the surface. Figure~\ref{fig:object}(a) shows three of the five Greeks. At the
autocall barrier cross-gamma nearly vanishes, making relative error a poor metric.

\subsubsection*{Contract extensions}

The factorization in Eq.~\eqref{eq:factorize} also accommodates a call lockout,
different coupon and call schedules, and step-down autocall levels. These change
the event map, with a panel edge added at each new level. Coupon memory needs an
additional state, illustrated below. Daily protection monitoring would require
a breach state and daily events; a barrier on an average would lose the rank-one
factorization; and an issuer's exercise right would require a maximum or minimum
on an initially unknown boundary. These three extensions are not tested here.

\subsubsection*{Term-sheet validation with historical inputs}

The benchmark of Sec.~\ref{sec:formulation} has memoryless coupons, and its model
equal volatilities and a single correlation. To test the solver beyond those
choices, we also priced an indicative three-stock worst-of Phoenix note on GOOGL,
JPM and WMT~\cite{MorganStanleyPhoenix2025}: $12$ quarterly coupons of $2.75\%$, a
$60\%$ coupon barrier, a $60\%$ European barrier at maturity, a $100\%$ autocall level
from the second observation, and catch-up of missed coupons. The value is affine
in the number $m$ of missed coupons, so the backward sweep carries only two
continuation surfaces,
$C_k(\x,m)=C_k^{(0)}(\x)+m\,C_k^{(1)}(\x)$, and the call lockout changes only the first
event map.

The dynamics are geometric Brownian motion with historical inputs, not the issuer's
model. We use
the one-year historical volatilities $(31.44,27.92,23.97)\%$ and pairwise correlations
$(0.402,0.233,0.405)$ up to the strike date,
1~August~2025~\cite{YahooFinanceHistorical2025}, trailing dividend yields
$(0.43,1.83,0.90)\%$ as constant-yield proxies, and
that day's three-year Treasury par yield, $3.67\%$, as a flat
rate~\cite{USTreasuryAugust2025}. Observation dates are equally spaced, shares
delivered at maturity are valued at their closing price, and payment lags are
ignored. The covariance has an exact one-factor decomposition,
$\Sigma=bb^\top+\diag(s_i^2)$, giving each asset its own kernel. At tolerance
$10^{-8}$ the largest ranks are $27$ for the base surface and $20$ for the memory
surface.

Table~\ref{tab:market_note} compares the solve with independently implemented Monte
Carlo paths that count their own missed coupons. Price and the three Greeks agree
within $1.2$ standard errors. A further check starts with three unpaid coupons,
testing the memory surface rather than only the zero-memory inception value;
its price agrees with $10^8$ Monte Carlo paths within one standard error. These
are implementation checks for coupon memory, a lockout and heterogeneous assets,
not a calibration to the indicative issue price or a cost comparison with tuned
Monte Carlo.

\begin{table}[t]
\caption{\textbf{A market term sheet.} Three-asset worst-of Phoenix note with the
indicative terms of Ref.~\cite{MorganStanleyPhoenix2025} ($12$ quarterly dates,
coupon memory, call lockout) under geometric Brownian motion with dated historical
inputs: price per unit notional and log-spot Greeks at inception, asset~1 GOOGL and
asset~2 JPM. The tensor carries two coupon-memory surfaces (tolerance
$10^{-8}$). Monte Carlo uses $4\times10^{8}$ antithetic paths in four
independent replicates, each path counting its own missed coupons, and its
uncertainty is the larger of the pair-based and between-replicate standard errors.}
\label{tab:market_note}
\centering
\renewcommand{\arraystretch}{1.2}
\begin{tabular}{?l|ccc?}
\thickhline
 & tensor & Monte Carlo & difference in standard errors \\
\hline
price & $0.9613261$ & $0.9613402\pm0.0000119$ & $-1.19$ \\
\hline
delta $\Delta_{1}$ & $0.1952999$ & $0.1951905\pm0.0001042$ & $+1.05$ \\
\hline
gamma $\Gamma_{11}$ & $-0.6534789$ & $-0.6563888\pm0.0046270$ & $+0.63$ \\
\hline
cross-gamma $\Gamma_{12}$ & $0.0566782$ & $0.0550480\pm0.0046790$ & $+0.35$ \\
\thickhline
\end{tabular}
\end{table}

\subsection{Geometric Brownian motion: costs and ratios}
\label{app:gbm_ratios}

\begin{figure}[t]
\centering
\includegraphics[width=\linewidth]{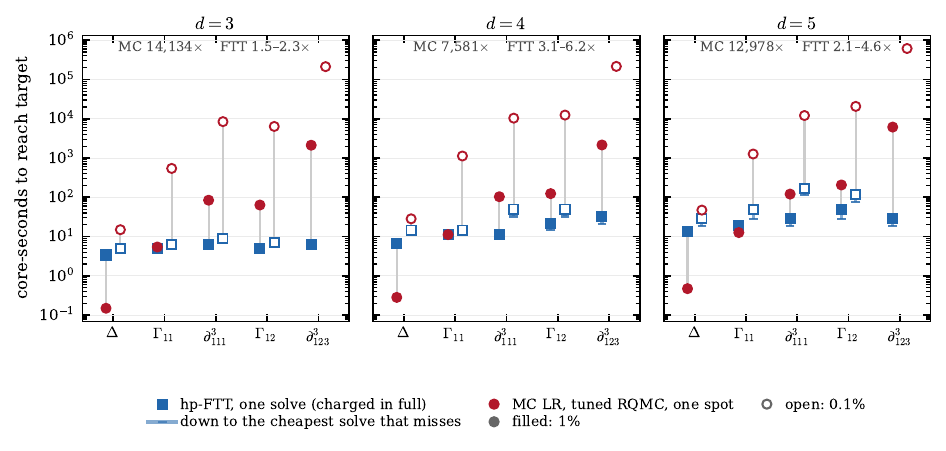}
\caption{\textbf{Monte Carlo pays for derivative order, the tensor far less.}
Core-seconds to reach $1\%$ (filled) and $0.1\%$ (open) relative error on each
Greek at $x_0=-0.10$, monthly observation, $d=3,4,5$ on a common vertical scale.
Monte Carlo is the tuned likelihood-ratio estimator, scaled by
Eq.~\eqref{eq:mc_cost} from $5.3\times10^{10}$ samples at $d=3$ and
$3.0\times10^{10}$ at $d=4,5$. Tensor squares charge the full solve at sustained
accuracy; bars extend to the next looser solve, half a tolerance decade away,
which misses the target. Cross-speed at $0.1\%$ is unresolved by the independent
reference and has no tensor point; its Monte Carlo cost is extrapolated beyond
the measured runs. All other targets shown lie within those runs.}
\label{fig:cost_dims}
\end{figure}

\begin{figure}[t]
\centering
\includegraphics[width=\linewidth]{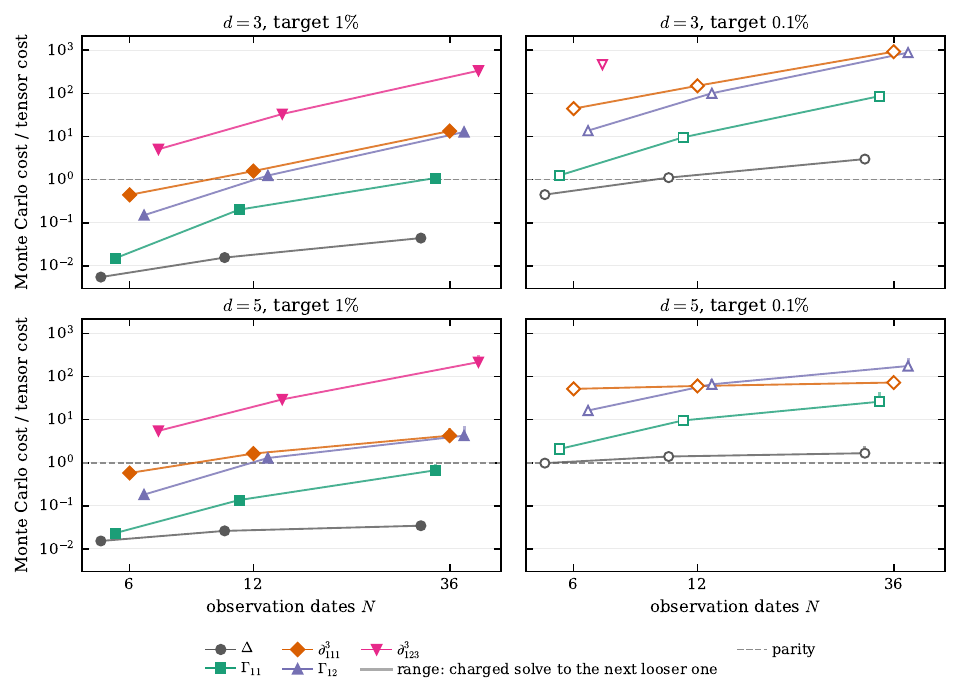}
\caption{\textbf{Observation frequency moves the ratio.} Monte Carlo cost over
tensor cost for each Greek at $x_0=-0.10$ under geometric Brownian motion, with
$6$, $12$ and $36$ observation dates (coupon $0.36/N$ per date), for three assets
(top) and five (bottom), at $1\%$ (left, filled) and $0.1\%$ (right, open). Each
marker uses the sustained tensor charge, bracketed by the next looser solve.
Monte Carlo is the tuned likelihood-ratio estimator at every $N$, with its own
deep runs supplying both the cost estimate and the reference. Cells unresolved
at two reference standard errors are omitted. Above the dashed line the tensor
is cheaper.}
\label{fig:dates}
\end{figure}

Figure~\ref{fig:cost_dims} separates the Monte Carlo and tensor costs behind the
monthly ratios in Table~\ref{tab:gbm_ratios}. It shows the contrast discussed in
Sec.~\ref{sec:lr}: derivative order changes sampling cost by orders of magnitude,
but changes the tensor charge much less. Figure~\ref{fig:dates} varies the
observation frequency while keeping the maturity and annual coupon fixed.

\begin{table}[t]
\caption{\textbf{Geometric Brownian motion, monthly observation: Monte Carlo cost
over tensor cost.} $N=36$, $x_0=-0.10$. Each entry is the range set by the tensor's
bracket, from its charged solve to the next looser one, which misses the target.
Above $1$ the tensor is cheaper. Monte Carlo is the tuned likelihood-ratio
estimator, scaled from its deepest runs, which also supply the independent
reference. n.s.: not scored because two reference standard errors exceed the
target. The three- to five-asset $1\%$ and $0.1\%$ rows correspond to
Fig.~\ref{fig:greeks}(a).}
\label{tab:gbm_ratios}
\centering
\renewcommand{\arraystretch}{1.2}
\begin{tabular}{?l|l|ccccc?}
\thickhline
 & & delta & gamma & speed & cross-gamma & cross-speed \\
 & target & $\Delta$ & $\Gamma_{11}$ & $\partial^3_{111}$ & $\Gamma_{12}$ & $\partial^3_{123}$ \\
\hline
 & $1\%$ & $0.044$--$0.056$ & $1.1$--$1.2$ & $13$--$17$ & $13$--$14$ & $336$--$424$ \\
\cline{2-7}
$d=3$ & $0.3\%$ & $0.38$--$0.49$ & $12$--$14$ & $130$--$149$ & $112$--$141$ & $3{,}269$--$3{,}736$ \\
\cline{2-7}
 & $0.1\%$ & $3.0$--$3.4$ & $86$--$109$ & $928$--$1{,}172$ & $884$--$1{,}011$ & n.s. \\
\hline
 & $1\%$ & $0.042$--$0.055$ & $1.0$--$1.3$ & $9.2$--$12$ & $5.9$--$8.4$ & $66$--$101$ \\
\cline{2-7}
$d=4$ & $0.3\%$ & $0.28$--$0.37$ & $8.5$--$11$ & $35$--$54$ & $42$--$65$ & n.s. \\
\cline{2-7}
 & $0.1\%$ & $1.9$--$2.5$ & $77$--$100$ & $207$--$318$ & $248$--$382$ & n.s. \\
\hline
 & $1\%$ & $0.035$--$0.045$ & $0.67$--$0.92$ & $4.2$--$6.4$ & $4.2$--$7.2$ & $215$--$323$ \\
\cline{2-7}
$d=5$ & $0.3\%$ & $0.28$--$0.38$ & $4.9$--$7.4$ & $18$--$27$ & $19$--$30$ & n.s. \\
\cline{2-7}
 & $0.1\%$ & $1.7$--$2.5$ & $26$--$44$ & $72$--$102$ & $175$--$273$ & n.s. \\
\hline
 & $1\%$ & $0.029$--$0.045$ & $0.84$--$1.3$ & $4.8$--$7.4$ & $4.3$--$6.8$ & $31$--$48$ \\
\cline{2-7}
$d=7$ & $0.3\%$ & $0.20$--$0.32$ & $3.2$--$5.7$ & $18$--$32$ & $30$--$48$ & n.s. \\
\cline{2-7}
 & $0.1\%$ & $1.0$--$1.8$ & $29$--$52$ & $104$--$164$ & n.s. & n.s. \\
\hline
 & $1\%$ & $0.018$--$0.031$ & $0.89$--$1.5$ & $1.5$--$2.6$ & $5.5$--$8.9$ & n.s. \\
\cline{2-7}
$d=10$ & $0.3\%$ & $0.070$--$0.11$ & $3.2$--$5.6$ & $5.9$--$10$ & n.s. & n.s. \\
\cline{2-7}
 & $0.1\%$ & $0.36$--$0.63$ & $4.1$--$6.9$ & n.s. & n.s. & n.s. \\
\thickhline
\end{tabular}
\end{table}

More observation dates cost the tensor roughly one operator application per date,
with modest changes in resolution and rank. From six dates to $36$ at three
assets, a solve at fixed tolerance grows about fivefold, and the charged second-
and third-order solves grow less than tenfold. Their Monte Carlo costs to $1\%$
grow by more than two orders of magnitude. With six dates, conditioning on the
common factor is a stronger sampling competitor, but the tensor is still cheaper
on every scored Greek at $0.1\%$ at three and five assets. Frequent observation
and tight targets therefore reinforce the same advantage for different reasons:
shorter steps raise sampling variance, and a tighter target requires far more
extra paths than extra tensor work.

Adding assets increases the tensor's cost mainly through rank, not spatial
resolution. Between three and five assets, each added asset multiplies solve
cost by less than two at loose tolerances and by roughly four to five at the tightest;
the tolerance needed for a fixed Greek target also tends to tighten. Beyond five
assets the growth slows. At tolerance $10^{-8}$ the largest ranks are $89$, $106$
and $130$ at five, seven and ten assets, and solve cost grows by about $1.2$ to
$1.8$ per added asset over the tested tolerances. The seven- and ten-asset runs
use the same scoring rule and tolerance ladder, though the robustness checks
above were not repeated for them. Their tightest solves agree with the independent
references on all five Greeks, with a largest discrepancy of $2.25$ standard errors.

The high-dimensional ratios in Table~\ref{tab:gbm_ratios} show where this rank cost
matters. At ten assets Monte Carlo is cheaper on delta even at $0.1\%$, while the
tensor retains a lead on gamma. Cross-speed is resolved at $1\%$ at seven assets,
but not at ten, where two reference standard errors are $2.9\%$. Few observation
dates favor sampling even more strongly: with six dates, tensor solves reaching
$1\%$ on all five Greeks take about $5$ and $16$ core-seconds at seven and ten
assets, whereas conditioning on the factor reaches that target in at most one
core-second. Dimension alone does not determine the cheaper method; the dates
and the requested accuracy remain decisive.

\subsection{Local volatility: the two-sub-step chain}
\label{app:lv}

The main local-volatility comparison uses the transformed monthly chain. To test
whether that choice determines the outcome, Tables~\ref{tab:lv_ratios}
and~\ref{tab:lv_ratios_d45} instead put both methods on two Euler sub-steps per
month. Monte Carlo is charged the cheapest applicable weight from
Appendix~\ref{app:lv_weights}. The reference pools independent first-sub-step
likelihood-ratio runs on this same chain, so agreement here tests the numerical
methods, not the approximation to the continuous diffusion.

Volatility rising into the left tail requires a wider domain than under GBM.
The production box is $[-4.5,4.5]^d$; widening it to $[-6,6]^d$ at three assets
and tolerance $10^{-9}$ changes no Greek by more than $0.0015\%$ at either test
spot and moves the price by less than one reference standard error. Across three
to five assets, the tightest solves agree with Monte Carlo on the price and all
five Greeks, with a largest discrepancy of $2.1$ standard errors.

Against the likelihood-ratio estimators, the dependence on order and accuracy is
unchanged: delta favors sampling at $1\%$ and the tensor at $0.1\%$, while all
scored higher-order Greeks favor the tensor. Conditioned bumps remain competitive:
at $1\%$ they are near parity with the tensor at four assets and on gamma at five,
and cheaper on cross-gamma at five. Their published bump cannot reach $1\%$ on
speed. The alternative chain therefore changes individual margins without
removing the dependence on derivative order and accuracy.

\begin{table}[t]
\caption{\textbf{Local volatility, three assets: Monte Carlo cost over tensor
cost.} Monthly observation, both methods on two Euler sub-steps per month
($n_{\rm sub}=2$). Brackets charge the tensor at sustained accuracy and at the
next looser solve; above $1$ the tensor is cheaper. Monte Carlo uses the cheapest
of the three weights in Appendix~\ref{app:lv_weights}, usually the split weight,
with costs scaled from its measured runs. The reference pools first-sub-step
runs. n.s.: unresolved at two reference standard errors.}
\label{tab:lv_ratios}
\centering
\renewcommand{\arraystretch}{1.2}
\begin{tabular}{?l|ccc|ccc?}
\thickhline
 & \multicolumn{3}{c|}{$x_0=-0.10$} & \multicolumn{3}{c?}{$x_0=0$ (autocall barrier)} \\
\cline{2-7}
 & $1\%$ & $0.3\%$ & $0.1\%$ & $1\%$ & $0.3\%$ & $0.1\%$ \\
\hline
delta $\Delta$ & $0.085$--$0.11$ & $0.69$--$0.94$ & $5.1$--$6.2$ & $0.30$--$0.36$ & $2.6$--$3.4$ & $11$--$14$ \\
\hline
gamma $\Gamma_{11}$ & $2.0$--$2.4$ & $22$--$27$ & $113$--$156$ & $1.1$--$1.4$ & $9.1$--$13$ & $82$--$114$ \\
\hline
speed $\partial^3_{111}$ & $65$--$83$ & $524$--$723$ & n.s. & $66$--$87$ & n.s. & n.s. \\
\hline
cross-gamma $\Gamma_{12}$ & $15$--$19$ & $134$--$170$ & $872$--$1{,}202$ & $63$--$87$ & n.s. & n.s. \\
\hline
cross-speed $\partial^3_{123}$ & $155$--$214$ & n.s. & n.s. & n.s. & n.s. & n.s. \\
\thickhline
\end{tabular}
\end{table}

\begin{table}[t]
\caption{\textbf{Local volatility, four and five assets: Monte Carlo cost over tensor
cost.} Monthly observation, both methods on two Euler sub-steps per month
($n_{\rm sub}=2$), $x_0=-0.10$.
Brackets and scoring are as in Table~\ref{tab:lv_ratios}; above $1$ the tensor is
cheaper. Monte Carlo uses the cheaper of the first-sub-step and split weights;
the quadrature weight was tested at three assets only. n.s.: unresolved at two
reference standard errors.}
\label{tab:lv_ratios_d45}
\centering
\renewcommand{\arraystretch}{1.2}
\begin{tabular}{?l|l|ccccc?}
\thickhline
 & & delta & gamma & speed & cross-gamma & cross-speed \\
 & target & $\Delta$ & $\Gamma_{11}$ & $\partial^3_{111}$ & $\Gamma_{12}$ & $\partial^3_{123}$ \\
\hline
 & $1\%$ & $0.072$--$0.094$ & $1.9$--$2.7$ & $34$--$47$ & $3.4$--$4.3$ & $73$--$105$ \\
\cline{2-7}
$d=4$ & $0.3\%$ & $0.41$--$0.58$ & $15$--$21$ & n.s. & $26$--$38$ & n.s. \\
\cline{2-7}
 & $0.1\%$ & $3.7$--$5.2$ & $71$--$91$ & n.s. & n.s. & n.s. \\
\hline
 & $1\%$ & $0.096$--$0.13$ & $1.7$--$2.6$ & $17$--$23$ & $11$--$17$ & $142$--$216$ \\
\cline{2-7}
$d=5$ & $0.3\%$ & $0.36$--$0.50$ & $19$--$29$ & $89$--$123$ & $62$--$91$ & n.s. \\
\cline{2-7}
 & $0.1\%$ & $3.3$--$4.5$ & $83$--$127$ & n.s. & n.s. & n.s. \\
\thickhline
\end{tabular}
\end{table}

\subsection{Local volatility: time steps and the continuous model}
\label{app:lv_routes}

\subsubsection*{The discretized models}

The Euler chain applies Eq.~\eqref{eq:lv_kernel} with
$\delta t=\Delta t/n_{\rm sub}$. The transformed chain used in the main comparisons
instead takes one Gaussian step per observation period in the Lamperti coordinate
$y=\tilde x$. Write $a(y)=\tilde\mu'(y)$ for the local slope of the transformed drift.
The implemented one-asset mean and variance are
\begin{equation}
m(y)=y+\tilde\mu(y)\,\frac{e^{a(y)\Delta t}-1}{a(y)},\qquad
v(y)=\frac{e^{2a(y)\Delta t}-1}{2a(y)},
\label{eq:lamperti_moments}
\end{equation}
with $m(y)=y+\tilde\mu(y)\,\Delta t$ and $v(y)=\Delta t$ when $a(y)=0$. These are the exact
moments of a step with the drift linearized at $y$. The chain couples
them through the same common-factor innovations as the benchmark, with
$Z,\xi_1,\dots,\xi_d$ independent standard normals:
\begin{equation}
\tilde x_i'=m(\tilde x_i)+\sqrt{v(\tilde x_i)}
\bigl(\sqrt\rho\,Z+\sqrt{1-\rho}\,\xi_i\bigr),\qquad i=1,\dots,d.
\label{eq:lamperti_chain}
\end{equation}
Thus the conditional cross-covariance is
$\rho\sqrt{v(\tilde x_i)v(\tilde x_j)}$ for $i\ne j$. This specifies the implemented
Gaussian chain, not an exact monthly transition of the coupled diffusion. Both Monte
Carlo and the tensor solver use this chain; the latter applies its density in log-spot
coordinates with the output-coordinate Jacobian $1/\sigma(x_i')$.

\subsubsection*{Time-step convergence}

On an Euler chain both methods work on a discretization of the diffusion, not on the
diffusion itself. The Euler scheme's weak error has an expansion in the step,
$\E f(\bar X_T)-\E f(X_T)=c_1\,\delta t+O(\delta t^2)$, for smooth functionals of the
terminal value~\cite{TalayTubaro1990} and for bounded measurable ones when the
diffusion is non-degenerate~\cite{BallyTalay1996}, so Richardson extrapolation in
$1/n_{\rm sub}$, $2V_{2n}-V_n$ with $V_n$ computed on $n$ sub-steps per period, cancels
the first-order term. We test this behavior for the contract recursion and its
Greeks rather than assume that those results apply directly. Deterministic
extrapolation costs two solves and adds no sampling noise; Monte Carlo would
also need to control the variance of the difference between levels.

\begin{table}[H]
\caption{\textbf{Time-step error under local volatility, three assets.} Error in
percent relative to the extrapolation from $8$ and $16$ sub-steps per month
(tolerance $10^{-9}$): at two sub-steps per month, the chain of
Table~\ref{tab:lv_ratios}, and after extrapolating from two and four sub-steps.}
\label{tab:lv_dt}
\centering
\renewcommand{\arraystretch}{1.2}
\begin{tabular}{?l|cc|cc?}
\thickhline
 & \multicolumn{2}{c|}{$x_0=-0.10$} & \multicolumn{2}{c?}{$x_0=0$} \\
\cline{2-5}
 & $n_{\rm sub}=2$ & from $2$ and $4$ & $n_{\rm sub}=2$ & from $2$ and $4$ \\
\hline
price & $-0.04$ & $-0.001$ & $-0.01$ & $0.000$ \\
\hline
delta $\Delta$ & $+0.28$ & $+0.003$ & $+0.43$ & $+0.002$ \\
\hline
gamma $\Gamma_{11}$ & $-0.85$ & $+0.002$ & $+0.20$ & $+0.007$ \\
\hline
speed $\partial^3_{111}$ & $+4.28$ & $-0.096$ & $-5.96$ & $-0.038$ \\
\hline
cross-gamma $\Gamma_{12}$ & $+0.79$ & $-0.001$ & $+2.49$ & $-0.016$ \\
\hline
cross-speed $\partial^3_{123}$ & $-0.21$ & $-0.031$ & $+4.26$ & $-0.069$ \\
\thickhline
\end{tabular}
\end{table}

At three assets and rounding tolerance $10^{-9}$, refinement through
$n_{\rm sub}=2,4,8,16$ nearly halves each quantity's error when the sub-step is
halved. Table~\ref{tab:lv_dt} shows that extrapolation removes most of the
several-percent error of the two-step chain on third-order Greeks. Extrapolating
from four and eight steps instead changes no Greek by more than $0.013\%$
relative to the eight- and sixteen-step result. Four and five assets show the
same behavior: two steps miss speed at the autocall barrier by up to about
$8\%$, and extrapolation from two and four reduces the largest discrepancy to
$0.12\%$. These measurements support first-order Euler error and second-order
extrapolated error; they do not prove either rate for this recursion.

The refinement solves use tolerance $10^{-9}$ throughout. A fixed looser
tolerance need not suffice as roundings accumulate: with sixteen sub-steps per
month, $10^{-8}$ and $10^{-9}$ differ by up to $0.34\%$. At four and five assets,
doubling the sub-steps more than doubles solve time. Finer chains also lengthen
Monte Carlo paths, but a whole-month weight could avoid the variance penalty
of differentiating the first short sub-step. We have built no such weight beyond
two sub-steps and therefore report no matched-cost ratio on the finer chains.

\subsubsection*{Accuracy and cost of the transformed chain}

The transformed monthly step avoids this repeated sub-stepping in both methods.
A one-asset check compares its barrier probabilities and first three spot
derivatives with a refined Euler extrapolation: at $x_0=-0.10$ they agree to
$0.004\%$. The larger relative discrepancy at $x_0=0$, $0.24\%$, occurs on a
nearly vanishing second derivative. For the full contract, the transformed-chain
Greeks lie within $0.2\%$ of extrapolation from eight and sixteen Euler steps at
three assets, and within $0.07\%$ of extrapolation from four and eight steps at
four and five assets. These are tests of the monthly chain's approximation to
the diffusion, separate from the agreement between methods on that chain.

The read-out uses $24$ factor nodes for the first month. Doubling them changes
no price or Greek by more than about $0.01\%$ at three to five assets. The
tightest solves also agree with independent likelihood-ratio Monte Carlo on
the chain: the largest discrepancies are $2.3$ standard errors at three assets
and $1.4$ at four and five. At tolerance $10^{-8}$ the transformed solve is
roughly twice as fast as two Euler sub-steps at each basket size. Monte Carlo
also gains, by factors of two to four on the second- and third-order Greeks
relative to its split-weight Euler estimator; both the chain and the weight
change in that comparison. Table~\ref{tab:lv_continuum} gives the resulting
cost ratios against the likelihood-ratio estimator, and Table~\ref{tab:published}
includes the conditioned bumps.

\begin{table}[t]
\caption{\textbf{Local volatility near the continuum limit: Monte Carlo cost over
tensor cost.} Both methods on the transformed chain of Sec.~\ref{sec:product}, one step
per month, $x_0=-0.10$. Each entry is the range set by the tensor's bracket, from its
charged solve to the next looser one, which misses the target. Above $1$ the tensor is
cheaper. Monte Carlo uses the exact whole-month likelihood-ratio weight; its
deep runs supply the cost estimate and independent reference. n.s.: unresolved
at two reference standard errors. The $1\%$ and $0.1\%$ rows are the blue markers
in Fig.~\ref{fig:greeks}(b).}
\label{tab:lv_continuum}
\centering
\renewcommand{\arraystretch}{1.2}
\begin{tabular}{?l|l|ccccc?}
\thickhline
 & & delta & gamma & speed & cross-gamma & cross-speed \\
 & target & $\Delta$ & $\Gamma_{11}$ & $\partial^3_{111}$ & $\Gamma_{12}$ & $\partial^3_{123}$ \\
\hline
 & $1\%$ & $0.051$--$0.054$ & $2.1$--$2.2$ & $54$--$59$ & $12$--$14$ & $132$--$151$ \\
\cline{2-7}
$d=3$ & $0.3\%$ & $0.51$--$0.57$ & $17$--$18$ & $522$--$596$ & $124$--$137$ & $1{,}288$--$1{,}471$ \\
\cline{2-7}
 & $0.1\%$ & $4.6$--$5.1$ & $150$--$166$ & $3{,}085$--$3{,}551$ & $976$--$1{,}114$ & n.s. \\
\hline
 & $1\%$ & $0.045$--$0.052$ & $1.9$--$2.3$ & $24$--$30$ & $10$--$13$ & $78$--$98$ \\
\cline{2-7}
$d=4$ & $0.3\%$ & $0.34$--$0.42$ & $21$--$26$ & $269$--$337$ & $57$--$71$ & $626$--$865$ \\
\cline{2-7}
 & $0.1\%$ & $2.5$--$3.1$ & $94$--$117$ & $1{,}869$--$2{,}418$ & $371$--$512$ & n.s. \\
\hline
 & $1\%$ & $0.044$--$0.053$ & $1.3$--$1.7$ & $15$--$19$ & $8.3$--$11$ & $181$--$233$ \\
\cline{2-7}
$d=5$ & $0.3\%$ & $0.25$--$0.32$ & $11$--$15$ & $167$--$215$ & $69$--$92$ & $1{,}510$--$2{,}012$ \\
\cline{2-7}
 & $0.1\%$ & $2.3$--$2.9$ & $102$--$131$ & $843$--$1{,}129$ & $319$--$442$ & n.s. \\
\thickhline
\end{tabular}
\end{table}

For a time-homogeneous model, a deterministic solver could also compose the
sub-steps into a compressed period operator and reuse it at every date. Its repeated
cost would depend on the operator bond rather than directly on the number of
sub-steps. This optimization is not implemented here.

\subsection{Exposure tails: reference and baselines}
\label{app:exposure}
\label{app:tails}

All exposure estimators are scored against the same month-$18$ quantile. The
definitions below distinguish the exposure being measured, the accuracy of its
reference and the approximation made by each estimator.

\subsubsection*{Estimand and accuracy scale}

The note's value on the exposure date includes that date's payments: $1+c$ on the paths that
the note calls there, and otherwise the coupon, if due, plus the continuation value.
The called states form an atom at $1+c$; the $97.5\%$ quantile lies above it,
among live states. Outer log-spots follow their unconditional GBM law at month
$18$, so this is the exposure of the note with its first call date at month $18$.
The equivalence uses memoryless coupons and protection monitored only at maturity.
For a note callable from month $1$, states on already-called paths would instead
have zero exposure~\cite{GlauPachonPoetz2021,DemeterfiGlauWunderlich2025}.
The comparison concerns one trade, without netting or collateral, under pricing-
measure scenarios rather than a historical scenario measure~\cite{BCBS_CRE50}.
Changing that measure changes the outer sampling law, not the pricing surface;
netting and collateral would additionally change the exposure mapping and are
not tested here.

Targets are RMSE relative to $\mathrm{PFE}-\mathrm{EE}$, not the whole PFE.
The gap is $0.108$ to $0.146$ of notional at three to five assets, and $0.161$
and $0.174$ at six and seven. EE is estimated on the $10^{-6}$ surface with
$2^{18}$ randomized Sobol states. At three to five assets, the targets $1\%$,
$0.3\%$ and $0.1\%$ of the gap correspond to about $11$--$15$, $3$--$4$ and
$1.1$--$1.5$ basis points of notional. As a sampling yardstick, $10^4$ independent
states valued exactly would give a quantile standard error of $0.12$--$0.15\%$
of the gap, from $\sqrt{\alpha(1-\alpha)/n}/f(q)$ with $f$ the value density at
the quantile~\cite{Glasserman2004}. A target relative to the whole PFE is much
looser: nested Monte Carlo reaches $0.1\%$ of it in under a core-second.

\subsubsection*{Reference construction and independent checks}

Table~\ref{tab:tail_ref} specifies the refined tensor surfaces and randomized Sobol
evaluations used for the reference. Their standard errors are less than a fifth
of the RMSE at the tensor's best scored points. Refining the surface at matched
outer states moves the quantile by at most $1.4\times10^{-6}$; this correction is
included.

An independent nested Monte Carlo oracle checks the reference without using a
tensor surface. It extrapolates in inverse inner-path count to remove the leading
inner bias, and agrees with the reference within $1.6$ standard errors at every
dimension. Its standard errors are of order $10^{-4}$, however, so it does not
independently resolve every tight scored error. The finer accuracy is supported
by surface refinement and deeper outer integration. At three and four assets,
panelwise Gauss--Legendre quadrature over the first $d-1$ assets, with the last
integrated analytically, agrees with the sampled reference within two reference
standard errors; further grid or surface refinement moves the quantile by less
than $3\times10^{-7}$. This last check tests outer integration on the same surface,
not an independent pricing approximation.

\begin{table}[t]
\caption{\textbf{The tail reference and its checks.} Potential future exposure at
$97.5\%$ at month $18$ of the monthly note under geometric Brownian motion.
Reference: the tensor's quantile on a refined surface (rounding tolerance $10^{-8}$
and $16$ replications of $131{,}072$ randomized Sobol points at $d=3,4$, and $10^{-7}$ and
$64$ replications at $d=5$--$7$), with its standard error (SE). It includes the shift
to a still tighter surface ($3\times10^{-9}$ at $d=3,4$, $10^{-8}$ at $d=5$--$7$)
evaluated at the same points, given under ``tighter''. Oracle: nested Monte Carlo
with no tensor ($20{,}000$ outer states, $12$ replications), extrapolated in
$1/N_{\rm in}$ from $N_{\rm in}=2{,}000$ and $4{,}000$, after also testing $1{,}000$.
The score $z$ is its difference from
the reference over the combined standard error.}
\label{tab:tail_ref}
\centering
\renewcommand{\arraystretch}{1.2}
\begin{tabular}{?c|ccc|ccc?}
\thickhline
$d$ & reference & SE & tighter & oracle $-$ reference & SE & $z$ \\
\hline
$3$ & $1.02251150$ & $2.1\times10^{-6}$ & $+1.2\times10^{-8}$ & $-2.16\times10^{-5}$ & $5.0\times10^{-5}$ & $-0.43$ \\
\hline
$4$ & $1.02120250$ & $4.2\times10^{-6}$ & $+2.0\times10^{-8}$ & $-1.20\times10^{-5}$ & $8.0\times10^{-5}$ & $-0.15$ \\
\hline
$5$ & $1.01994045$ & $5.3\times10^{-6}$ & $+7.8\times10^{-7}$ & $-6.69\times10^{-5}$ & $1.38\times10^{-4}$ & $-0.48$ \\
\hline
$6$ & $1.01873245$ & $6.1\times10^{-6}$ & $+1.4\times10^{-6}$ & $+1.54\times10^{-4}$ & $1.20\times10^{-4}$ & $+1.28$ \\
\hline
$7$ & $1.01753783$ & $6.9\times10^{-6}$ & $+5.8\times10^{-7}$ & $+1.98\times10^{-4}$ & $1.23\times10^{-4}$ & $+1.61$ \\
\thickhline
\end{tabular}
\end{table}

\subsubsection*{Estimator configurations}

Table~\ref{tab:exposure_config} gives the tested settings. At total inner budget
$B$, nested Monte Carlo uses $N_{\rm in}=\max\{4,\operatorname{round}(aB^{1/3})\}$
paths at each of $N_{\rm out}=\lfloor B/N_{\rm in}\rfloor$ states, varying the split
multiplier $a$. The cube-root scale balances bias and variance for the plain
quantile estimator~\cite{GordyJuneja2010}; the comparison selects the lowest measured
RMSE across the tested splits and the plain and jackknifed estimates at each budget.

\begin{table}[t]
\caption{\textbf{Exposure estimator configurations.} Month-$18$ PFE under GBM at
$97.5\%$, three to seven assets. Settings are varied within each method; the
comparison uses their measured error-cost envelopes. Replications are independent.}
\label{tab:exposure_config}
\centering
\renewcommand{\arraystretch}{1.15}
\begin{tabular}{?l|l|l?}
\thickhline
\parbox[t]{0.18\linewidth}{method} & \parbox[t]{0.43\linewidth}{tuning} &
\parbox[t]{0.29\linewidth}{budgets and replications} \\
\hline
\parbox[t]{0.18\linewidth}{Nested Monte Carlo} &
\parbox[t]{0.43\linewidth}{Split multipliers $1,3,6,10,20,40$, supplemented by
$1/4$ and $1/2$ near the crossover. At the deepest budgets only the best few
splits are tested. Plain and jackknife outputs share paths.} &
\parbox[t]{0.29\linewidth}{$10^5$--$10^8$ inner samples, extended to $3\times10^8$ at
$d=3,4$; $18$--$54$ replications per point.} \\
\hline
\parbox[t]{0.18\linewidth}{Multilevel Monte Carlo} &
\parbox[t]{0.43\linewidth}{Finest level $L=2$--$7$. Pilot: $128$ inner paths at
$8{,}192$ states, enlarged to $512$ at $d\ge4$ or $1{,}024$ at $d\ge5$, both
at $40{,}000$ states.} &
\parbox[t]{0.29\linewidth}{$M_0=500$--$10^6$ outer states on level zero;
$8$--$9$ replications per configuration.} \\
\hline
\parbox[t]{0.18\linewidth}{Regression Monte Carlo} &
\parbox[t]{0.43\linewidth}{Total polynomial degree $4$--$12$, evaluated at
$262{,}144$ fresh outer states.} &
\parbox[t]{0.29\linewidth}{$3.2\times10^4$--$2\times10^6$ regression paths;
$16$ replications at $d=3$--$5$, $4$ at $d=6,7$.} \\
\thickhline
\end{tabular}
\end{table}

Inner noise biases the plain quantile upward here. The Gordy--Juneja half-sample
jackknife cancels its leading $1/N_{\rm in}$ term without new paths: if $q_n$ is
the quantile from $n$ inner paths per state and $q_{n/2}^{(1)},q_{n/2}^{(2)}$ use
the two halves, the corrected estimate is
$2q_n-\tfrac12(q_{n/2}^{(1)}+q_{n/2}^{(2)})$, with size-adjusted weights for unequal
halves. A residual of order $1/n^2$ requires a higher-order smoothness expansion;
the original bound is $O(n^{-3/2})$. The correction
reduces measured RMSE by roughly a factor of two at the tensor's cost.

The multilevel estimator~\cite{Giles2008} uses $32\,2^\ell$ inner paths at level
$\ell=0,\ldots,L$. Above level zero, each correction subtracts the mean of the
two coarse half-sample CDF indicators from the fine indicator. Outer counts are
$M_\ell=\max\{300,\operatorname{round}(M_0\,2^{-0.725\ell})\}$, varying $M_0$ and
the finest level. A separate nested pilot estimates a quantile $q_0$ and a local
density $f_0$ from order-statistic spacing over a $\pm2\%$ probability window.
Its cost is included. One Newton step gives $q_0+(\alpha-\widehat F(q_0))/f_0$
from the multilevel CDF. Inner budgets are fixed
within a level, unlike the adaptive method in Ref.~\cite{GilesHajiAli2019}.
Increasing levels alone does not remove pilot-induced bias, so pilot size is
also varied.

Regression Monte Carlo~\cite{LongstaffSchwartz2001} fits the continuation value,
not the discontinuous event-mapped value. It uses ordinary least squares on
products of probabilists' Hermite polynomials of bounded total degree, with each
log-spot standardized by its training-sample mean and standard deviation. Each
training state contributes one simulated continuation payoff. The coupon and
call events at month $18$ are then applied exactly, and the fitted value is
evaluated on fresh outer states. Polynomial degree and training budget control
approximation bias and sampling error separately.

Tensor surfaces use tolerances $10^{-3}$ to $10^{-6}$ in half-decades and
$512$ to $131{,}072$ randomized Sobol states, with matched outer points across
tolerances and $16$--$64$ independent replications. Table~\ref{tab:tail_ladder}
shows how rounding changes build cost and quantile bias. The reused-surface
comparison fixes tolerance $10^{-6}$ and uses $8{,}192$--$131{,}072$ outer states
with $24$--$48$ replications.

Surface builds and outer evaluations are charged separately. Target costs and
equal-compute errors use log-log interpolation between measured points.
Targets looser than the error of the smallest run are charged that run;
unattained targets give lower cost bounds, not extrapolated costs.
Untimed nested splits use the measured per-path cost at the same budget;
recalibration changes equal-compute margins by at most $12\%$.

\begin{table}[H]
\caption{\textbf{The tolerance ladder of the tail surface.} Build of the
month-$18$ surface in core-seconds, its largest rank, and the
bias of the quantile it gives, in $\%$ of $\mathrm{PFE}-\mathrm{EE}$, from $16$
replications of $131{,}072$ randomized Sobol points (standard error $0.002$ to
$0.007\%$).}
\label{tab:tail_ladder}
\centering
\renewcommand{\arraystretch}{1.2}
\begin{tabular}{?c|ccc|ccc|ccc?}
\thickhline
& \multicolumn{3}{c|}{$d=3$} & \multicolumn{3}{c|}{$d=4$} & \multicolumn{3}{c?}{$d=5$} \\
tolerance & build & rank & bias & build & rank & bias & build & rank & bias \\
\hline
$10^{-3}$ & $1.05$ & $10$ & $+5.11$ & $2.09$ & $12$ & $+1.00$ & $3.09$ & $13$ & $+0.49$ \\
$3\times10^{-4}$ & $1.50$ & $15$ & $+0.37$ & $2.81$ & $17$ & $+0.50$ & $4.30$ & $18$ & $-0.02$ \\
$10^{-4}$ & $1.85$ & $18$ & $-0.21$ & $3.56$ & $20$ & $-0.21$ & $5.30$ & $21$ & $-0.19$ \\
$3\times10^{-5}$ & $2.35$ & $21$ & $0.00$ & $4.63$ & $22$ & $+0.02$ & $6.90$ & $23$ & $-0.02$ \\
$10^{-5}$ & $2.91$ & $24$ & $+0.02$ & $5.71$ & $25$ & $0.00$ & $8.51$ & $26$ & $-0.01$ \\
$3\times10^{-6}$ & $3.59$ & $26$ & $0.00$ & $6.64$ & $28$ & $0.00$ & $10.55$ & $31$ & $-0.01$ \\
$10^{-6}$ & $3.87$ & $28$ & $0.00$ & $8.31$ & $34$ & $0.00$ & $14.79$ & $40$ & $-0.01$ \\
\thickhline
\end{tabular}
\end{table}

\subsection{Exposure tails: comparisons and controls}
\label{app:exposure_cmp}

\subsubsection*{Standalone and reused surfaces}

The surface build sets the loose-accuracy disadvantage in Fig.~\ref{fig:exposure}(a).
At five assets even the coarsest build costs about three core-seconds, while nested
Monte Carlo reaches $1\%$ of $\mathrm{PFE}-\mathrm{EE}$ in about $0.2$. Reducing rank
does not eliminate the backward sweep from maturity to month $18$. Once the
surface exists, however, more outer states improve the quantile without inner
simulation. The build is then spread over cheap evaluations, and the error falls
rapidly with little additional compute until surface error limits the accuracy.

Table~\ref{tab:tail_target_costs} summarizes the target comparisons in
Fig.~\ref{fig:exposure}(b). The standalone crossover is near $0.3\%$ of the gap.
At three assets, regression sets the Monte Carlo cost at this target; nested
Monte Carlo is the cheaper baseline at higher dimensions. The lower bound at
three assets and $0.1\%$ reflects the measured range of the other baselines,
not an extrapolation of their error curves.

\begin{table}[H]
\caption{\textbf{Which method reaches an exposure target more cheaply.} Each entry
names the cheaper method and its cost advantage as a factor. Targets are RMSE in
$\%$ of $\mathrm{PFE}-\mathrm{EE}$; tensor costs include the surface build and outer
sampling. The last column gives the target at which costs cross, in the same units.}
\label{tab:tail_target_costs}
\centering
\renewcommand{\arraystretch}{1.2}
\begin{tabular}{?l|ccc|c?}
\thickhline
 & $1\%$ & $0.3\%$ & $0.1\%$ & crossover ($\%$) \\
\hline
$d=3$ & MC, $10$ & hp-FTT, $1.1$ & hp-FTT, $\ge11$ & $0.31$ \\
$d=4$ & MC, $13$ & MC, $1.9$ & hp-FTT, $6.0$ & $0.24$ \\
$d=5$ & MC, $15$ & MC, $1.1$ & hp-FTT, $4.3$ & $0.29$ \\
\thickhline
\end{tabular}
\end{table}

The reference selects the tensor tolerance, as in the Greek comparison. An
empirical check with a second build half a decade tighter would add $1.2$ to
$1.4$ times the first build's cost. When the surface is already available from a
Greek solve, that investment is shared and only outer sampling is charged.
At three to five assets this makes the tensor at least an order of magnitude
cheaper at $0.3\%$ and about seventy times or more at $0.1\%$. This reuse applies
only to the same contract and model parameters. Regression could also reuse
simulated paths across exposure dates; a date profile is not benchmarked here.

\subsubsection*{Dimensional and accuracy scaling}

Additional assets raise the build cost more rapidly than the cost of nested
sampling. The build is about four fifths of total tensor cost at its best measured
points and grows ninefold from three to seven assets. At those points, beyond
the tightest target, the equal-compute accuracy advantage falls from about
seventeenfold at three assets to about threefold at six and seven. At the scored
targets the crossover pattern nevertheless persists: at six and seven assets,
Monte Carlo is about twenty times cheaper at $1\%$ of the gap, while the tensor
is about $3.5$ times cheaper at $0.1\%$, its build included. Reusing the surface
raises the latter advantage to about thirtyfold.

Accuracy has a stronger effect than dimension over this range. Nested error
falls as cost$^{-0.40}$ to cost$^{-0.47}$, close to the cost$^{-2/5}$ balance
after the jackknife removes the leading bias. Tightening the target by a decade
therefore raises nested cost by more than a hundredfold, whereas a tighter
surface and more outer evaluations raise tensor cost only a fewfold. At much
tighter targets the outer sampling, rather than the build, would dominate.
Regression avoids inner sampling but retains basis error: from five assets up,
its bias is comparable to or larger than its sampling spread at its best measured
point. More training paths alone do not remove this approximation limit.

\subsubsection*{Sampling and scoring controls}

Two controls separate the gain from the stored surface from that of randomized
Sobol sampling. Giving the nested estimator the same outer points, with
pseudo-random inner paths, changes its best RMSE by a factor of $0.81$--$1.24$
at the two budgets bracketing tensor cost, over $108$ replications per estimator.
Conversely, using pseudo-random outer states on the tensor surface still gives
roughly two to six times lower RMSE than the best baseline at equal compute,
at three to seven assets. Randomized points improve the surface method but do
not create its advantage.

Shifting the reference by one standard error changes equal-compute margins by
at most $7\%$. Selecting the nested split on half the replications and scoring
on the remainder raises those margins by up to $16\%$; using the same
replications for selection and scoring therefore favors the baseline.

\end{document}